\documentclass[aps,prd,longbibliography,reprint,twocolumn,superscriptaddress,
nofootinbib]{revtex4-2}

\usepackage{graphicx}
\usepackage{dcolumn}
\usepackage{enumitem}
\usepackage{latexsym,bm,hyperref,amsmath,amssymb,hyperref,physics}

\usepackage{float}
\floatstyle{plaintop}
\restylefloat{table}

\begin{document}

\title{Quasinormal mode frequencies of Kerr black holes from Regge trajectories}

\author{Antoine Folacci}%
 \email{folacci@univ-corse.fr}
\affiliation{Equipe Physique Th{\'e}orique,\\
SPE, UMR 6134 du CNRS et de l’Universit{\'e} de Corse,\\ Universit{\'e} de Corse, Facult{\'e} des Sciences, \\
BP 52, F-20250 Corte, France}

\author{Aditya Tamar}
\email{adityatamar@gmail.com}
 \affiliation{Independent Researcher, Delhi, India}

\date{\today}

\begin{abstract}

The response of a Kerr black hole to an external excitation is dominated, at intermediate time scale, by a characteristic damped ringing involving the complex frequencies of its quasinormal modes. The associated spectrum encodes the mass and spin of the black hole and can be considered as its fingerprint. A large portion of the studies concerning the quasinormal mode frequencies of a Kerr black hole have focused only on achieving higher numerical accuracy with limited emphasis on providing their physical interpretation. In this article, we partially address this issue by computing the quasinormal mode frequency spectrum of a Kerr black hole using the theory of Regge poles. By considering the retarded Green's function of the Teukolsky equation, we establish for scalar, electromagnetic and gravitational perturbations an equation linking the Regge poles to the quasinormal frequencies and we solve it in the high-frequency regime to get ``semiclassical'' relations permitting us to obtain the complex frequencies of the weakly damped quasinormal modes from the Regge trajectories. Numerical results concerning gravitational perturbations ($s=-2$) are displayed. They are in excellent agreement with the ``exact'' ones in the eikonal regime $(\ell \gg 1)$ and in very good agreement even for lower values of $\ell$. Moreover, the splitting of each Regge pole of the Schwarzschild black hole into an infinite number of Kerr Regge poles explains the breaking of the azimuthal degeneracy of the quasinormal frequencies of the Schwarzschild black hole due to rotation. Our work is a first step to extend to Kerr black holes the approach developed for static spherically symmetric black holes which allowed, from a geometrical interpretation of the Regge poles in terms of the properties of the unstable circular null geodesics lying on the photon sphere, to derive accurate analytical formulas for the Regge trajectories and, as a by-product, for the complex frequencies of the weakly damped quasinormal modes.

\end{abstract}

\maketitle

\section{\label{sec:Introduction} Introduction}

\subsection{Prologue}

Rotating uncharged black holes (BHs) are probably ubiquitous in the Universe and, under the near certain assumption that Einstein's general theory of relativity is the correct theory of gravitation, we must consider they are described by the Kerr metric \cite{Kerr:1963ud,Chandrasekhar:1985kt,BarrettONiell1995,Teukolsky:2014vca}. As a result, most analyses carried out to interpret the recent observations of binary BH mergers by gravitational wave detectors (see Ref.~\cite{Abbott:2016blz} for the first event detected) as well as the first images of the shadow of a supermassive BH produced by the Event Horizon Telescope \cite{Akiyama:2019cqa} involve this particular Ricci-flat solution of Einstein's equations.

The Kerr BH, like all other BHs, is a resonant object and its vibrational modes, the so-called quasinormal modes (QNMs), encode information about its mass and spin independent of any other astrophysical considerations. From the physical point of view, the gravitational QNMs of the Kerr BH permit one to describe theoretically the ringdown phase occurring after the merger of a binary BH, a phase associated with the relaxation of the highly distorted final BH which has now been observed \cite{Abbott:2016blz}. Moreover, it seems there exists a connection between the electromagnetic QNMs of the Kerr BH and its shadow radius \cite{Jusufi:2020dhz} and it is interesting to note that these electromagnetic resonant modes might even be observed with advanced versions of the Event Horizon Telescope \cite{Chesler:2020gtw}. Indeed, from a theoretical point of view, they generate echoes in the coherent temporal autocorrelation function of the electric field measured. Therefore, there is no doubt that QNMs will be subjected to continued study in the near future because of their importance in interpreting observations but also due to their interest to test possible extensions of general relativity and, more generally, to probe fundamental physics with the next generations of gravitational wave detectors (see, e.g., Refs.~\cite{Nakamura:2016gri,Chirenti:2017mwe,Berti:2018vdi,Barack:2018yly,Baibhav:2018rfk,Baibhav:2019rsa,
Giesler:2019uxc,Isi:2019aib,Bhagwat:2019dtm,Barausse:2020rsu}) and interferometric arrays (see, e.g., Refs.~\cite{Chesler:2020gtw,Johnson:2019ljv,Gralla:2020pra}).

QNMs are solutions of the equations governing the linearized perturbations of a BH background with boundary conditions of being purely ingoing at the event horizon and purely outgoing at spatial infinity. They are characterized by complex frequencies, the so-called quasinormal frequencies, and are exponentially decaying in time. These solutions were first discovered by Vishveshwara \cite{Vishveshwara:1970zz,Vishveshwara:1970cc} and there exist excellent reviews of the subject \cite{Kokkotas:1999bd,Nollert:1999ji,Berti:2009kk,Konoplya:2011qq}. Over the years several methods have been developed to construct QNMs and compute quasinormal frequencies, most of which are given in Refs.~\cite{Kokkotas:1999bd,Nollert:1999ji,Berti:2009kk,Konoplya:2011qq}. An extremely accurate and numerically efficient method was given by Leaver \cite{Leaver:1985ax} (the so-called Leaver's method). It is based on techniques used for the study of the electronic spectrum of the hydrogen molecular ion (see Ref.~\cite{LeaverJMP1986} and references therein), wherein the wave equations are analytically solved as generalized spheroidal wave equations, providing a solution for QNM frequencies in terms of continued fractions. It was further improved by Nollert \cite{Nollert:1993zz}, Onazawa and collaborators \cite{Onozawa:1995vu,Onozawa:1996ux} and many others. Thus, the underlying numerical accuracy of the data has been well studied and established. For the Kerr BH, QNM data and Mathematica notebooks based on Leaver's method are even available \cite{Berti:2009kk,RingdownBerti}. However, as has been noted in various papers (see, e.g., Ref.~\cite{Kokkotas:1999bd}), Leaver's method offers very little physical insight.

\subsection{Physical motivations and Regge theory}
In this regard, almost immediately after the publication of the first papers dealing with QNMs,
an appealing physical interpretation for the QNMs of a Schwarzschild BH of
mass $M$ was proposed by Goebel \cite{Goebel1972}: He suggested they are
generated by gravitational waves in spiral orbits close
to the unstable circular photon orbit at $r=3M$ (the so-called light ring),
i.e., close to the Schwarzschild photon sphere, which decay by radiating away energy.
Over the following years, various implementations of the Goebel
interpretation where null circular unstable geodesics play a crucial role have been developed (see, e.g.,
Refs.~\cite{Ferrari:1984ozr,Ferrari:1984zz,Mashhoon:1985cya,Stewart89,
Andersson:1996xw,Vanzo:2004fy,Cardoso:2008bp,Dolan:2009nk,Hod:2009td,
Yang:2012he,Glampedakis:2017dvb}). They concern more general field theories and BH backgrounds and are mainly based on the concepts of geodesic and bundle of
geometrical rays. They allowed to obtain analytical approximations
for the leading-order terms of the complex
frequencies of the weakly damped QNMs from an interpretation in terms of
massless particles ``trapped'' near unstable circular null geodesics, approximations which are valid in the eikonal regime ($\ell \gg 1$). In this limit, the real part of the quasinormal frequencies is associated with the light ring angular frequency
and the imaginary part with the Lyapunov exponent which measures the rate of divergence of the bundle of null geodesics (see, e.g., Ref.~\cite{Cardoso:2008bp}).
In fact, in the particular case of the Kerr BH, it seems to us that the relation between null unstable geodesics and quasinormal frequencies is not as obvious as for the Schwarzschild BH or for static spherically symmetric BHs and cannot be considered as well-established. Indeed, in that case, the calculations existing in the literature and dealing with a geometrical interpretation of the quasinormal frequencies (see, e.g., Refs.~\cite{Ferrari:1984ozr,Ferrari:1984zz,Dolan:2010wr,Yang:2012he})
have been obtained under too restrictive approximations, such as ignoring the spin of the field, or lead to dubious numerical results for astrophysically relevant situations.

An alternative and much richer implementation of the Goebel
interpretation of BH QNMs has also been formulated
\cite{Decanini:2002ha,Decanini:2009dn,Decanini:2009mu,Decanini:2010fz,Decanini:2011eh}
(see also Ref.~\cite{Dolan:2009nk}). It is not limited to purely geometrical
considerations but is based on wave theory (see also Ref.~\cite{Hod:2009td} for a wave analysis of QNMs from a different point of view) and goes beyond leading-order terms. In some sense, it provides a ``rigorous'' basis to discuss the physically
intuitive interpretation of Goebel. It uses complex angular
momentum (CAM) techniques and the Regge pole machinery, both dating back to the work of Poincar\'e \cite{Poincare1910}, Watson \cite{Watson1918} and Sommerfeld \cite{Sommerfeld1949}. These techniques also played a central role in resonant scattering theory \cite{Poincare1910,Watson1918,Sommerfeld1949,deAlfaro:1965zz,Newton:1982qc,Nussen1992,Grandy2000} as well as in high-energy physics \cite{Collins:1977jy,Barone:2002cv,Donnachie:2002en,Gribov:2003nw}. It is interesting to recall that the success of CAM techniques is due to their ability to provide effective resummations of partial wave sums and to extract from them the hidden information, which leads to a clear description of physical processes. In the context of BH physics, these techniques have permitted to make significant contributions in scattering dynamics \cite{Andersson:1994rk,Andersson:1994rm,Decanini:2011xi,Macedo:2013afa,Folacci:2019cmc,
Folacci:2019vtt,OuldElHadj:2019kji} and to revisit from an alternative point of view radiation phenomenons \cite{Folacci:2018sef,Folacci:2020ekl}.

In this article, we make the first step with the intention of partially extending to the Kerr BH the Regge pole analysis of QNMs which was developed for static symmetrically spherical BHs\footnote{It is worth noting that, some years ago, Glampedakis and Andersson have considered the Regge poles of the Kerr BH for scalar perturbations and developed a ``quick and dirty'' method to compute them numerically \cite{Glampedakis:2003dn}. Their purpose was actually different from ours. We will come back to their results in Sec.~\ref{RTQNF_num}.}. Let us recall that, in this context, it was possible to establish that the complex frequencies of the weakly damped QNMs of the Schwarzschild BH of
mass $M$ are Breit-Wigner-type resonances generated by a family of
``surface waves''\footnote{Formally, the term surface wave which is borrowed from the literature concerning resonant scattering in optics and in acoustics is in fact inadequate for BHs. Indeed, unlike Regge modes in optics and in acoustics, the Regge modes of BHs are not, in general, exponentially decaying transversally to their direction of propagation, i.e., transversally to the photon sphere.} propagating on/close to its photon sphere at $r=3M$ and exponentially damped in the direction of propagation
\cite{Decanini:2002ha}. This was carried out by noting that each
surface wave can be associated with a Regge pole of the ${\cal S}$-matrix of
the Schwarzschild BH
\cite{Andersson:1994rk,Andersson:1994rm,Decanini:2002ha} and allowed to
construct numerically from the Regge trajectories, i.e.,
from the curves traced out in the CAM plane by the Regge poles as a function of the frequency $\omega \in \mathbb{R}$, the frequency spectrum of the weakly damped QNMs. This work was later generalized to derive ``simple'' but very accurate analytical expressions for the Regge poles (i) of the Schwarzschild BH taking moreover into account the spin dependence of the fields considered \cite{Decanini:2009mu} and (ii) of static spherically symmetric BHs of arbitrary dimension with a photon sphere \cite{Decanini:2010fz}. This was achieved by adapting to the determination of the Regge poles the WKB techniques developed in the context of the
determination of the QNMs by Schutz and Will and generalized by various authors \cite{Schutz:1985km,Iyer:1986np,Iyer:1986nq,Will:1988zz,Seidel:1989bp,Konoplya:2003ii}. It should be noted that the Regge trajectories are very accurately described by the analytical expressions obtained, can be interpreted in terms of dispersion relation and damping of the surface waves lying close to the photon sphere and that, from Bohr-Sommerfeld--type resonance conditions, they provide analytical formulas
beyond the leading-order terms for the complex frequencies
of the weakly damped QNMs.

\subsection{Layout of the paper}
Our paper is organized as follows. In Sec.~\ref{RGF}, we consider the retarded Green's function associated with a linear perturbation of a Kerr BH (that encompasses scalar, electromagnetic and gravitational perturbations) and we provide its expression by using the Teukolsky formalism \cite{Teukolsky:1972my,Teukolsky:1973ha}. It is important to recall that this two-point function is of fundamental interest because it permits one to construct theoretically, by convoluting it with the source of the perturbation, the time response of an excited BH. In Sec.~\ref{RGFapprox}, from the expression of the retarded Green's function we extract two useful approximations: (i) the so-called QNM approximation which involves the QNMs and their complex frequencies and (ii) the so-called Regge pole approximation which is obtained by using the CAM machinery and involves the Regge modes and the Regge poles. It is worth pointing out that the former permits one to identify, in the time-response, the ringdown of the BH. The status of the latter is less clear. This is due to the fact that it has been the subject of relatively few studies. However, for the Schwarzschild BH, we know that it provides an effective resummation of the partial wave expansion defining the retarded Green's function which can reproduce with very good agreement the quasinormal ringdown without requiring a starting time, as well as, in certain circumstances, the tail of the signal and the pre-ringdown phase with rather good agreement \cite{Folacci:2018sef,Folacci:2020ekl}. In Sec.~\ref{RTQNF_thnum}, we first establish an equation linking the Regge poles and the quasinormal frequencies and we then solve it in the high-frequency regime to get ``semiclassical'' relations permitting us to obtain the complex frequencies of the weakly damped QNMs from the Regge trajectories. We finally describe the methodology used for obtaining numerically the Regge poles and the Regge trajectories. In Sec.~\ref{RTQNF_res}, we provide a series of numerical results in the case of gravitational perturbations. (Numerical results concerning scalar and electromagnetic perturbations are available upon request from the authors. They are not displayed in this article because they can be simply obtained by changing the value of the field spin parameter $s$ in our method, which can be done without any difficulties.) In particular, we exhibit the splitting, due to rotation, of each Regge pole of the Schwarzschild BH into an infinite number of Kerr Regge poles and we compare the quasinormal frequency spectrum of the Kerr BH obtained semiclassically with the ``exact'' one. In the Conclusion, we briefly discuss potential avenues for future researches and, more particularly, the possibility to interpret geometrically the Kerr Regge poles in order to derive simple analytical formulas allowing us to describe the associated Regge trajectories as well as the complex frequencies of the weakly damped QNMs.

\subsection{Notations}
Throughout this article, we adopt units such that $G = c = 1$ and we consider Kerr spacetime in Boyer-Lindquist coordinates $(t,r,\theta,\varphi )$ for which the line element takes the form  \cite{Boyer:1966qh,Misner:1974qy}
\begin{eqnarray} \label{Kerr_metric}
& & ds^2 = -\left(1-\frac{2Mr}{\Sigma}\right) dt^2 - \frac{4Ma \, r \sin^2 \theta}{\Sigma} \, dt d\varphi + \frac{\Sigma}{\Delta} \, dr^2 \nonumber \\
& & \quad  + \Sigma \, d\theta^2 + \left[(r^2+a^2) + \frac{2Ma^2 r \sin^2 \theta}{\Sigma} \right] \sin^2 \theta \, d\varphi^2
\end{eqnarray}
where $\Delta = r^2 - 2 M r + a^2$ and $\Sigma = r^2 + a^2 \cos^2\theta$. Here $M$ and $a$ are respectively the mass of the BH and its angular momentum per unit mass and we assume $0 \le a < M $. The outer event horizon is located at $r_{+} = M + \sqrt{M^2 - a^2}$, the largest root of $\Delta$, while its second root $r_{-} = M - \sqrt{M^2 - a^2}$ defines the inner Cauchy horizon. Of course, when $a=0$, the metric (\ref{Kerr_metric}) reduces to the Schwarzschild metric. In the following, we shall also use the so-called tortoise coordinate $r_\ast$ defined from the
radial coordinate $r$ by $dr_\ast/dr=(r^2 + a^2)/\Delta$ which permits one to move the position of the event horizon from $r = r_{+}$ to  $r_\ast = - \infty$, provides a bijection from $r \in ]r_{+},+\infty[$ to $r_\ast \in ]-\infty,+\infty[$ and is given by
\begin{equation}\label{Kerr_TortoiseCoord}
r_\ast =  r+ \frac{2Mr_{+}}{r_{+} - r_{-}}\ln \left(\frac{r-r_{+}}{2M}\right) - \frac{2Mr_{-}}{r_{+}-r_{-}}\ln \left(\frac{r-r_{-}}{2M}\right).
\end{equation}
In this article, we only consider the exterior of the Kerr BH and a spacetime point $x$ is therefore identified by its coordinates $(t,r,\theta,\varphi)$ with $r>r_{+}$ or $(t,r_\ast,\theta,\varphi)$.

\section{The retarded Green's function for the Teukolsky equation}
\label{RGF}

In this section, we review the spectral decomposition of the retarded Green's function associated with the linear perturbations of the Kerr BH.

We recall that the study of the linear perturbations of the Kerr BH leads to focus on the behavior of a massless field $_{s}\psi$ (here $s$ is the spin weight of the field which takes the values $s = 0, \pm 1, \pm 2$ for scalar, electromagnetic and gravitational perturbations respectively) governed by the Teukolsky master equation \cite{Teukolsky:1972my,Teukolsky:1973ha}, a gauge invariant equation \cite{GaugeInv}, which is obtained by making use of the Newman-Penrose formalism \cite{Newman:1961qr}. It should be noted that, even if neutrino field perturbations also satisfy the Teukolsky master equation for $s=\pm 1/2$, we do not consider this case in our article in order to avoid some complications in the notations as well as difficulties linked to fermionic fields. By working with Boyer-Lindquist coordinates and in the Kinnersley tetrad \cite{Kinnersley:1969zza}, the Teukolsky equation is written as
\begin{equation} \label{eq:TeukolskyEqn}
\mathcal{L}_x^{(s)}   \, _{s}\psi(x)=  \Sigma \, _{s}T (x)
\end{equation}
where $\mathcal{L}_x^{(s)} $ is the Teukolsky differential operator given by
\begin{eqnarray}\label{eq:TeukolskyEqnOp}
& & \mathcal{L}_x^{(s)} = \bigg[ \frac{(r^2 + a^2)^2}{ \Delta} - a^2 \sin^2 \theta \bigg] \frac{ \partial^2 }{\partial t^2} +\bigg(\frac{4Mar}{\Delta}\bigg) \frac{\partial^2 }{\partial t \partial \varphi}  \nonumber \\
&& +
 \bigg[ \frac{a^2}{\Delta} - \frac{1}{\sin^2 \theta} \bigg] \frac{\partial^2  }{\partial \varphi^2}  - \Delta^{-s} \frac{\partial}{\partial r} \bigg( \Delta^{s+1} \frac{\partial }{\partial r} \bigg) \nonumber \\
 && - \frac{1}{\sin \theta} \frac{\partial}{\partial \theta}  \bigg( \sin \theta \frac{ \partial }{\partial \theta} \bigg)  -2s \bigg[ \frac{a(r-M)}{\Delta} + \frac{i \cos \theta}{\sin^2 \theta} \bigg] \frac{\partial }{\partial \varphi} \nonumber \\
 & & -2s \bigg[ \frac{M(r^2 - a^2)}{\Delta} - r - ia\cos \theta \bigg] \frac{\partial }{\partial t} + (s^2 \cot^2 \theta - s) \nonumber \\
 & &
\end{eqnarray}
and where $_{s}T(x)$ is a source term built from the energy-momentum tensor \cite{Teukolsky:1972my,Teukolsky:1973ha}. It should be noted that the Teukolsky equation (\ref{eq:TeukolskyEqn}) reduces to the Bardeen-Press equation (see Ref.~\cite{Bardeen:1973xb}) in the non-rotating $(a = 0)$ case.

The Teukolsky equation with source (\ref{eq:TeukolskyEqn}) can be solved by using the machinery of Green’s functions \cite{Chrzanowski:1974nr,Chrzanowski:1975wv}. We consider the retarded Green's function ${_{s}G}_\mathrm{ret}(x,x')$ solution of
 \begin{equation} \label{eq:TeukolskyEqnGreenF}
\mathcal{L}_x^{(s)}  \, {_{s}G}_\mathrm{ret}(x,x') =  \Sigma \, \delta^{(4)}(x,x')
\end{equation}
for any two spacetime points $x$ and $x'$ that vanishes if $x'$ lies outside the causal future of $x$ (in Boyer-Lindquist coordinates, if $t<t'$). Here $\delta^{(4)} (x,x') = \delta (x-x')/\sqrt{-g(x)}$ is the invariant Dirac delta distribution, $g(x)=-\Sigma^{2}\sin^{2}\theta$ is the metric determinant for the Kerr BH and we have therefore $\Sigma \, \delta^{(4)}(x,x') = \delta (t-t')\delta (r-r')\delta (\cos \theta- \cos \theta')\delta (\varphi-\varphi')$.

The Teukolsky formalism \cite{Teukolsky:1972my,Teukolsky:1973ha} permits one to construct the solution of (\ref{eq:TeukolskyEqnGreenF}) by separation of variables. We can write \cite{Casals:2016soq,Casals:2019vdb}
\begin{eqnarray}\label{eq:GreenFexp}
&& {_{s}G}(x,x') = -\frac{(\Delta')^s}{2\pi}\sum_{\ell=|s|}^{\infty}\sum_{m=-\ell}^{+\ell}
\int_{-\infty+ic}^{+\infty+ic}d\omega \, e^{-i\omega t + im\varphi} \nonumber \\
& & \qquad\quad \times  {_{s}G}_{\ell m}(r,r';\omega) \, {_{s}S}_{\ell m}(\theta,a\omega) {_{s}{S}}^\ast_{lm}(\theta ',a\omega)
\end{eqnarray}
with $c>0$. Here, $\Delta'= r'^2 - 2 M r' + a^2$ while the functions ${_{s}G}_{\ell m}(r,r';\omega)$ are the Fourier modes of the retarded Green's function arising from the radial part of the Teukolsky equation (see below) and the functions ${_{s}S}_{\ell m}(\theta,a\omega)$ are the usual spin-weighted spheroidal harmonics (see, e.g., Ref.~\cite{Fackerell1977}). It should be noted that, in Eq.~(\ref{eq:GreenFexp}), we have set $t'=0$ and $\varphi ' =0$, an assignment which does not affect the computations due to stationary and axisymmetric nature of Kerr spacetime. We recall that the spin-weighted spheroidal harmonics ${_{s}S}_{\ell m}(\theta,a\omega)$ are the solutions of the Teukolsky angular equation
\begin{eqnarray} \label{eq:TAE}
&& \bigg[ \frac{1}{\sin \theta} \frac{d}{d\theta}\left(\sin \theta \frac{d}{d\theta} \right) + \left( a^2 \omega^2 \cos^2 \theta - 2 s a\omega \cos \theta  \phantom{\frac{(m)^2}{\sin^2 \theta}} \right. \nonumber \\
&& \quad  \left. - \frac{(m + s \cos \theta)^2}{\sin^2 \theta}+ s  + {_{s}A}_{\ell m}(a\omega)
 \right)\bigg] {_{s}S}_{\ell m}(\theta,a\omega) = 0 \nonumber \\
 & &
\end{eqnarray}
which are regular at $\theta =0$ and $\theta =\pi $ and which satisfy the normalization condition
\begin{equation} \label{eq:TAE_norm}
\int_{0}^{\pi} d\theta \sin  \theta \, \vert {_{s}S}_{\ell m}(\theta,a\omega) \vert^2 =1.
\end{equation}
In Eq.~(\ref{eq:TAE}), ${_{s}A}_{\ell m}(a\omega)$ are angular separation constants which can be also considered as the eigenvalues of the differential operator involved in Eq.~(\ref{eq:TAE}) associated with the eigenfunctions ${_{s}S}_{\ell m}(\theta,a\omega)$. It is important to recall that, in Ref.~\cite{Leaver:1985ax}, Leaver provided an algorithm for the numerical determination of the spin-weighted spheroidal harmonics ${_{s}S}_{\ell m}(\theta,a\omega)$ and of the angular separation constants ${_{s}A}_{\ell m}(a\omega)$. It is also important to note that, for $a \omega \ll 1$, we have ${_{s}A}_{\ell m}(a\omega) \approx \ell (\ell +1) - s (s+1)$ and that, extending a previous work by Fackerell and Crossman \cite{Fackerell1977} (see also Ref.~\cite{Breuer1977}), Seidel obtained the first seven terms of a series expansion of ${_{s}A}_{\ell m}(a\omega)$ in powers of $a\omega$ \cite{Seidel:1988ue}. This last result will be used in Sec.~\ref{RTQNF_num} and Sec.~\ref{RTQNF_res}.

As far as the functions ${_{s}G}_{\ell m}(r,r';\omega)$ are concerned, they are in fact Green's functions of the Teukolsky radial equation with a delta source, i.e., they satisfy
\begin{eqnarray} \label{eq:TREdelta}
& & \left[  \Delta^{-s} \frac{d}{dr} \left( \Delta^{s+1} \frac{d}{dr} \right) + \frac{K^2 -2is(r-M)K}{\Delta} +  4is \omega r \right.  \nonumber \\
& &  \left.  \phantom{\left(\frac{d}{dr}\right)} - a^2\omega^2 + 2m a\omega -  {_{s}}A_{\ell m}(a\omega)   \right] \, {_{s}G}_{\ell m}(r,r';\omega) \nonumber \\
&&  \qquad \qquad \qquad \qquad \qquad \qquad \qquad \qquad = -\delta (r-r')
\end{eqnarray}
where
\begin{equation}\label{eq:TRE_K}
K \equiv (r^2 + a^2)\omega -ma.
\end{equation}
They can be expressed in the form \cite{Casals:2016soq,Casals:2019vdb}
\begin{equation}\label{eq:TRE_solret}
{_{s}G}_{\ell m}(r,r';\omega) = \frac{{_{s}R}^\mathrm {in}_{\ell m}(r_<,\omega) \,  {_{s}R}^\mathrm {up}_{\ell m}(r_>,\omega)}{{_{s}W}_{\ell m}(\omega)}
\end{equation}
[here $r_< \equiv \mathrm{min} (r,r')$ and $r_> \equiv \mathrm{max} (r,r')$] where ${_{s}R}^\mathrm {in}_{\ell m}(r,\omega)$ and ${_{s}R}^\mathrm {up}_{\ell m}(r,\omega)$ are linearly independent solutions of the Teukolsky radial equation, i.e., of the homogeneous version of (\ref{eq:TREdelta}), given by
 \begin{eqnarray} \label{eq:TREhom}
& & \left[  \Delta^{-s} \frac{d}{dr} \left( \Delta^{s+1} \frac{d}{dr} \right) + \frac{K^2 -2is(r-M)K}{\Delta} +  4is \omega r \right.  \nonumber \\
& &  \left.  \phantom{\left(\frac{d}{dr}\right)} - a^2\omega^2 + 2m a\omega -  {_{s}}A_{\ell m}(a\omega)   \right] \, {_{s}R}_{\ell m}(r,\omega) = 0, \nonumber \\
& &
\end{eqnarray}
satisfying particular boundary conditions at the event horizon and at spatial infinity while ${_{s}W}_{\ell m}(\omega)$ denotes their constant ``Wronskian''. More precisely, when $\mathrm{Im} (\omega) \ge 0$, the function ${_{s}R}^\mathrm {in}_{\ell m}(r,\omega)$ is uniquely defined by its purely ingoing behavior at the event horizon $r=r_{+}$, i.e., for $r_\ast \to -\infty$, where we have
\begin{subequations}
\label{bc_in}
\begin{equation}\label{bc_1_in}
{_{s}R}^\mathrm {in}_{\ell m}(r,\omega) \scriptstyle{\underset{r_\ast \to -\infty}{\sim}} \displaystyle{\Delta^{-s}e^{-ikr_\ast}}
\end{equation}
with $k = \omega - ma/(2Mr_{+})$ [here we assume a harmonic time dependence $\exp (-i\omega t)$ for all fields] while, at spatial infinity, i.e., for $r \to +\infty$ or $r_\ast \to +\infty$, it has an
asymptotic behavior of the form
\begin{eqnarray}\label{bc_2_in}
& & {_{s}R}^\mathrm {in}_{\ell m}(r,\omega) \scriptstyle{\underset{r_\ast \to +\infty}{\sim}} \,\,
\displaystyle{{ _{s}{\cal A}}^{(-)}_{\ell m} (\omega) r^{-1} e^{-i\omega r_\ast} } \nonumber \\
& & \qquad\qquad\qquad\qquad  + {_{s}{\cal A}}^{(+)}_{\ell m} (\omega) r^{-2s-1} e^{i\omega r_\ast}.
\end{eqnarray}
\end{subequations}
Similarly, when $\mathrm{Im} (\omega) \ge 0$, the function ${_{s}R}^\mathrm {up}_{\ell m}(r,\omega)$ is uniquely defined by its purely outgoing behavior at spatial infinity where we have
\begin{subequations}
\label{bc_up}
\begin{equation}\label{bc_1_up}
{_{s}R}^\mathrm {up}_{\ell m}(r,\omega) \scriptstyle{\underset{r_\ast \to +\infty}{\sim}}
\displaystyle{r^{-2s-1} e^{+i\omega r_\ast}}
\end{equation}
and, at the horizon, it has an asymptotic behavior of the form
\begin{eqnarray}\label{bc_2_up}
& & {_{s}R}^\mathrm {up}_{\ell m}(r,\omega)
\scriptstyle{\underset{r_\ast \to -\infty}{\sim}} \,\,
\displaystyle{ {_{s}{\cal B}}^{(-)}_{\ell m} (\omega) \Delta^{-s} e^{-ik r_\ast} } \nonumber \\
& & \qquad\qquad\qquad\qquad + {_{s}{\cal B}}^{(+)}_{\ell m} (\omega) e^{+ik r_\ast}.
\end{eqnarray}
\end{subequations}
In the previous expressions, the coefficients ${_{s}{\cal A}}^{(-)}_{\ell m} (\omega)$, ${_{s}{\cal A}}^{(+)}_{\ell m} (\omega)$, ${_{s}{\cal B}}^{(-)}_{\ell m} (\omega)$ and ${_{s}{\cal B}}^{(+)}_{\ell m} (\omega)$ are complex amplitudes. The Wronskian of the functions ${_{s}R}^\mathrm {in}_{\ell m}(r,\omega)$ and ${_{s}R}^\mathrm {up}_{\ell m}(r,\omega)$ is defined by
\begin{eqnarray}\label{eq:TRE_Wronsk}
& & {_{s}W}_{\ell m}(\omega) =\Delta^{s+1} \left( {_{s}R}^\mathrm {in}_{\ell m}(r,\omega) \frac{d}{dr} \, {_{s}R}^\mathrm {up}_{\ell m}(r,\omega) \right. \nonumber \\
& & \qquad\qquad\qquad\qquad  \left. -  {_{s}R}^\mathrm {up}_{\ell m}(r,\omega) \frac{d}{dr} \, {_{s}R}^\mathrm {in}_{\ell m}(r,\omega) \right)
\end{eqnarray}
and its evaluation at $r_\ast \to -\infty$ and $r_\ast \to +\infty$ provides
\begin{subequations}\label{Wellm}
\begin{eqnarray}
& & {_{s}W}_{\ell m}(\omega) =(2i\omega) {_{s}{\cal A}}^{(-)}_{\ell m} (\omega)  \label{Wellm_a}\\
& & \phantom{{_{s}W}_{\ell m}(\omega)}= 2i[(2Mr_{+})k \nonumber \\
& & \qquad\qquad\qquad -i(r_{+}-M)s] \, {_{s}{\cal B}}^{(+)}_{\ell m} (\omega).\label{Wellm_b}
\end{eqnarray}
\end{subequations}

Finally, by using Eqs.~(\ref{eq:TRE_solret}) and (\ref{Wellm_a}) in Eq.~(\ref{eq:GreenFexp}), we get the expression for the retarded Green's function as
\begin{eqnarray}\label{eq:GreenFexp_def}
&& {_{s}G}(x,x') = - \frac{(\Delta')^s}{2\pi}\sum_{\ell=|s|}^{\infty}\sum_{m=-\ell}^{+\ell}\int_{-\infty+ic}^{+\infty+ic}d\omega \, e^{-i\omega t + im\varphi} \nonumber \\
& & \quad \times  \frac{{_{s}R}^\mathrm {in}_{\ell m}(r_<,\omega) \, {_{s}R}^\mathrm {up}_{\ell m}(r_>,\omega)}{(2i\omega) \, {_{s}{\cal A}}^{(-)}_{\ell m} (\omega)}  \, {_{s}S}_{\ell m}(\theta,a\omega) {_{s}{S}}^\ast_{lm}(\theta ',a\omega). \nonumber \\
& &
\end{eqnarray}

\section{Quasinormal and Regge pole approximations of the retarded Green's function}
\label{RGFapprox}

In this section, from the spectral decomposition (\ref{eq:GreenFexp_def}) of the retarded Green's function associated with the linear perturbations of the Kerr BH, we extract two useful approximations: (i) the QNM approximation which involves the QNMs and their complex frequencies and (ii) the Regge pole approximation which is obtained by using the CAM machinery and involves the Regge modes and the Regge poles.

\subsection{The quasinormal approximation of the retarded Green's function}
\label{RGFapproxQNM}

It is important to keep in mind that, even if the boundary conditions (\ref{bc_in}) and (\ref{bc_up}) fix the modes ${_{s}R}^\mathrm {in}_{\ell m}(r,\omega)$ and ${_{s}R}^\mathrm {up}_{\ell m}(r,\omega)$ only in the region $\mathrm{Im} (\omega) \ge 0$, these modes can be in fact defined also for $\mathrm{Im} (\omega) < 0$ by analytic continuation. As a consequence, the integrand in Eq.~(\ref{eq:GreenFexp_def}) can be considered in the full complex $\omega$ plane and we can perform a deformation of the path integration in Eq.~(\ref{eq:GreenFexp_def}) to obtain an alternative expression of the retarded Green's function (see Ref.~\cite{Leaver:1986gd} for the use of this method in Schwarzschild spacetime). However, it is then necessary to carefully take into account all the singularities of the integrand. For a thorough discussion on this topic in Kerr spacetime, we refer more particularly to Sec.~IV of the article by Casals, Kavanagh and Ottewill \cite{Casals:2016soq}. In their article, these authors consider not only the simple poles lying in the lower part of the complex $\omega$ plane that correspond to QNMs but also the branch point at the origin $(\omega=0)$ which leads to a branch cut along the negative imaginary axis and the spurious branch points away from the origin due to the spin-weighted spheroidal harmonics.

In this article, we are interested in the resonant behaviour of the Kerr BH. It can be described by considering that part of the retarded Green's function which is constructed from the QNMs and which is obtained by collecting, via Cauchy's theorem, the simple poles lying in the lower complex $\omega$ plane. These poles are the zeros of the Wronskian (\ref{Wellm}) or, equivalently, of the complex amplitudes ${_{s}{\cal A}}^{(-)}_{\ell m} (\omega)$ and ${_{s}{\cal B}}^{(+)}_{\ell m} (\omega)$, i.e., the solutions of the equation
\begin{equation} \label{qnmfreq}
 {_{s}{\cal A}}^{(-)}_{\ell m} (\omega_{lmn})=0
\end{equation}
where $\omega_{lmn}$ denotes the complex frequency of the QNM labelled by the three indices $(\ell, m, n)$  (here $n=0,1...$ is the overtone index). A very important fact to be noted is that all these zeros having negative imaginary parts, the corresponding QNMs, i.e., the modes ${_{s}R}^\mathrm {in}_{\ell m}(r,\omega_{lmn})$ and ${_{s}R}^\mathrm {up}_{\ell m}(r,\omega_{lmn})$, are exponentially damped modes which are, in addition, ingoing at the horizon and outgoing at infinity and linearly dependent [see Eqs.~(\ref{bc_in}) and (\ref{bc_up})].

Now, using Cauchy's theorem, we extract from (\ref{eq:GreenFexp_def}) the contribution of the QNMs of the Kerr BH. We obtain
\begin{eqnarray}\label{eq:GreenFexp_QNM}
&& {_{s}G}^\text{\tiny{QNM}}(x,x') = (\Delta')^s \sum_{\ell=|s|}^{\infty}\sum_{m=-\ell}^{+\ell} \sum_{n=0}^{+\infty}  e^{-i \omega_{lmn} t + im\varphi} \nonumber \\
& &
\qquad \times \left(\frac{{_{s}\alpha}_{\ell m n}}{2 \omega_{lmn}} \right) \, {_{s}R}^\mathrm {in}_{\ell m}(r_<,\omega_{lmn}) \, {_{s}R}^\mathrm {up}_{\ell m}(r_>,\omega_{lmn})  \nonumber \\
& &
\qquad \times {_{s}S}_{\ell m}(\theta,a\omega_{lmn}) {_{s}{S}}^\ast_{lm}(\theta ',a\omega_{lmn})
\end{eqnarray}
where
 \begin{equation} \label{res_AatOM}
{_{s}\alpha}_{\ell m n} = 1 \Big/ \left[ \frac{d}{d\omega} \, {_{s}{\cal A}}^{(-)}_{\ell m} (\omega) \right]_{\omega = \omega_{lmn}}
\end{equation}
denotes the residue of the function $1/{_{s}{\cal A}}^{(-)}_{\ell m} (\omega)$ at $\omega= \omega_{lmn}$. It should be noted that it would have been possible to slightly simplified the expression of this QNM retarded Green's function by using some symmetry properties of the various functions involved in (\ref{eq:GreenFexp_def}) in the transformation $(m \to -m, \omega \to -\omega^\ast)$ followed by complex conjugation (see, e.g., Ref.~\cite{Casals:2018eev} for these symmetry properties). This can be especially interesting in the case of the scalar field ($s=0$) but, in general, it presents a limited interest. It is however important to note that, as a consequence of this symmetry, we have
\begin{equation}
{_{s}{\cal A}}^{(-)}_{\ell, -m}(r,-\omega^\ast)=[ {_{s}{\cal A}}^{(-)}_{\ell m}(\omega)]^\ast
\end{equation}
which implies for the solutions of Eq.~(\ref{qnmfreq})
\begin{equation}\label{symprop_qn}
\omega_{\ell, -m, n}=-\omega_{\ell m n}^\ast
\end{equation}
and therefore the symmetry of the quasinormal frequency spectrum with respect to the imaginary $\omega$ axis. This permits one to limit the search for the quasinormal frequencies to the fourth quadrant of the complex $\omega$ plane, those lying in the third one being obtain by symmetry.

\subsection{The Regge pole approximation of the retarded Green's function}
\label{RGFapproxRP}

In order to derive a CAM representation of the retarded Green's function, we first rewrite the double sum over the ``angular momentum number''\footnote{It should be noted that, formally, in Kerr spacetime, $\ell$ cannot be considered as the angular momentum number. In fact, for $s$, $a \omega$ and $m$ given, it serves to label the possible values of the separation constant ${_{s}A}_{\ell m}(a\omega)$ \cite{Teukolsky:1973ha}. It is only in the limit $a \to 0$, i.e., in the Schwarzschild case, that $\ell $ is a true angular momentum number.} $\ell$ and the ``magnetic number'' (or azimuthal number) $m$ appearing in Eq.~(\ref{eq:GreenFexp_def}) in the form
\begin{equation}\label{comm_sum_lm}
\sum_{m=-\infty}^{+\infty} \,\, \sum_{\ell=\ell_\mathrm{min}}^{+\infty} \quad \mathrm{where} \quad \ell_\mathrm{min} = \mathrm{max}(|m|,|s|)
\end{equation}
and by means of the Sommerfeld-Watson transformation \cite{Watson1918,Sommerfeld1949,Newton:1982qc} we then replace the discrete sum over the ordinary angular momentum $\ell \in \mathbb{N}$ by a contour integral in the
complex $\lambda$ plane (i.e., in the complex $\ell$ plane with $\lambda = \ell +1/2$) to obtain
\begin{eqnarray}\label{eq:GreenFexp_SW}
&& {_{s}G}(x,x') = - \frac{(\Delta')^s}{2\pi}\sum_{m=-\infty}^{+\infty}\int_{-\infty+ic}^{+\infty+ic}d\omega \, e^{-i\omega t + im\varphi} \nonumber \\
& & \quad \times  \int_{\cal C'} d\lambda \frac{e^{i\pi (\lambda-1/2)}}{\cos (\pi \lambda)} \nonumber \\
& & \quad \times \frac{{_{s}R}^\mathrm {in}_{\lambda-1/2, m}(r_<,\omega) \, {_{s}R}^\mathrm {up}_{\lambda-1/2, m}(r_>,\omega)}{4\omega \, {_{s}{\cal A}}^{(-)}_{\lambda-1/2, m} (\omega)}  \nonumber \\
& & \quad \times {_{s}S}_{\lambda-1/2, m}(\theta,a\omega) {_{s}{S}}^\ast_{\lambda-1/2, m}(\theta ',a\omega).
\end{eqnarray}
In Eq.~(\ref{eq:GreenFexp_SW}), we have taken ${\cal C'}=]+\infty +i\epsilon,\ell_\mathrm{min}+i\epsilon] \cup [\ell_\mathrm{min}+i\epsilon,\ell_\mathrm{min}-i\epsilon] \cup [\ell_\mathrm{min}-i\epsilon, +\infty -i\epsilon[$
with $\epsilon \to 0_+$.  We can recover (\ref{eq:GreenFexp_def}) from (\ref{eq:GreenFexp_SW}) by using
Cauchy's residue theorem and by noting that the poles of the integrand in
(\ref{eq:GreenFexp_SW}) are the zeros of $\cos (\pi \lambda)$ that are enclosed into ${\cal C'}$, i.e., the semi-integers $\lambda = \ell + 1/2$ with $\ell = \ell_\mathrm{min}, \ell_\mathrm{min}+1,\dots  $

It should be noted that, in Eq.~(\ref{eq:GreenFexp_SW}), we have introduced ``the'' analytic extensions in the complex $\lambda$ plane of the functions ${_{s}R}^\mathrm {in}_{\ell m}(r,\omega)$ and ${_{s}R}^\mathrm {up}_{\ell m}(r,\omega)$ and ${_{s}S}_{\ell m}(\theta,a\omega)$. In theory, analytic extensions of these various functions can be constructed directly from the Teukolsky radial and angular equations taking into account appropriate boundary conditions or, equivalently, from the expansions existing in the literature of these functions in terms of special functions. However, it is well known that the uniqueness of such analytic extensions is one of the difficulties of the CAM method: Indeed, there are infinitely many ways of constructing an analytic function taking prescribed values for integers but, as noted by Newton, the ``right'' one is justified by the results (see Ch.~13 of Ref.~\cite{Newton:1982qc}). In this article, we adhere to this philosophy. Linked to this difficulty, we are facing the important problem of the determination of the angular separation constant ${_{s}A}_{\lambda-1/2, m}(a\omega)$ which appears in both the Teukolsky radial and angular equations. For given values of $s$, $m$ and $\omega$, the possible values of ${_{s}A}_{\ell m}(a\omega)$ are obtained by asking for the regularity of the spin-weighted spheroidal harmonic ${_{s}S}_{\ell m}(\theta,a\omega)$ at $\theta =0$ and $\theta =\pi $ and are labelled by $\ell$, the smallest value of $\ell$ corresponding to $\ell= \ell_\mathrm{min}$ \cite{Teukolsky:1973ha}. The transition from $\ell \in \mathbb{N}$ to $\lambda \in \mathbb{C}$ is therefore far from obvious, especially since ${_{s}S}_{\lambda-1/2, m}(\theta,a\omega)$ cannot be regular both at $\theta =0$ and $\theta =\pi $. We encounter this fact even for $a=0$ and $s=0$ (i.e., for the massless scalar field in the Schwarzschild BH - see, e.g., Ref.~\cite{Decanini:2002ha}): Indeed, in that case, the angular function is the Legendre function of first kind $P_{\lambda -1/2} (\cos \theta)= F[1/2-\lambda,1/2+\lambda;1;(1-\cos \theta)/2]$ which is the analytic extension of the Legendre polynomial $P_\ell (\cos \theta)$ and which is irregular at $\theta =\pi $. However, in that case, we have ${_{s}A}_{\ell m}(0) = \ell (\ell +1)$ and we consider that ${_{s}A}_{\lambda-1/2, m}(0) = (\lambda-1/2) (\lambda+1/2)$, a relation which permits us in addition to fix $\lambda$ from the angular separation constant. For the Kerr BH, we will consider that the transition from $\ell \in \mathbb{N}$ to $\lambda \in \mathbb{C}$ can be achieved similarly, i.e., by first extending analytically ${_{s}A}_{\ell m}(a\omega)$ by just replacing in its formal expression $\ell$ by $\lambda-1/2$ and then by inverting the expression of ${_{s}A}_{\lambda-1/2, m}(a\omega)$ to obtain $\lambda$. We can already note that this approach will allow us to obtain the numerical results of Sec.~\ref{RTQNF_num}.

In order to now collect the contribution of the Regge poles, we first move the contour ${\cal C'}$ appearing in (\ref{eq:GreenFexp_SW}) to the left so that it coincides with the contour ${\cal C}=]+\infty +i\epsilon, +i\epsilon] \cup
[+i\epsilon,-i\epsilon] \cup [-i\epsilon, +\infty -i\epsilon[$. As a consequence, we introduce when $\ell_\mathrm{min} >0 $ ``spurious'' poles at the semi-integers $\lambda=1/2,\dots, \ell_\mathrm{min} -1/2$ coming from the term $1/\cos (\pi \lambda)$. A similar problem has been encountered in the Regge pole description of scattering of electromagnetic and gravitational waves by a Schwarzschild BH \cite{Folacci:2019cmc,Folacci:2019vtt}. In fact, such poles have nothing to do with the resonant behavior of the BH. We then deform the contour ${\cal C}$ in the right part of the CAM plane in order to collect, by using Cauchy's residue theorem, the Regge pole contributions, i.e., the contributions of the zeros of the coefficient ${_{s}{\cal A}}^{(-)}_{\lambda-1/2, m} (\omega)$. This is achieved by following, {\it mutatis mutandis}, the approach developed in Refs.~\cite{Folacci:2018sef,Folacci:2019cmc,Folacci:2019vtt}. We drop here, in addition to the spurious pole contributions, the background integrals coming from the paths $\lambda \in [0,+i \infty[$ and $\lambda \in [0,-i \infty[$ as well as the contributions coming from the quarter circles at infinity. We just retain that part of the retarded Green's function involving the Kerr Regge poles. It is given by
\begin{eqnarray}\label{eq:GreenFexp_RP}
&& {_{s}G}^\text{\tiny{RP}}(x,x') = i (\Delta')^s \sum_{m=-\infty}^{+\infty}\sum_{n=0}^{+\infty}\int_{-\infty+ic}^{+\infty+ic}d\omega \, \nonumber \\
& & \quad \times  e^{-i\omega t + im\varphi} \left(\frac{{_{s}r}_{n m}(\omega)}{4\omega} \right) \frac{e^{i\pi [\lambda_{n m} (\omega)-1/2]}}{\cos [\pi \lambda_{n m} (\omega)]} \nonumber \\
& & \quad \times {_{s}R}^\mathrm {in}_{\lambda_{n m} (\omega)-1/2, m}(r_<,\omega) \, {_{s}R}^\mathrm {up}_{\lambda_{n m} (\omega)-1/2, m}(r_>,\omega)  \nonumber \\
& & \quad \times {_{s}S}_{\lambda_{n m} (\omega)-1/2, m}(\theta,a\omega) {_{s}{S}}^\ast_{\lambda_{n m} (\omega)-1/2, m}(\theta ',a\omega). \nonumber \\
& &
\end{eqnarray}
In this expression\footnote{It is important to realize that the Regge pole approximation ${_{s}G}^\text{\tiny{RP}}(x,x')$ given by (\ref{eq:GreenFexp_RP}) encodes not only the contribution of the QNMs of the Kerr BH as we shall show in Secs.~\ref{RTQNF_thnum} and \ref{RTQNF_res} but also the tail of the time-response to an excitation of the BH. Indeed, it is well known (see, e.g., Refs.~\cite{Leaver:1986gd,Casals:2016soq} and references therein) that this tail is theoretically associated with the branch cut due to the behavior of the radial $\mathrm {up}$-modes in the complex $\omega$ plane. The contribution of this branch cut can be captured by deforming the path of integration over $\omega$ in Eq.~(\ref{eq:GreenFexp_RP}). This provides a compact expression for the late-time tail of the retarded Green's function which includes all the multipolar contributions and, in addition, this partially explains the superiority of the Regge pole approximation over the quasinormal approximation, a fact already noted in Refs.~\cite{Folacci:2018sef,Folacci:2020ekl}.}, we have introduced the Regge poles $\lambda_{n m} (\omega)$ satisfying
\begin{equation}\label{PR_def_Am}
{_{s}{\cal A}}^{(-)}_{\lambda_{n m} (\omega)-1/2, m} (\omega) =0.
\end{equation}
For each value of $m \in \mathbb{Z}$, they constitute an infinite family indexed by $n=0, 1, 2,\dots $. For $\mathrm{Re} (\omega) \ge 0$, they lie in the first quadrant of the CAM plane (the lower Regge pole corresponding to $n=0$ \footnote{It should be noted that, in previous works \cite{Decanini:2002ha,Decanini:2009mu}, we labelled the Regge poles of the Schwarzschild BH from $n=1$.}) and for $\mathrm{Re} (\omega) \le 0$, they have migrated in the fourth one. It should be noted that the Regge poles lying in the second and third quadrants (they are symmetrical with respect to the origin to those lying in the first and fourth quadrants) do not contribute to the expression (\ref{eq:GreenFexp_RP}). In Eq.~(\ref{eq:GreenFexp_RP}), we have also introduced the residue of the function $1/{_{s}{\cal A}}^{(-)}_{\lambda-1/2, m} (\omega)$ at $\lambda=\lambda_{n m} (\omega)$ given by
\begin{equation} \label{res_AatRP}
{_{s}r}_{n m}(\omega) = 1 \Big/ \left[ \frac{d}{d\omega} \, {_{s}{\cal A}}^{(-)}_{\lambda-1/2 m} (\omega) \right]_{\lambda = \lambda_{n m} (\omega)}.
\end{equation}
The modes ${_{s}R}^\mathrm {in}_{\lambda_{n m} (\omega)-1/2, m}(r,\omega)$ and ${_{s}R}^\mathrm {up}_{\lambda_{n m} (\omega)-1/2, m}(r,\omega)$ appearing in Eq.~(\ref{eq:GreenFexp_RP}) are the so-called Regge modes. Due to (\ref{PR_def_Am}), they are ingoing at the horizon and outgoing at infinity and linearly dependent as the QNMs but unlike those they are not exponentially damped for $r_\ast \to \pm \infty$.

\section{From Regge trajectories to quasinormal frequencies: Theoretical and numerical considerations}
\label{RTQNF_thnum}

In this section, we first establish an equation linking the Regge poles to the quasinormal frequencies and we solve it by deriving semiclassical relations, formally valid in the high-frequency regime, allowing us to obtain the complex frequencies $\omega_{\ell m n}$ of the weakly damped QNMs from the Regge trajectories $\lambda_{n m}=\lambda_{n m} (\omega)$ with $\omega >0$. We also describe the numerical method used to obtain the Regge poles and the associated Regge trajectories.

\subsection{Theoretical considerations}
\label{RTQNF_th}

Let us first note that we could deform the path integration over $\omega$ in the expression (\ref{eq:GreenFexp_RP}) to obtain an alternative expression of the Regge pole approximation of the retarded Green's function. It would be necessary to take into account all the singularities of the integrand and, in particular, its poles lying in the lower part of the complex $\omega$ plane coming from the term $1/\cos [\pi \lambda_{n m} (\omega)] $ and therefore solutions of the equation
 \begin{equation}\label{RPtoQNF}
\lambda_{n m} (\omega) = \ell +1/2
\end{equation}
with $\ell \ge \ell_\mathrm{min}$ [we recall that the Regge pole $\lambda_{n m} (\omega)$ lies in the right part of the CAM plane and are solutions of Eq.~(\ref{PR_def_Am})]. If we assume that (\ref{eq:GreenFexp_RP}) encodes the resonant behavior of the Kerr BH, it is then natural to consider that the solutions of Eq.~(\ref{RPtoQNF}) are the quasinormal frequencies $\omega = \omega_{\ell m n}$.

In order to solve Eq.~(\ref{RPtoQNF}), we write  $\omega_{\ell m n}$ in the form
\begin{equation}\label{scQNMI}
\omega_{\ell m n}=\omega ^\text{\tiny{R}}_{\ell m n}- i \, \omega ^\text{\tiny{I}}_{\ell m n}
\end{equation}
where $\omega ^\text{\tiny{R}}_{\ell m n} >0$ and $\omega ^\text{\tiny{I}}_{\ell m n} >0$ (here we focus on the quasinormal frequencies lying in the fourth quadrant of the complex $\omega$ plane, those lying in the third one being obtained using Eq.~(\ref{symprop_qn}). We moreover assume that
\begin{equation}\label{hyp1_scQNM}
\omega ^\text{\tiny{R}}_{\ell m n} \gg \omega ^\text{\tiny{I}}_{\ell m n}
\end{equation}
(here we only consider the frequencies of the weakly damped QNMs). We also impose the condition
\begin{equation}\label{hyp2_scQNM}
\mathrm{Re} [\lambda_{n m} (\omega)] \gg \mathrm{Im} [\lambda_{n m} (\omega)],
\end{equation}
a hypothesis that is valid for high frequencies and low values of $n$. Under these assumptions, we can solve (\ref{RPtoQNF}) perturbatively and we obtain the
semiclassical relations
\begin{subequations} \label{sc12}
\begin{equation}\label{sc1}
\mathrm{Re}  \left[ \lambda_{n m} \left( \omega ^\text{\tiny{R}}_{\ell m n} \right) \right]= \ell
+ 1/2   \qquad \ell \ge \ell_\mathrm{min}
\end{equation}
and
\begin{equation}\label{sc2}
\omega ^\text{\tiny{I}}_{\ell m n} =
\left.  \frac{\mathrm{Im} \left[ \lambda_{n m}
(\omega ) \right]}{d/d\omega \,  \mathrm{Re} \left[ \lambda_{n m} (\omega ) \right] }
\right|_{\omega =\omega ^\text{\tiny{R}}_{\ell m n}}
\end{equation}
\end{subequations}
by working successively at the first and second order. It is worth pointing out that these formulas involve the Regge trajectories, i.e.,
the curves traced out in the CAM plane by the Regge poles $\lambda_{n m} (\omega )$ as a
function of the frequency $\omega >0$. We also recall that analogous semiclassical relations have been used in Refs.~\cite{Decanini:2002ha,Decanini:2009mu,Decanini:2010fz,Decanini:2011eh} to obtain analytical formulas for the quasinormal frequencies of the Schwarzschild BH and of various static spherically symmetric BHs and to interpret them in term of unstable circular null geodesics.

\subsection{Numerical methodology}
\label{RTQNF_num}

In the next section, we shall compare the quasinormal frequency spectrum of the Kerr BH obtained semiclassically from the Regge trajectories with the ``exact'' one. Numerical methods permitting us to obtain the quasinormal frequencies $\omega_{\ell m n}$ of the Kerr BH with a high precision have been developed since the seminal paper by Leaver \cite{Leaver:1985ax} (see, in particular, Refs.~\cite{Nollert:1993zz,Onozawa:1995vu,Berti:2005ys,Berti:2009kk,Cook:2014cta}). Let us recall that Leaver's method requires to solve simultaneously the Teukolsky radial equation (\ref{eq:TREhom}) and the Teukolsky angular equation (\ref{eq:TAE}) [these equations are coupled through the unknown parameters $\omega$ and $_{s}A_{lm}(a \omega)$] taking into account the appropriate boundary conditions defining the QNMs: The radial modes are ingoing at the horizon and outgoing at infinity while the angular modes are regular at $\theta=0$ and $\theta=\pi$. Given a value for $_{s}A_{lm}(a \omega)$, we can solve (\ref{eq:TREhom}) for the complex frequency $\omega$ which can be obtained as a root of a continued fraction equation and starting from this value of $\omega$, we can solve (\ref{eq:TAE}) as an eigenvalue problem for $_{s}A_{lm}(a \omega)$ which can be obtained as a root of an other continued fraction equation. This back and forth process can be stopped when a sufficient precision on $\omega$ and $_{s}A_{lm}(a \omega)$ has been reached.

As far as the determination of Regge modes and Regge poles is concerned, the situation is a little bit different: $\omega>0$ is now fixed and we need to determine $_{s}A_{\lambda(\omega)-1/2, m}(a \omega)$ and $\lambda(\omega)$ but, while the radial modes are here again ingoing at the horizon and outgoing at infinity, we cannot fix the separation constant $_{s}A_{\lambda(\omega)-1/2, m}(a \omega)$ by asking for the regularity of the angular modes at $\theta=0$ and $\theta=\pi$ (see also the discussion in Sec.~\ref{RGFapproxRP}). To overcome this problem, we have considered the series expansion of ${_{s}A}_{\ell m}(a\omega)$ in powers of $a\omega$ provided by Seidel and given by
\begin{equation} \label{expanSeidel}
{_{s}A}_{\ell m}(a\omega) = - s (s+1) + \sum_{p=0}^6 \, {_{s}f}_{\ell m}^{(p)} \, (a\omega)^p
\end{equation}
(the coefficients ${_{s}f}_{\ell m}^{(p)}$ appear in Eq.~(10) of Ref.~\cite{Seidel:1988ue}) and we have continued analytically this expression in the CAM plane by replacing $\ell \in \mathbb{N}$ by $\lambda (\omega)-1/2 \in \mathbb{C}$. Then, it just remains to solve the Teukolsky radial equation (\ref{eq:TREhom}) where ${_{s}A}_{\ell m}(a\omega)$ is replaced by ${_{s}A}_{\lambda (\omega)-1/2, m}(a\omega)$ by focusing on the Regge modes, i.e., by asking for the radial modes an ingoing behavior at the horizon and an outgoing behavior at infinity. {\it Mutatis mutandis}, Leaver's method permitting one to determine the radial part of the QNMs can be easily adapted and we can obtain the Regge poles $\lambda (\omega)$ as the roots of a continued fraction equation. In fact, we have implemented a
slightly modified version of Leaver's method due to Majumdar and Panchapakesan
and based on the Hill determinant \cite{Majumdar:1989tzg} (see also Refs.~\cite{Decanini:2002ha,Decanini:2009mu,Decanini:2010fz,Decanini:2011eh}). In our opinion, this alternative method is more numerically tractable than Leaver's to scan the CAM plane. More precisely and in summary, we have obtained the Regge poles $\lambda (\omega)$ as the solutions of the equation
\begin{equation} \label{HillDet}
\left|
 \begin{array}{cccccc} \beta_0 & \alpha_0 & \cdot & \cdot & \cdot & \cdots
\\
  \gamma_1&\beta_1 & \alpha_1 & \cdot& \cdot & \cdots \\
  \cdot&\gamma_2 &\beta_2 & \alpha_2 &\cdot &\cdots \\
    \cdot&\cdot&\gamma_3 &\beta_3 & \alpha_3 &\cdots  \\
  \vdots&\vdots&\vdots &\ddots&\ddots &\ddots\\
  \end{array}
 \right| =0
\end{equation}
where the coefficients $\alpha_k$, $\beta_k$ and $\gamma_k$ with $k \in \mathbb{N}$ are given by Eqs.~(25) and (26) of Ref.~\cite{Leaver:1985ax} and in this last equation, we have replaced ${_{s}A}_{\ell m}(a\omega)$ by ${_{s}A}_{\lambda (\omega)-1/2, m}(a\omega)$ obtained from the expansion (\ref{expanSeidel}) .

It should be pointed out that our numerical method presents instabilities when the expansion of ${_{s}A}_{\lambda (\omega)-1/2, m}(a\omega)$ obtained from (\ref{expanSeidel}) does not converge. This occurs in particular when we have simultaneously $a\omega \ll 1$ and $|\lambda (\omega)| \ll 1$ but also when $a/M \to 1$, i.e., for the near-extremal Kerr BH. It is moreover important to note that there exists, in the literature, various expansions (asymptotic expansions, uniform asymptotic expansions) which could be used as a replacement for the series expansion (\ref{expanSeidel}) to explore the CAM plane (see, e.g., Refs.~\cite{Breuer1977,Casals:2004zq,Berti:2005gp,Casals:2018cgx}). Maybe, they could permit us to avoid the instabilities encountered in this article.

It is interesting to recall that, in Ref.~\cite{Glampedakis:2003dn}, Glampedakis and Andersson dealing with the case of scalar perturbations ($s=0$) developed two ``quick and dirty'' schemes based on the usual phase-amplitude method and the complexification of the radial coordinate $r_\ast$ permitting one to compute the quasinormal frequencies of the Schwarzschild and Kerr BHs as well as their Regge poles. These authors, unlike us, did not seek to establish a relation between QNMs and Regge modes but they have however, for a particular value of the frequency $\omega$, exhibit the splitting of the Schwarzschild Regge poles due to rotation. We have checked the results they displayed and, in particular, their Table 5 in Ref.~\cite{Glampedakis:2003dn}. However, it seems to us that these authors did not realize that each Schwarzschild Regge pole splits into an infinity of Kerr Regge poles.

\section{From Regge trajectories to quasinormal frequencies: Results and remarks}
\label{RTQNF_res}

In this section, we provide a series of numerical results which concern the astrophysically relevant case of $s=-2$ describing gravitational perturbations by means of the Newman-Penrose scalar $\Psi_4$. In particular, we exhibit the splitting due to rotation of each Regge pole of the Schwarzschild BH into an infinite number of Kerr Regge poles and we compare the QNM spectrum of the Kerr BH obtained semiclassically with the ``exact'' one. Numerical results concerning scalar and electromagnetic perturbations are available upon request from the authors. We have not displayed them in this article because they are not really different from those obtained for gravitational perturbations.

\subsection{The splitting of Regge poles}
\label{RTQNF_res1}

\begin{figure}
\centering
\includegraphics[height=6cm,width=8.5cm]{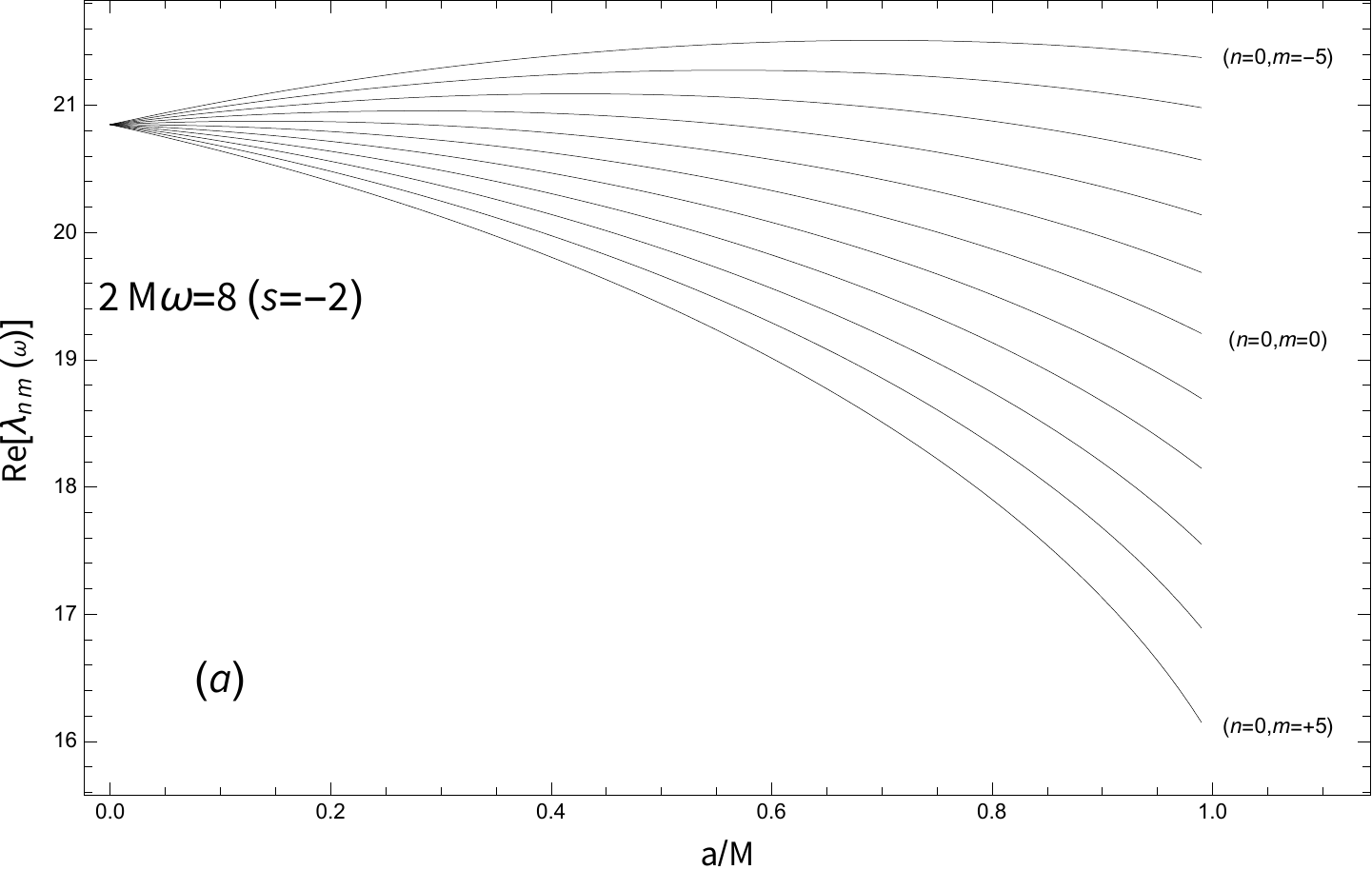}
\centering
\vspace{0.2cm}
\includegraphics[height=6cm,width=8.5cm]{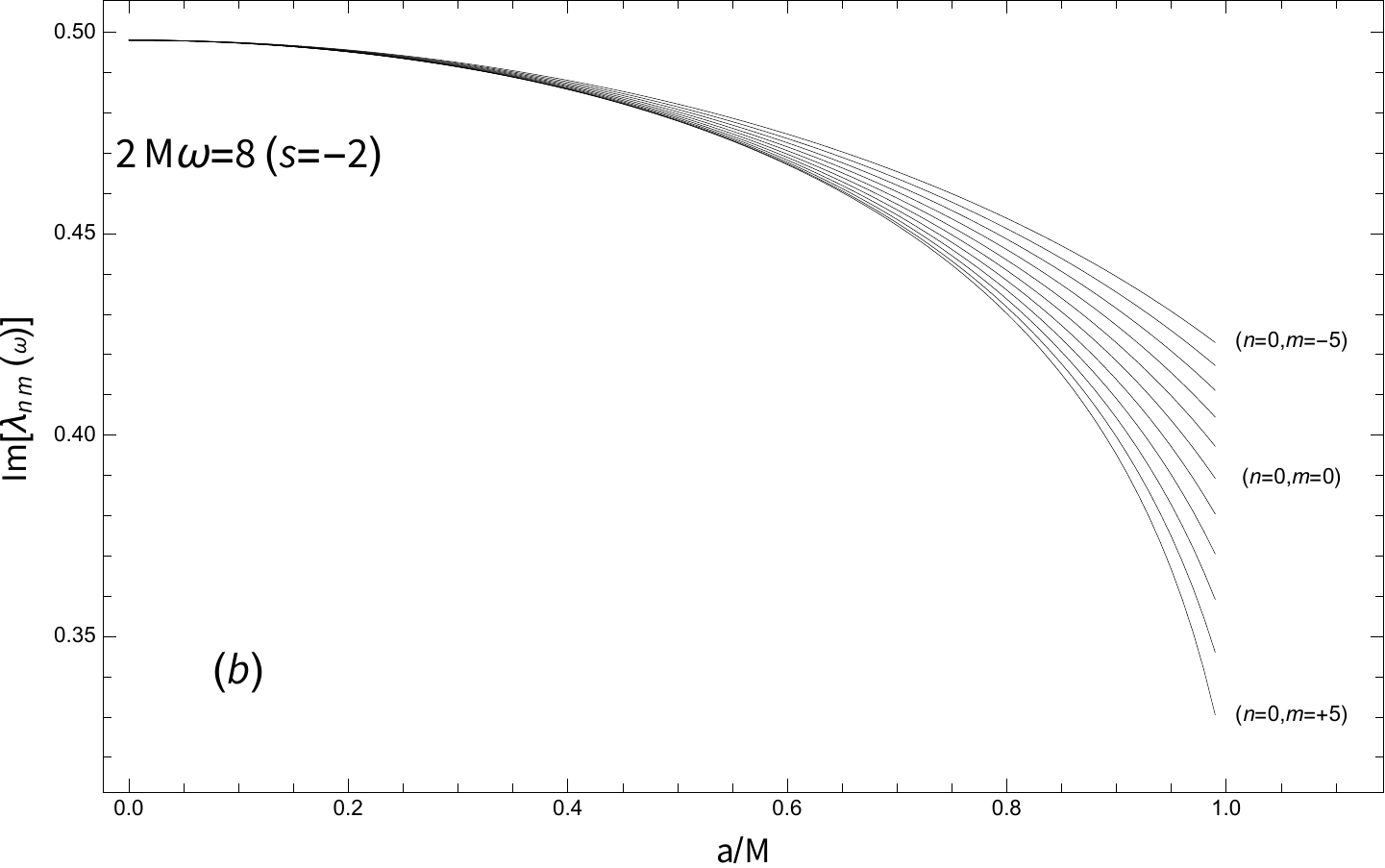}
\centering
\vspace{0.2cm}
\includegraphics[height=5cm,width=8.5cm ]{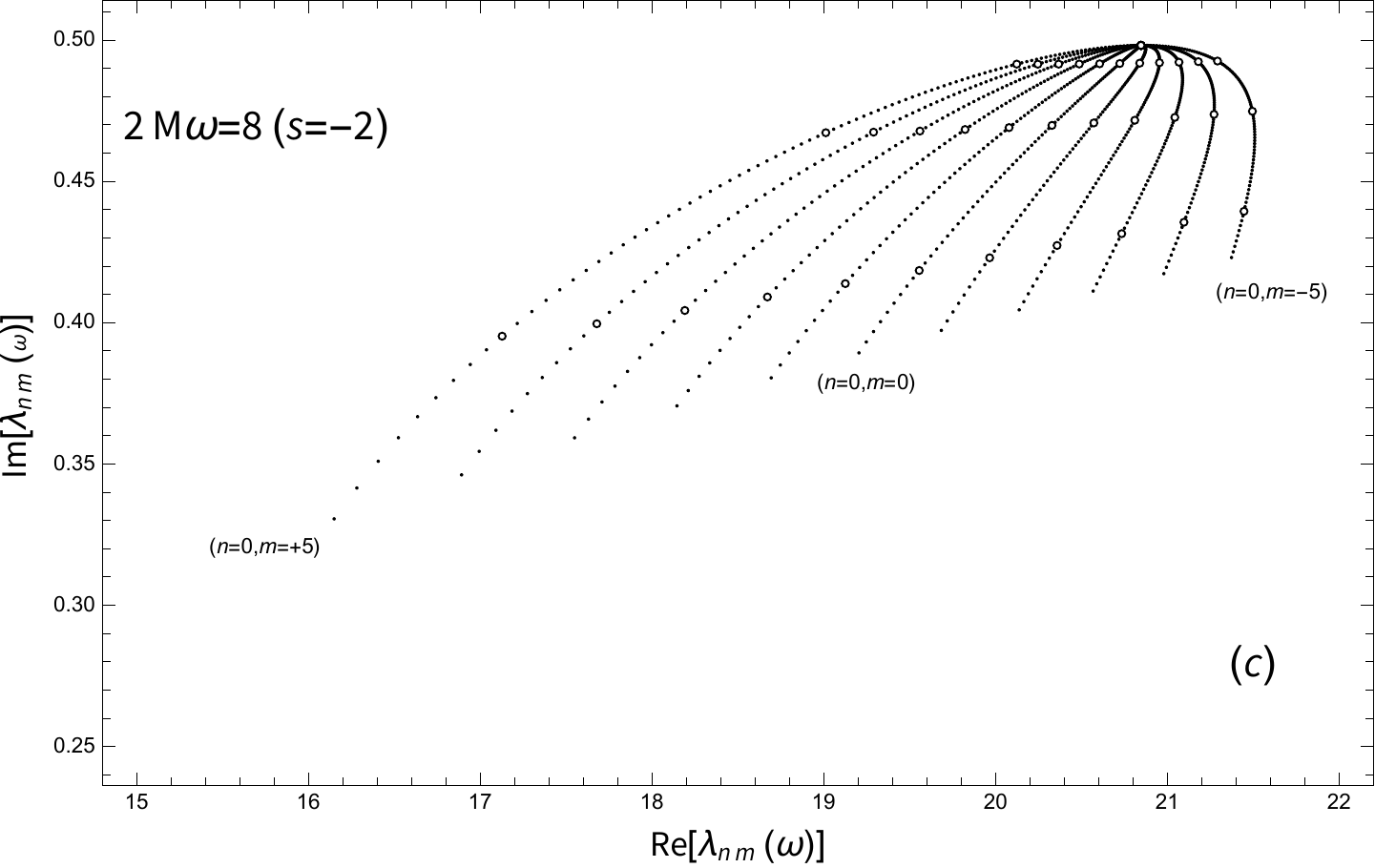}
\centering
\vspace{0.2cm}
\caption{\label{fig:SplittingPR1} The splitting of the first Regge pole of the Schwarzschild BH due to the rotation of the Kerr BH is considered for $2M\omega=8$ in the case of gravitational perturbations ($s=-2$). The behavior of the corresponding split poles $\lambda_{n=0,m}(\omega)$ with $m \in \mathbb{Z}$ of the Kerr BH depends on the rotation rate $a/M$. Here, we display it for $m=-5,\dots,+5$ and the associated Regge poles of the Kerr BH are followed for rotation rates $a/M \in [0,0.99]$. (a) and (b) $\mathrm{Re}[\lambda_{n=0,m}(\omega)]$ and $\mathrm{Im}[\lambda_{n=0,m}(\omega)]$ are plotted as functions of $a/M$. (c) $\mathrm{Im}[\lambda_{n=0,m}(\omega)]$ as a function of $\mathrm{Re}[\lambda_{n=0,m}(\omega)]$ is plotted for rotation rates $a/M=0, 0.01, 0.02,\dots,0.99$. Open dots spot the position of the Regge poles for $a/M=0, 0.30, 0.60, 0.90$.}
\end{figure}

\begin{figure}
\centering
\includegraphics[height=6cm,width=8.5cm]{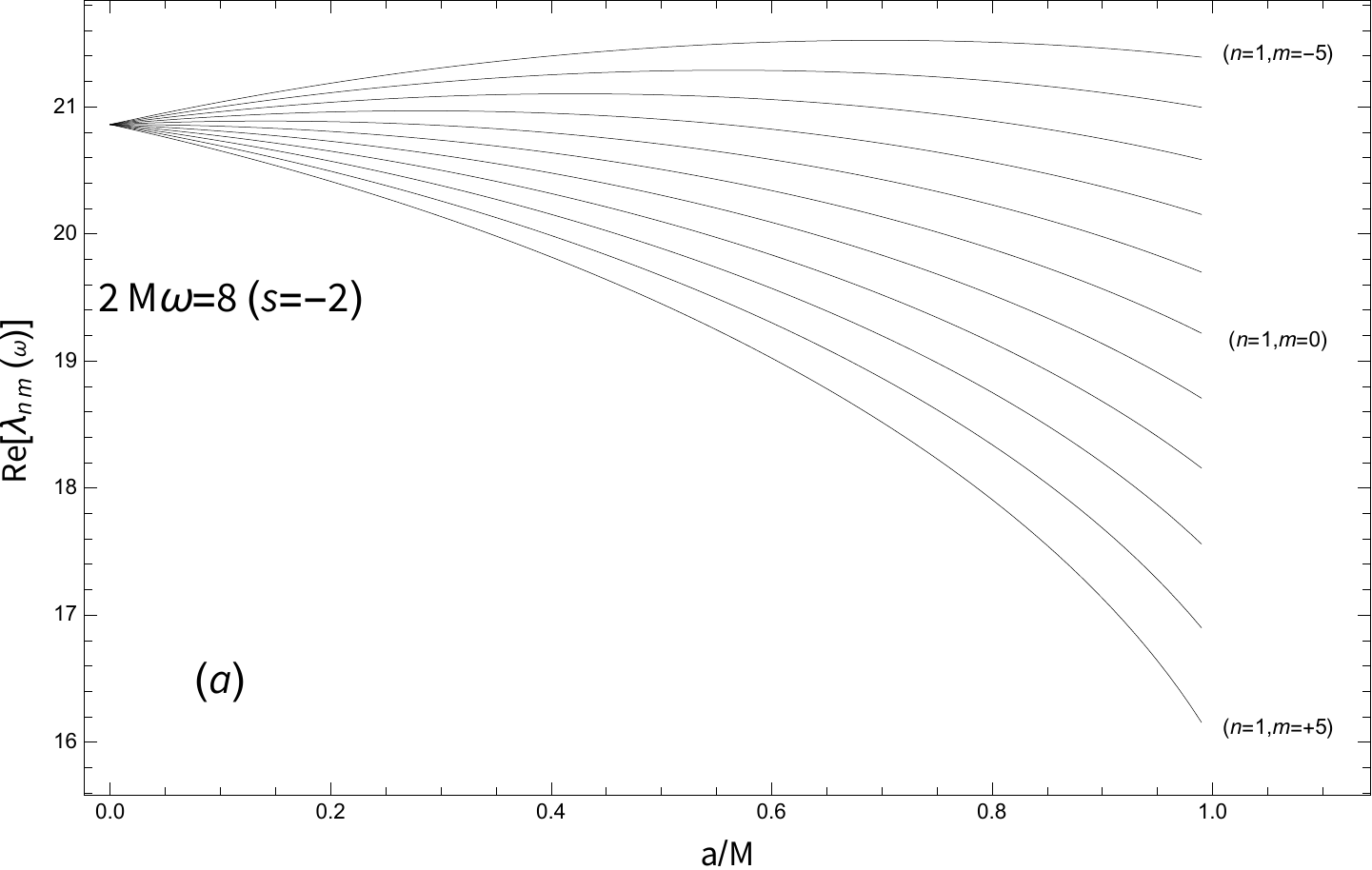}
\centering
\vspace{0.2cm}
\includegraphics[height=6cm,width=8.5cm]{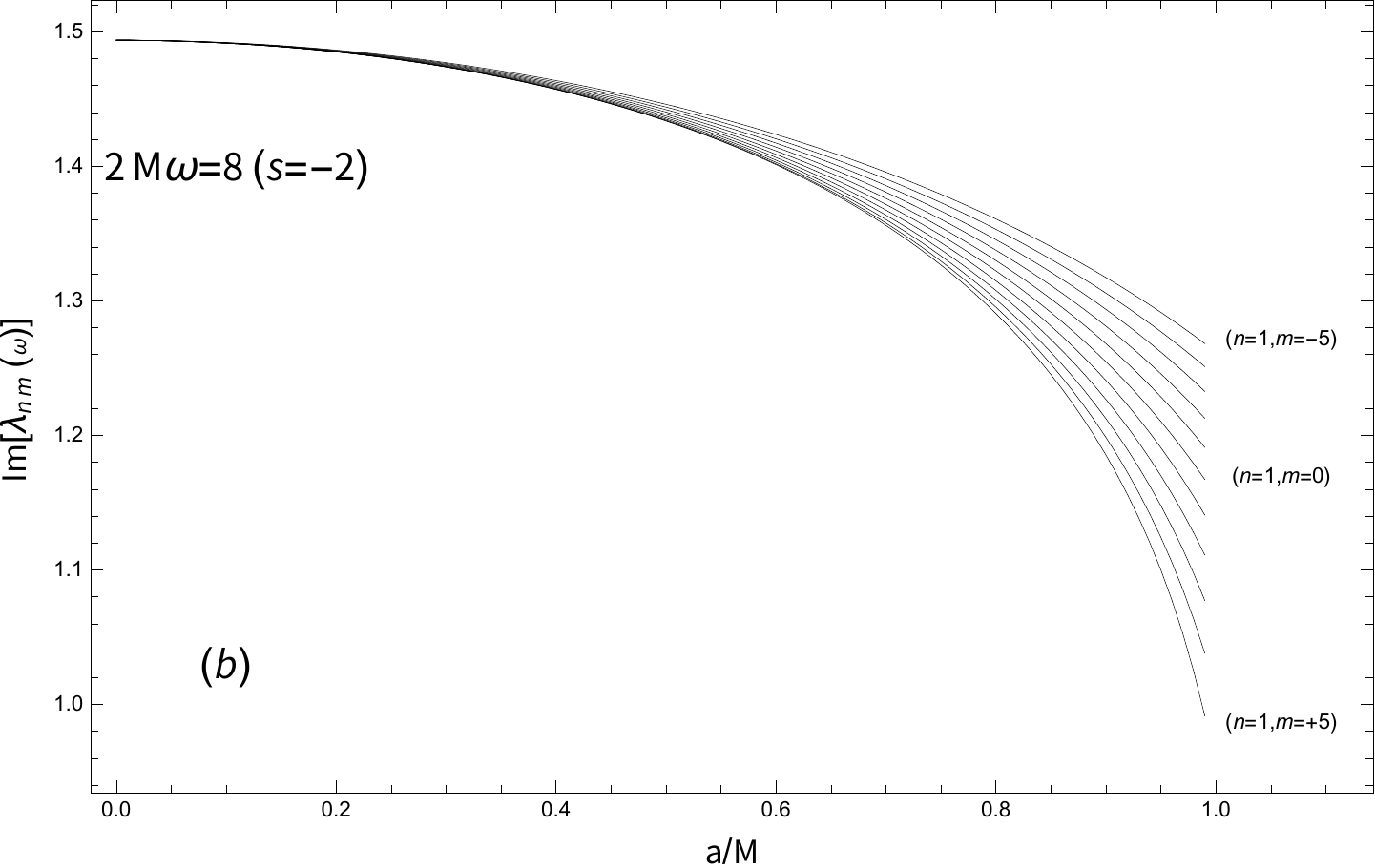}
\centering
\vspace{0.2cm}
\includegraphics[height=5cm,width=8.5cm ]{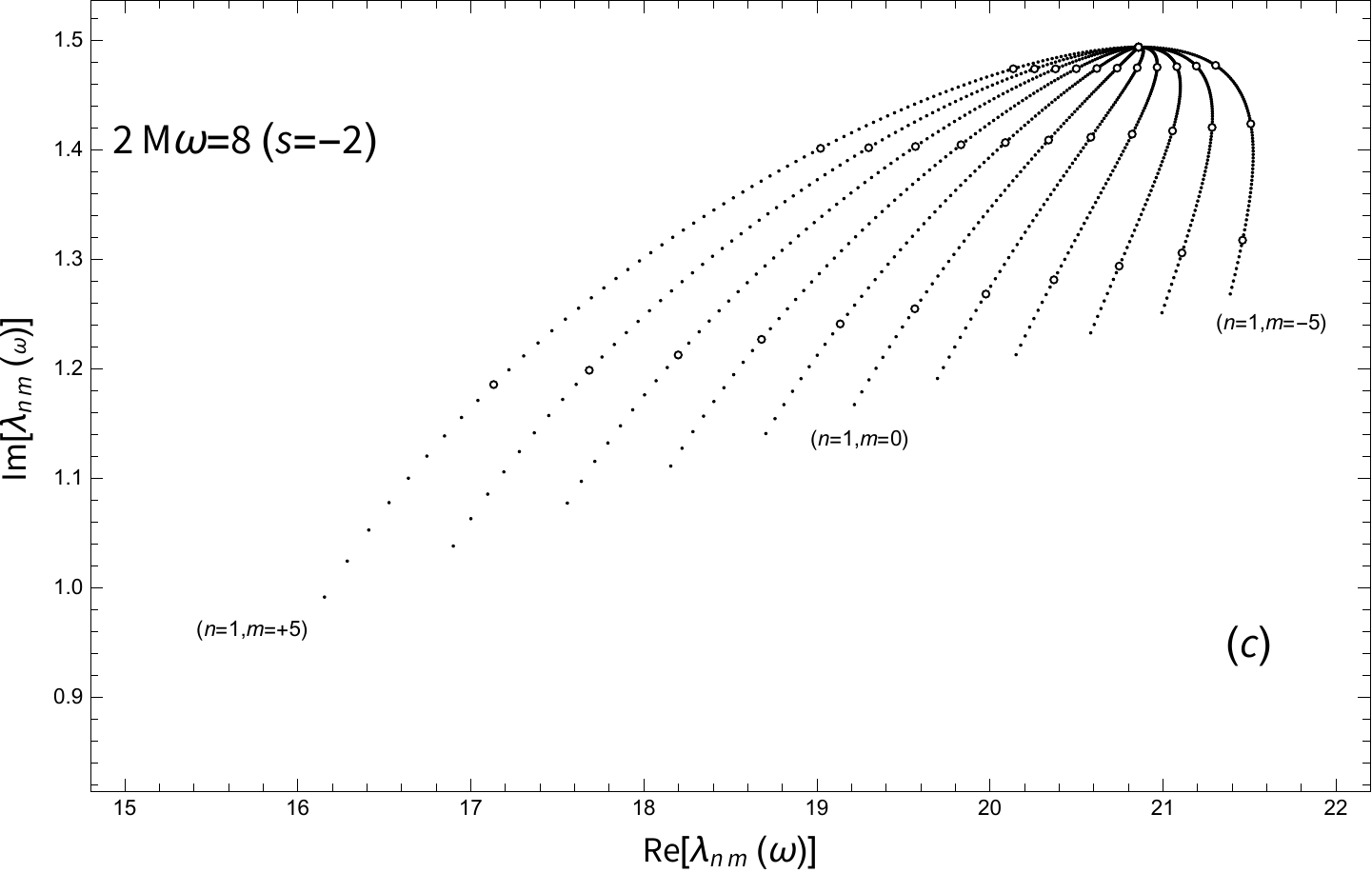}
\centering
\vspace{0.2cm}
\caption{\label{fig:SplittingPR2} The splitting of the second Regge pole of the Schwarzschild BH due to the rotation of the Kerr BH is considered for $2M\omega=8$ in the case of gravitational perturbations ($s=-2$). The behavior of the corresponding split poles $\lambda_{n=1,m}(\omega)$ with $m \in \mathbb{Z}$ of the Kerr BH depends on the rotation rate $a/M$. Here, we display it for $m=-5,\dots,+5$ and the associated  Regge poles of the Kerr BH are followed for rotation rates $a/M \in [0,0.99]$. (a) and (b) $\mathrm{Re}[\lambda_{n=0,m}(\omega)]$ and $\mathrm{Im}[\lambda_{n=0,m}(\omega)]$ are plotted as functions of $a/M$. (c) $\mathrm{Im}[\lambda_{n=0,m}(\omega)]$ as a function of $\mathrm{Re}[\lambda_{n=0,m}(\omega)]$ is plotted for rotation rates $a/M=0, 0.01, 0.02,\dots,0.99$. Open dots spot the position of the Regge poles for $a/M=0, 0.30, 0.60, 0.90$.}
\end{figure}

In Figs.~\ref{fig:SplittingPR1} and \ref{fig:SplittingPR2}, we have displayed, for the dimensionless frequency $2M\omega=8$, the splitting of the first two Regge poles of the Schwarzschild BH. Indeed, due to rotation of the Kerr BH, the spacetime spherical symmetry and therefore the azimuthal degeneracy are broken and each Regge pole of the Schwarzschild BH splits into an infinity of Kerr Regge poles $\lambda_{n m}(\omega)$ with $m \in \mathbb{Z}$. More precisely, we have followed, for rotational rates $a/M  \in [0,0.99]$, the behavior of the first two Schwarzschild Regge pole in the CAM plane for $m=-5,\dots, +5$. It should be noted that, by working here with a high dimensionless frequency $2M\omega$, we have avoided the occurrence of numerical instabilities and have been able to follow the Kerr Regge poles even for $a/M \approx 1$.

\subsection{Regge trajectories}
\label{RTQNF_res2}

In Figs.~\ref{fig:RTPR1M0M2Mm2} and \ref{fig:RTPR2M0M2Mm2}, we  have considered the Regge trajectories of the Kerr Regge poles $\lambda_{n=0,m=0}(\omega)$, $\lambda_{n=0,m=2}(\omega)$, $\lambda_{n=0,m=-2}(\omega)$, $\lambda_{n=1,m=0}(\omega)$, $\lambda_{n=1,m=2}(\omega)$ and $\lambda_{n=1,m=-2}(\omega)$ for the rotation rates $a/M= 0, 0.30, 0.60, 0.90$. This allows us to highlight the respective roles of the index $m$ and of the rotation rate $a/M$ and to compare the behavior of the Schwarzschild Regge poles with that of the corresponding Kerr ones. It is important to note that, for low dimensionless frequency $2M\omega$, we have encountered numerical instabilities when $m \ge 0$. They are due to the lack of convergence, in this particular regime, of the expansion of ${_{s}A}_{\lambda (\omega)-1/2, m}(a\omega)$ obtained from (\ref{expanSeidel}). They are all the more important that $a/M$ increases. Their existence prevents us from analyzing when $a/M \to 1$ the behavior of the ``low'' quasinormal frequencies by considering the Regge trajectories.

\subsection{Quasinormal frequencies: Exact versus semiclassical results}
\label{RTQNF_res3}

Tables \ref{tab:tablePR1m0}-\ref{tab:tablePR2mm2} present a sample of Kerr quasinormal frequencies $\omega_{\ell n m}$ calculated from the trajectories of the Regge poles $\lambda_{n m} (\omega)$ with $n=\{0, 1\}$ and $m=\{-2, -1, 0, 1, 2, 4\}$ by using the semiclassical formulas (\ref{sc1}) and (\ref{sc2}) for $2 \leq  \ell \leq 7$ as well as $l={30, 50}$ to describe the eikonal $(\ell \gg 1)$ regime\footnote{Therefore, in our parameter range, we have accommodated the crucial $(\ell=2,m=2,n=0)$-QNM that has been noted in literature to be the dominant mode for the ringdown after a binary BH merger and we have also touched upon the overtones, whose incorporation into the ringdown spectrum leads to improved estimation of the mass and spin parameters of the remnant BH \cite{Giesler:2019uxc} and contribute towards tests of the no-hair theorem from gravitational wave observations \cite{Isi:2019aib,Bhagwat:2019dtm}.}. Here again, we have considered the rotation rates $a/M= 0, 0.30, 0.60, 0.90$. The values obtained are compared with the ``exact'' quasinormal frequencies determined numerically by Leaver's method \cite{Leaver:1985ax}. The exact values for $2 \leq  \ell \leq 7$ have been taken on Berti’s ``Ringdown'' website \cite{RingdownBerti} while those for $l={30, 50}$ have been calculated by the authors.

Broadly speaking, the semiclassical values possess a high degree of numerical accuracy for both real and imaginary parts when being compared to the numerical values obtained from Leaver's method. A common trend across all parameter ranges is the improved accuracy as we reach the eikonal limit wherein the agreement with Leaver's values ranges from $10^{-5}-10^{-8}$ . Even for the lower values of $\ell$, our results are in agreement with Leaver's with an accuracy of $10^{-1}-10^{-3}$. Moreover, the numerical accuracy is retained for slowly spinning ($a/M=0.3$) as well as rapidly rotating Kerr BHs ($a/M=0.9$).

\begin{figure*}
\centering
\includegraphics[height=5.8cm,width=8.2cm]{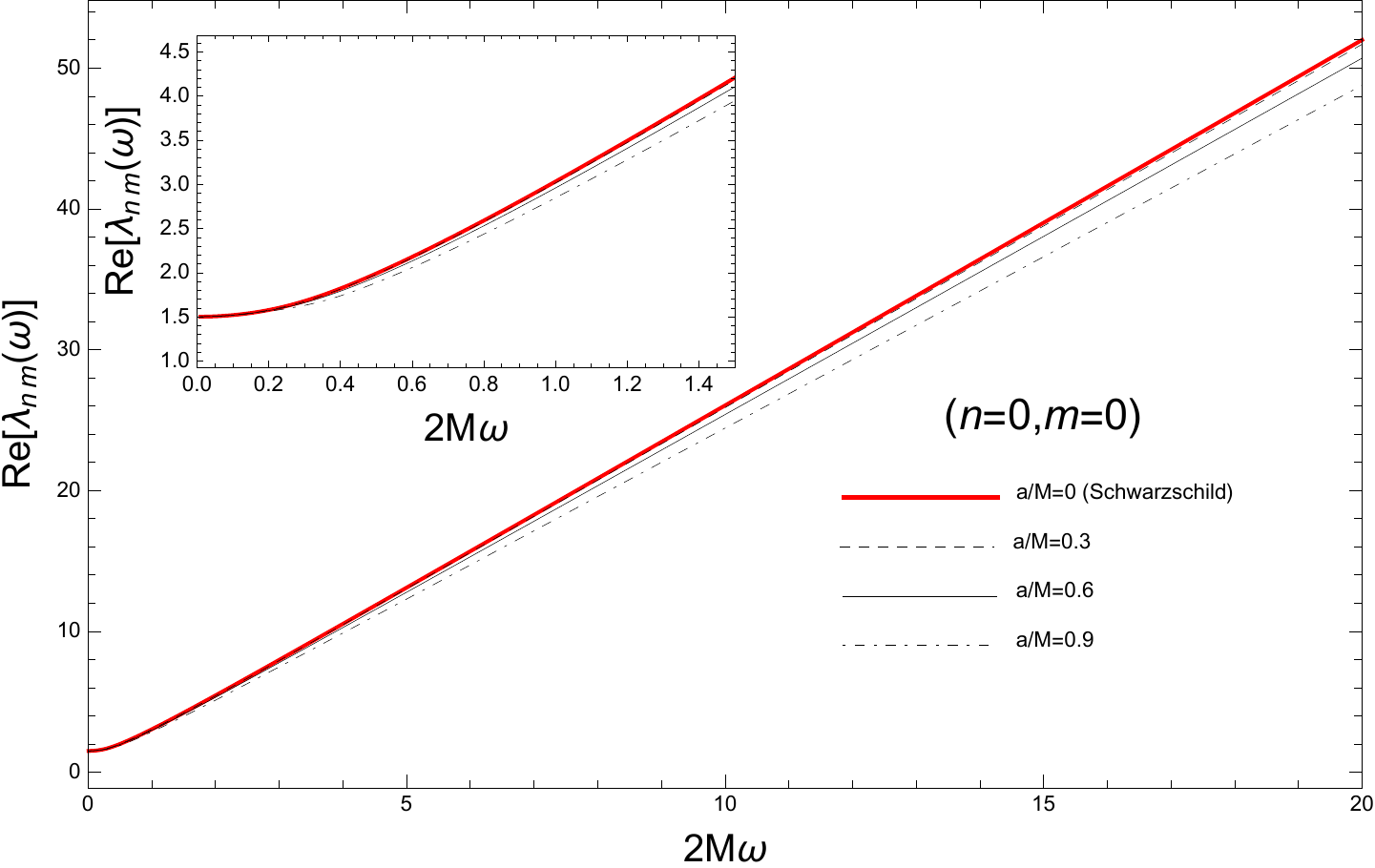}
\centering
\vspace{0.2cm}
\includegraphics[height=5.8cm,width=8.2cm]{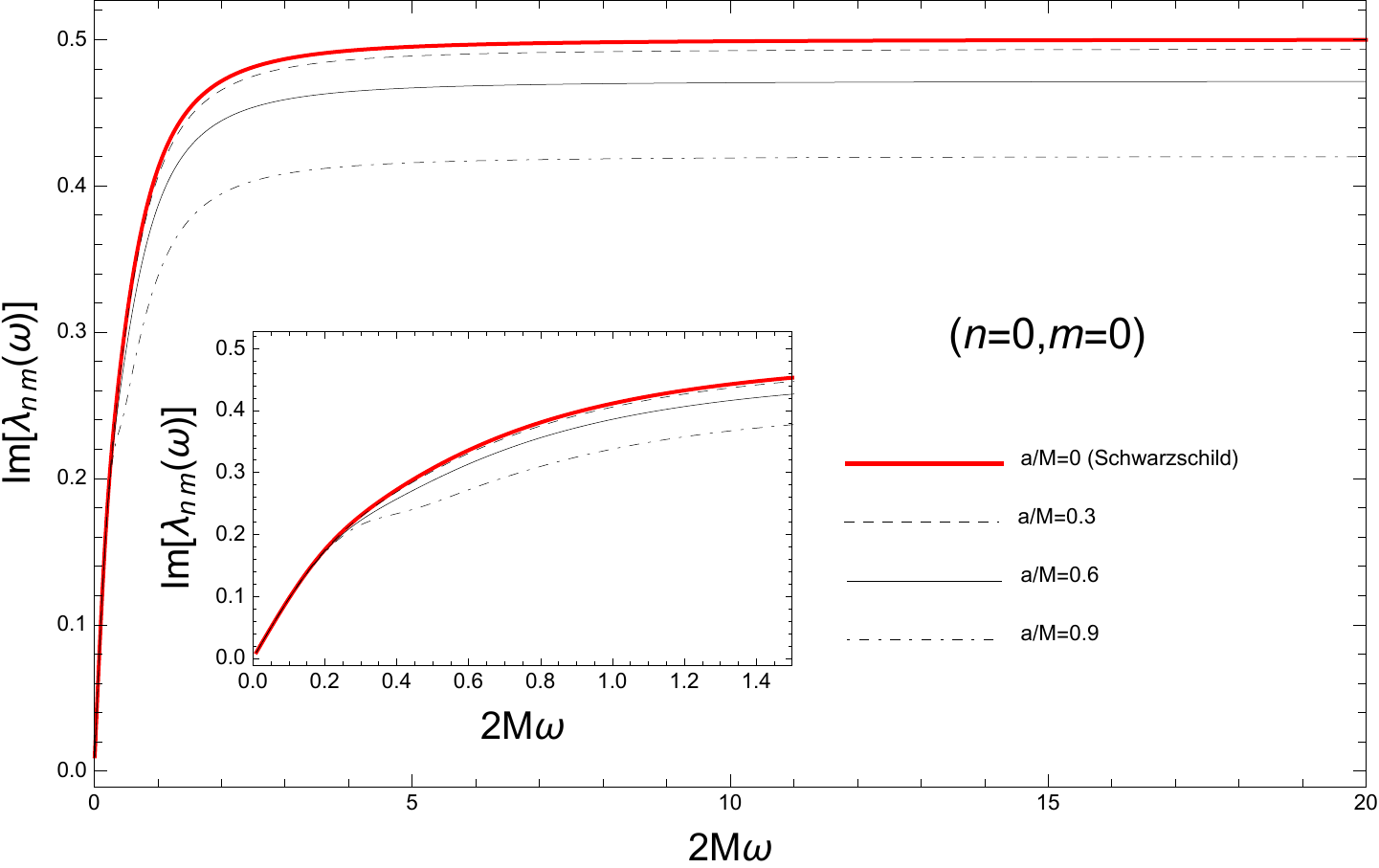}
\centering
\vspace{0.2cm}
\centering
\includegraphics[height=5.8cm,width=8.2cm]{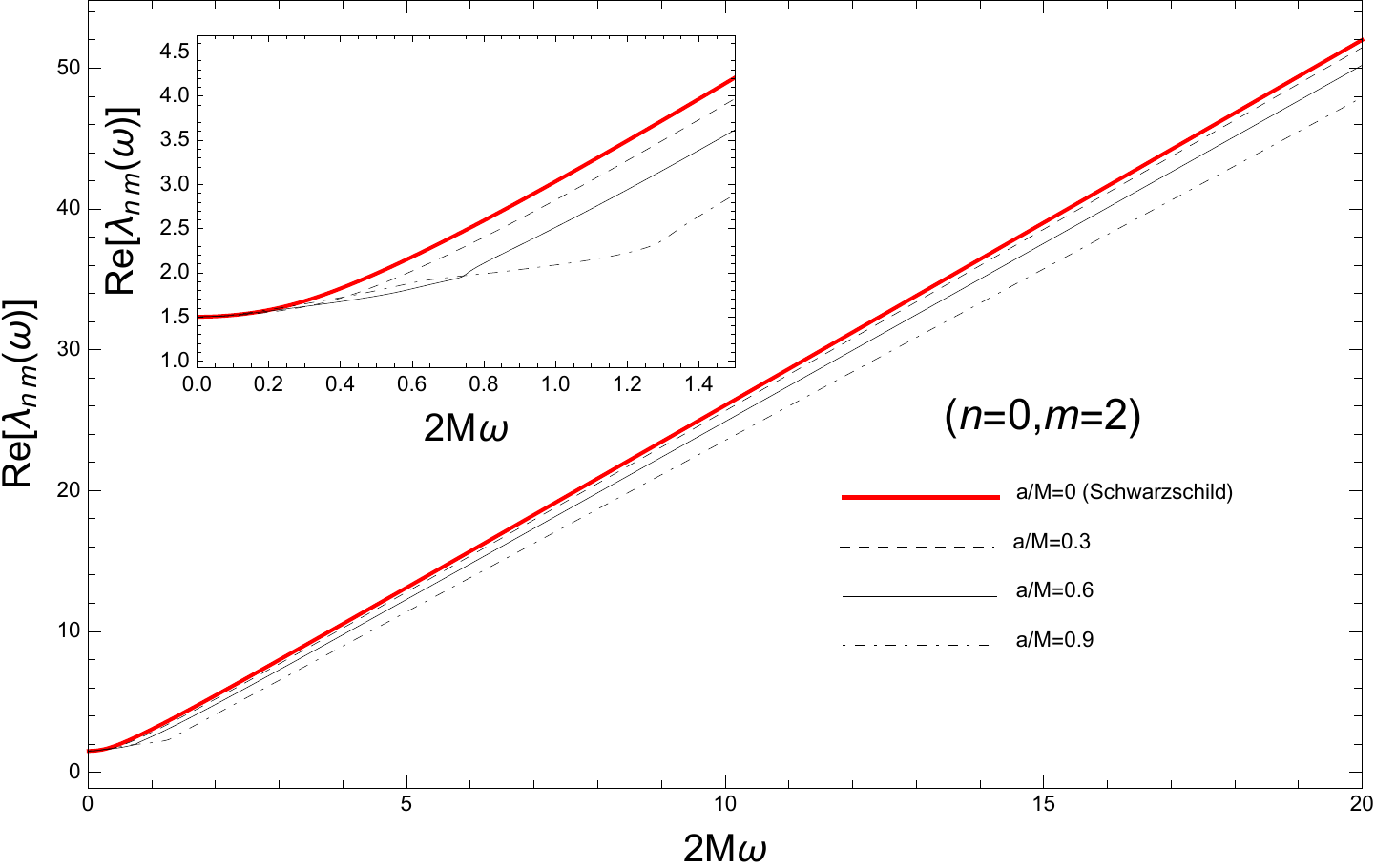}
\centering
\vspace{0.2cm}
\includegraphics[height=5.8cm,width=8.2cm]{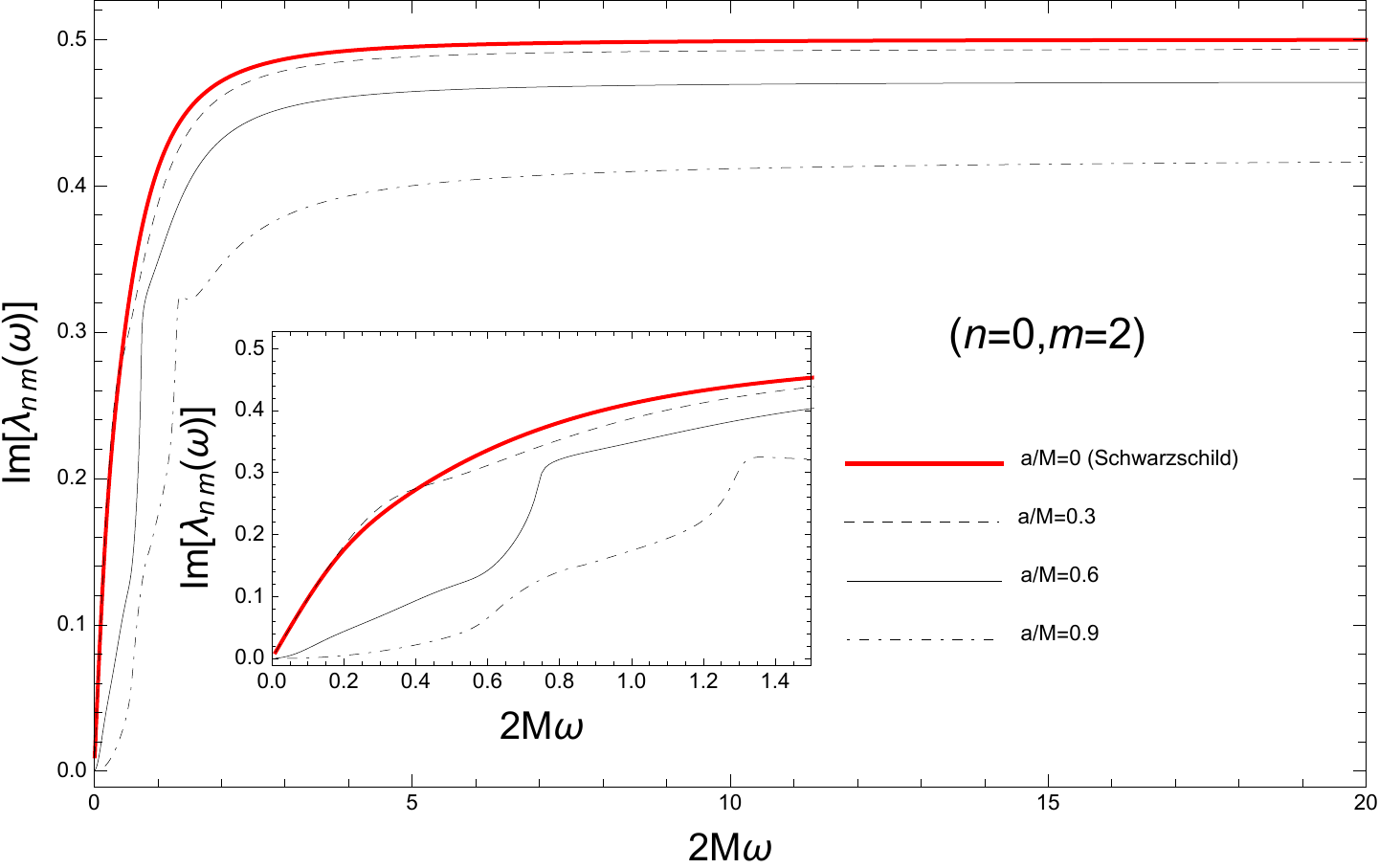}
\centering
\vspace{0.2cm}
\centering
\includegraphics[height=5.8cm,width=8.2cm]{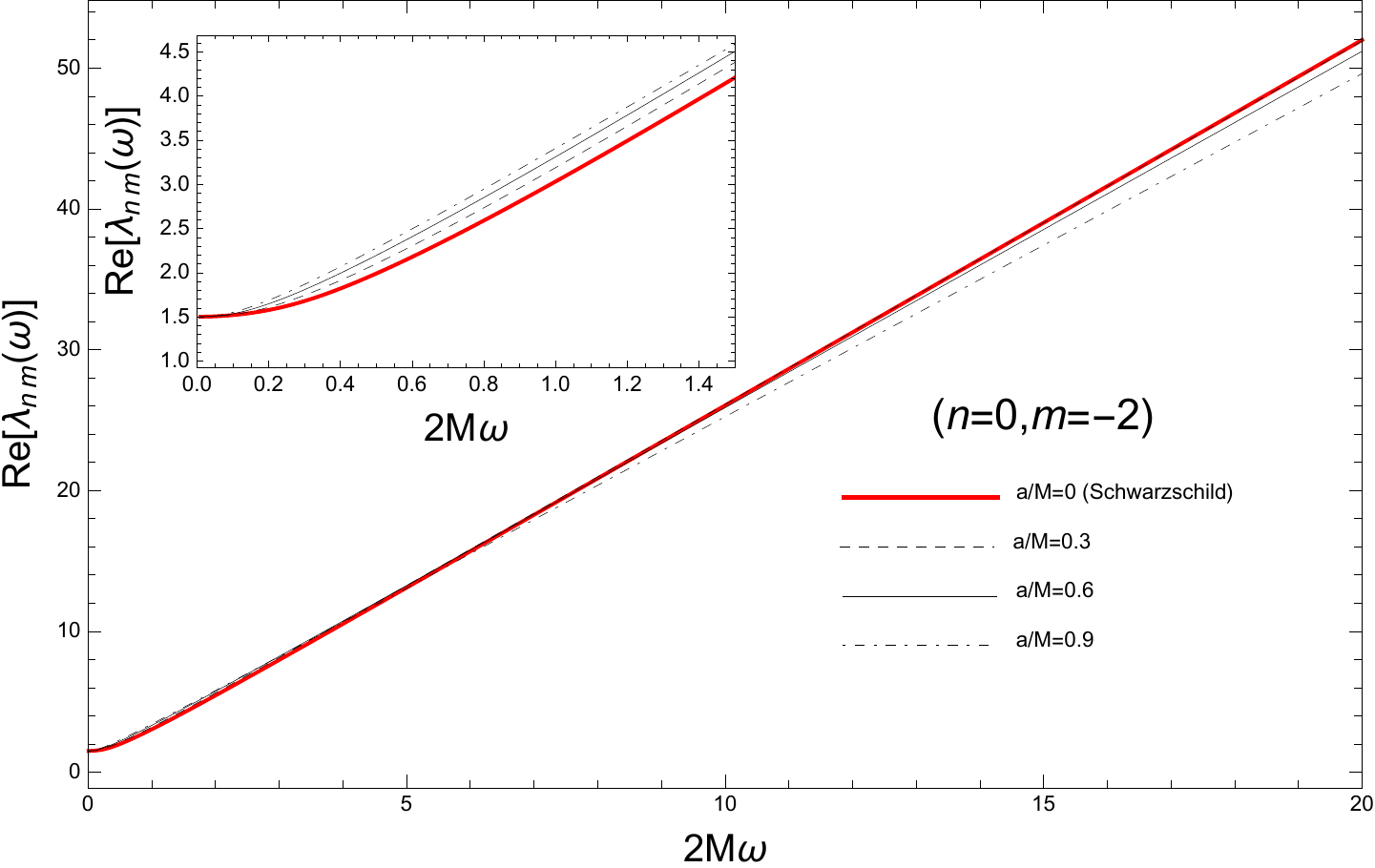}
\centering
\vspace{0.2cm}
\includegraphics[height=5.8cm,width=8.2cm]{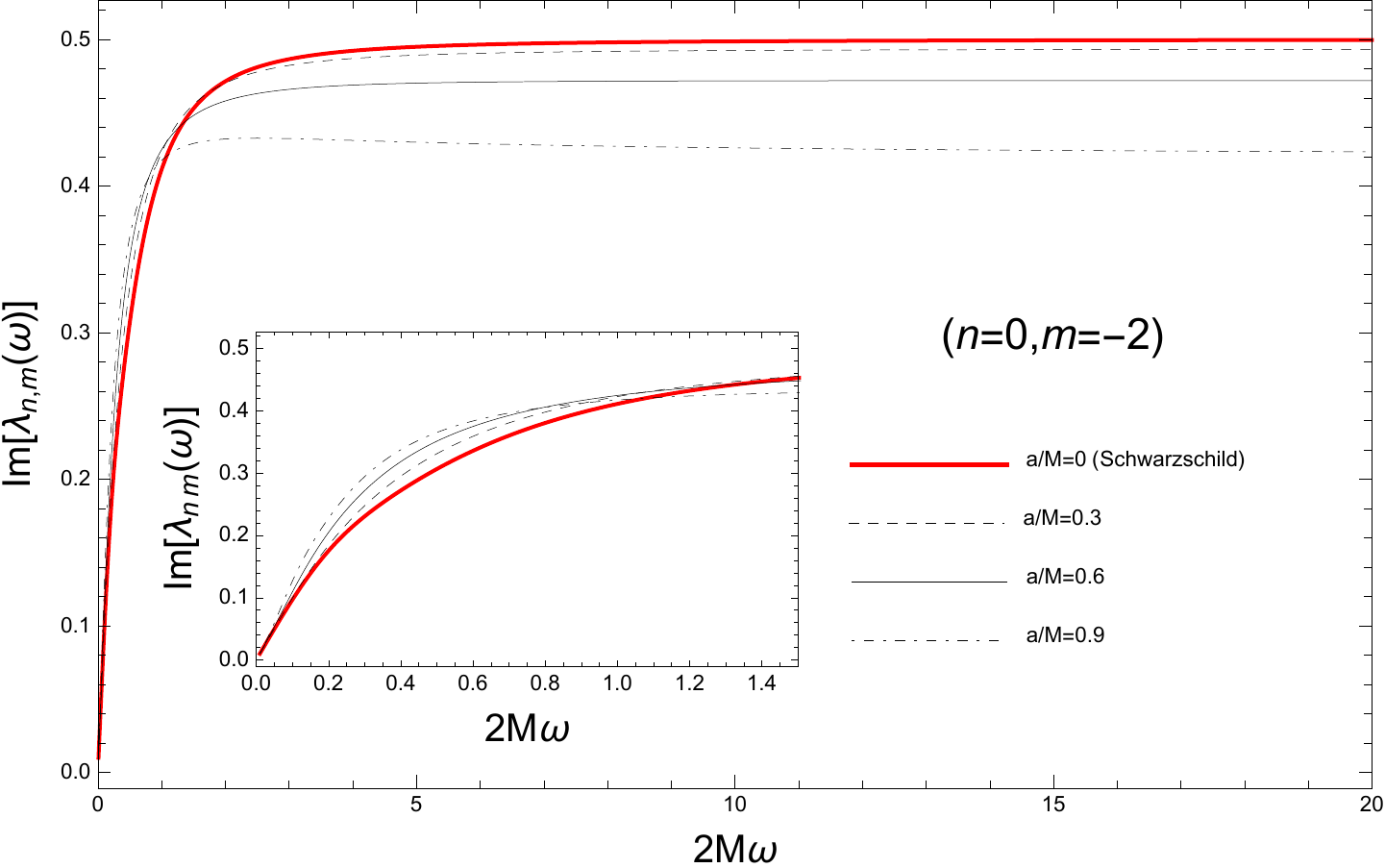}
\centering
\vspace{0.2cm}
\caption{\label{fig:RTPR1M0M2Mm2} Regge trajectories of the Regge poles $\lambda_{n=0,m=0}(\omega)$, $\lambda_{n=0,m=2}(\omega)$ and $\lambda_{n=0,m=-2}(\omega)$ of the Kerr BH for the rotation rates $a/M= 0, 0.30, 0.60, 0.90$. We consider the gravitational perturbations ($s=-2$) of the BH. The behavior of $\lambda_{n=0,m=0}(\omega)$ and $\lambda_{n=0,m=2}(\omega)$ is typical of the Kerr Regge poles $\lambda_{n=0, m}(\omega)$ with $m \ge 0$ while the behavior of $\lambda_{n=0,m=-2}(\omega)$ is typical of the Kerr Regge poles $\lambda_{n=0, m}(\omega)$ with $m < 0$. In particular, it should be noted that, for a given value of $a/M$, the Regge trajectories of $\mathrm{Re}[\lambda_{n=0,m=0}(\omega)]$ and $\mathrm{Re}[\lambda_{n=0,m=2}(\omega)]$ lie below that of the corresponding Schwarzschild Regge pole and those with lower rotation rates while the Regge trajectory of $\mathrm{Re}[\lambda_{n=0,m=-2}(\omega)]$ lies below that of the corresponding Schwarzschild Regge pole and those with lower rotation rates only for high frequencies. The results obtained for $m \ge 0$ are affected for low frequencies by numerical instabilities (see explanations and discussions in the text).}
\end{figure*}

\begin{figure*}
\centering
\includegraphics[height=5.8cm,width=8.2cm]{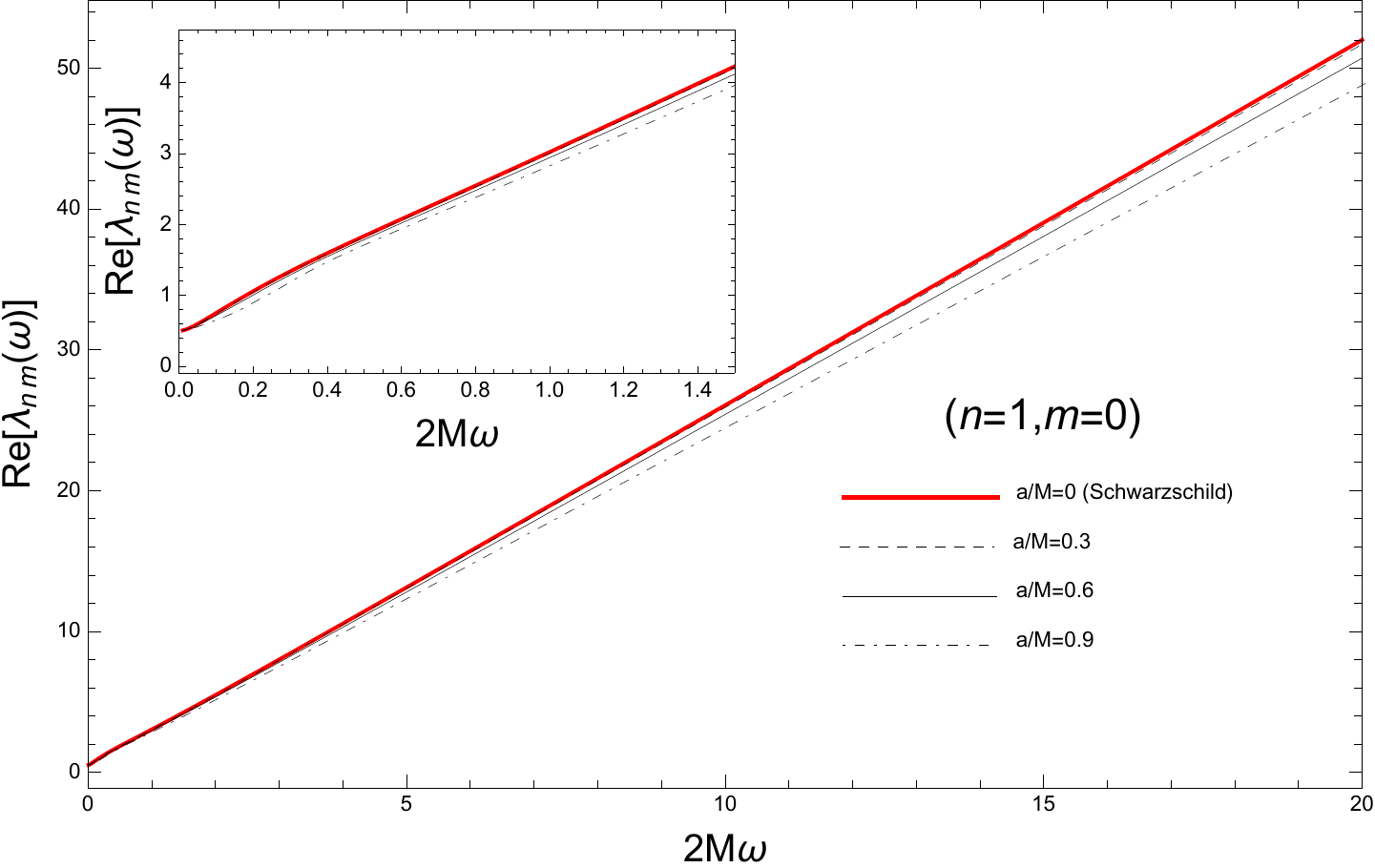}
\centering
\vspace{0.2cm}
\includegraphics[height=5.8cm,width=8.2cm]{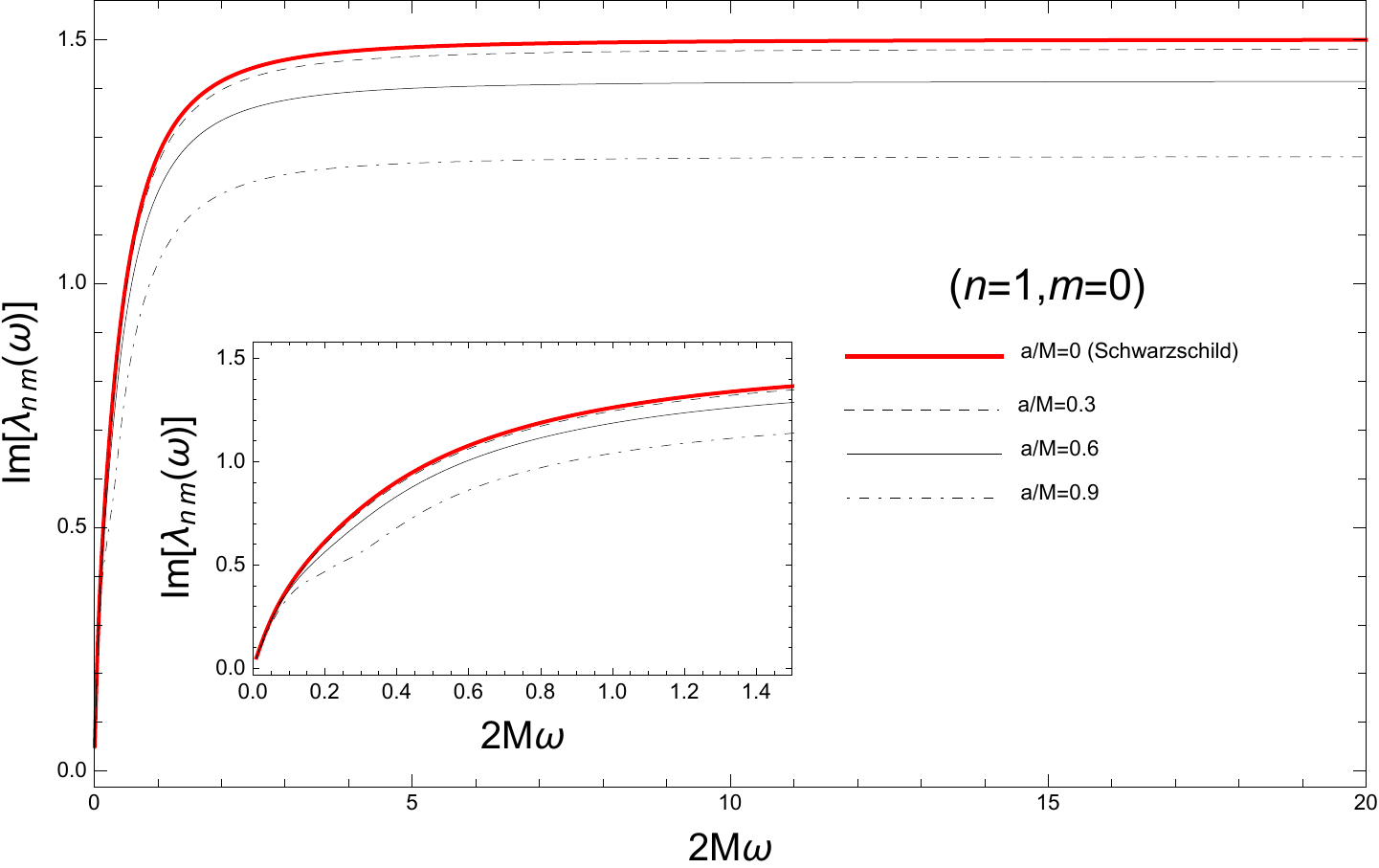}
\centering
\vspace{0.2cm}
\centering
\includegraphics[height=5.8cm,width=8.2cm]{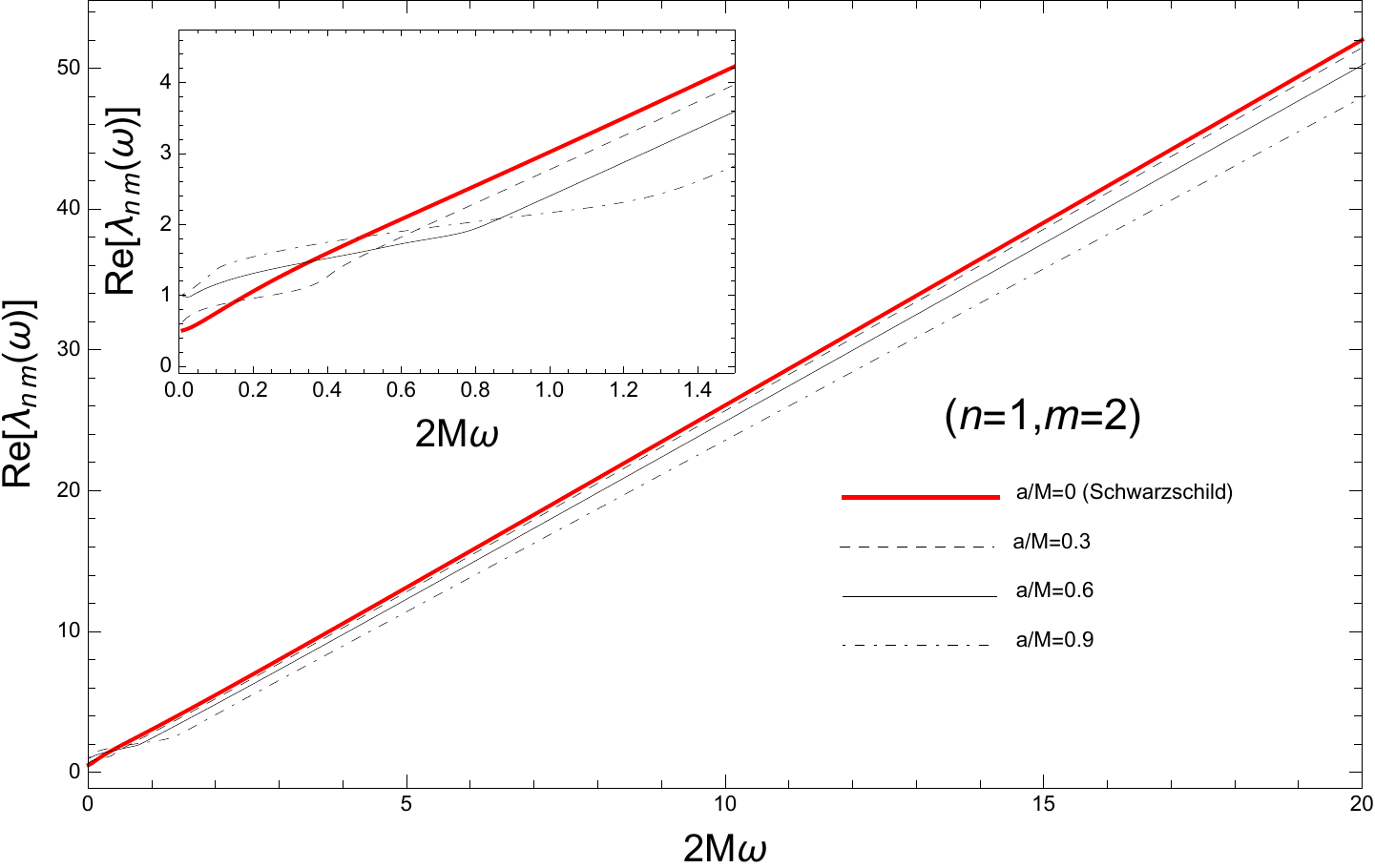}
\centering
\vspace{0.2cm}
\includegraphics[height=5.8cm,width=8.2cm]{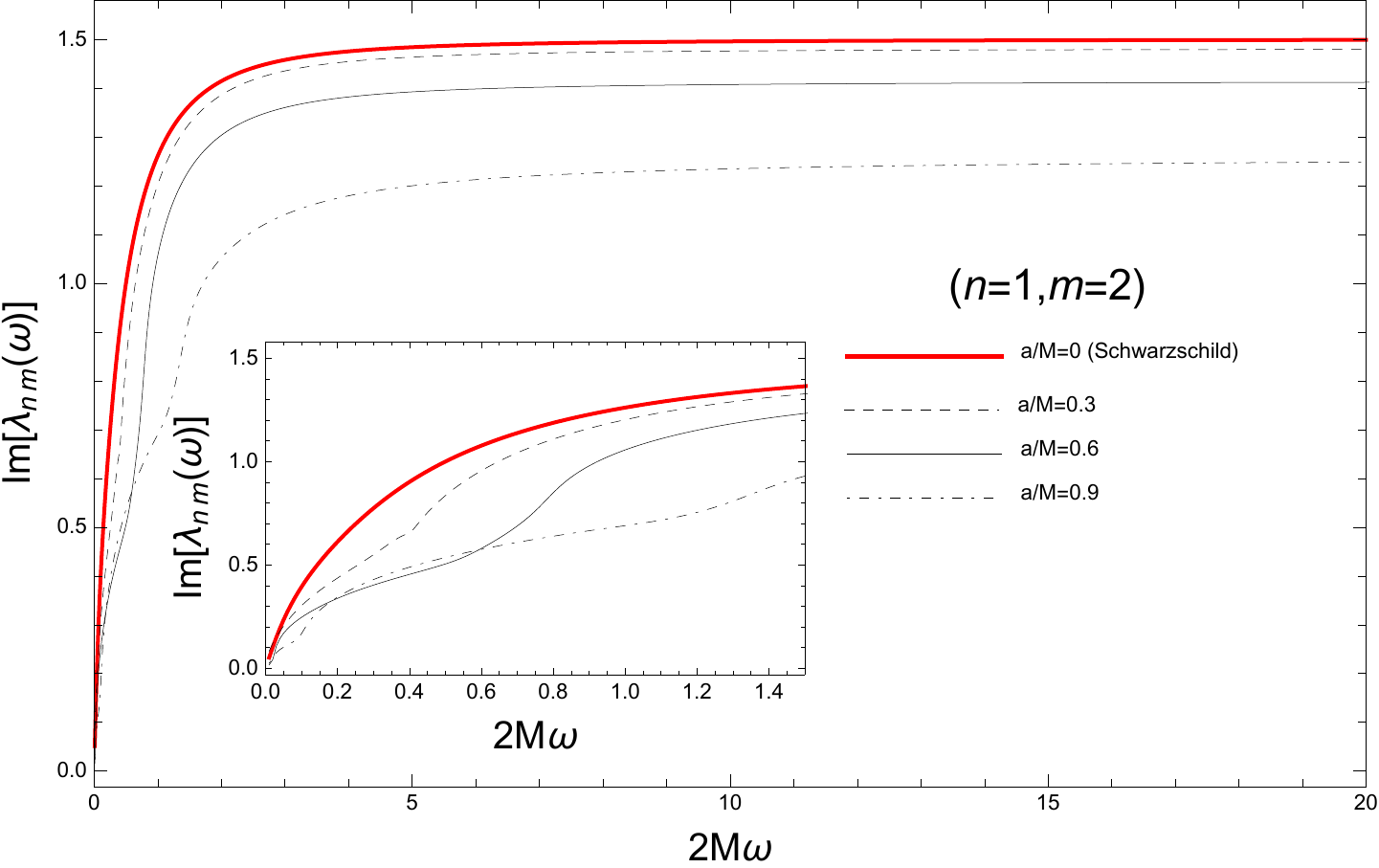}
\centering
\vspace{0.2cm}
\centering
\includegraphics[height=5.8cm,width=8.2cm]{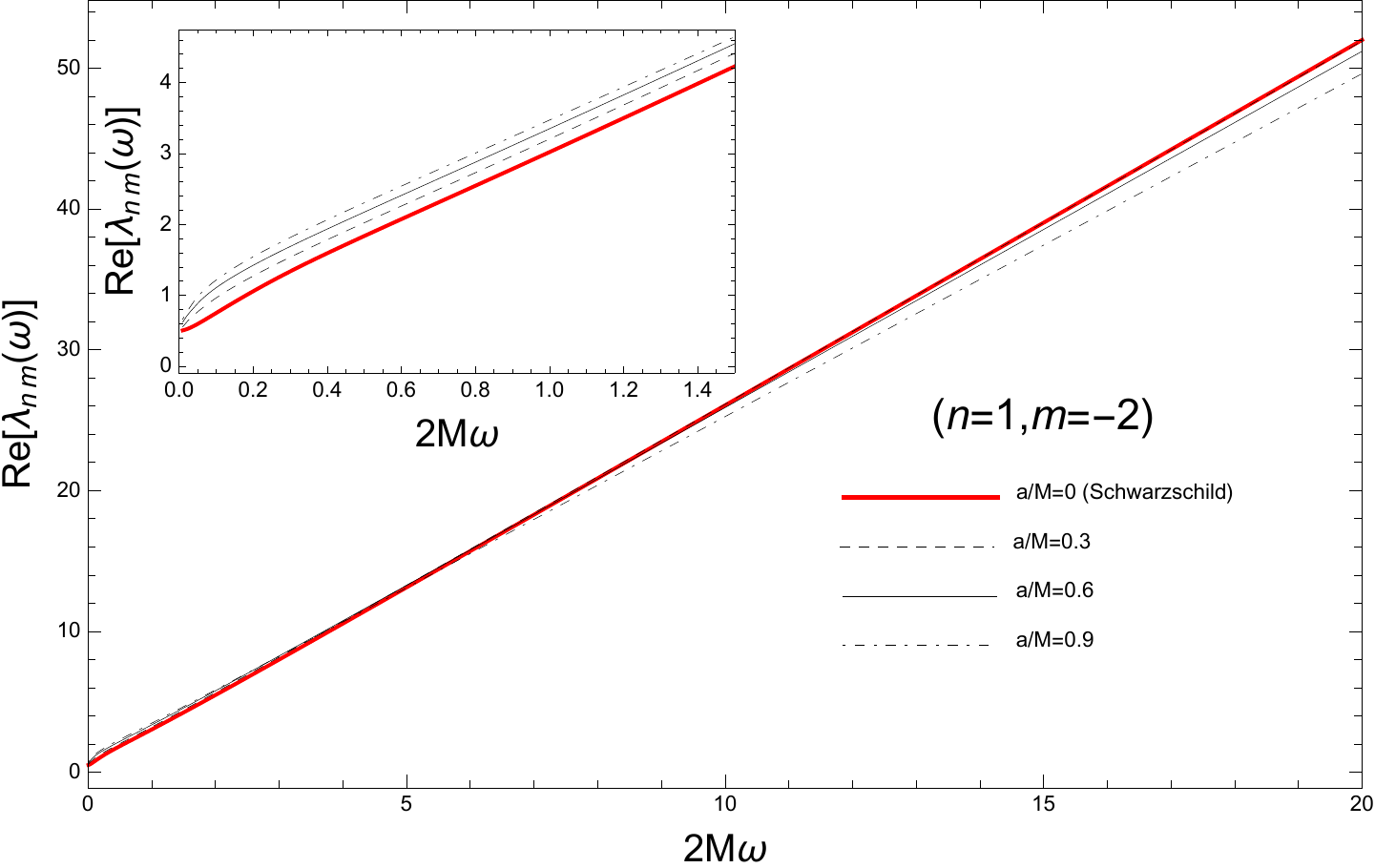}
\centering
\vspace{0.2cm}
\includegraphics[height=5.8cm,width=8.2cm]{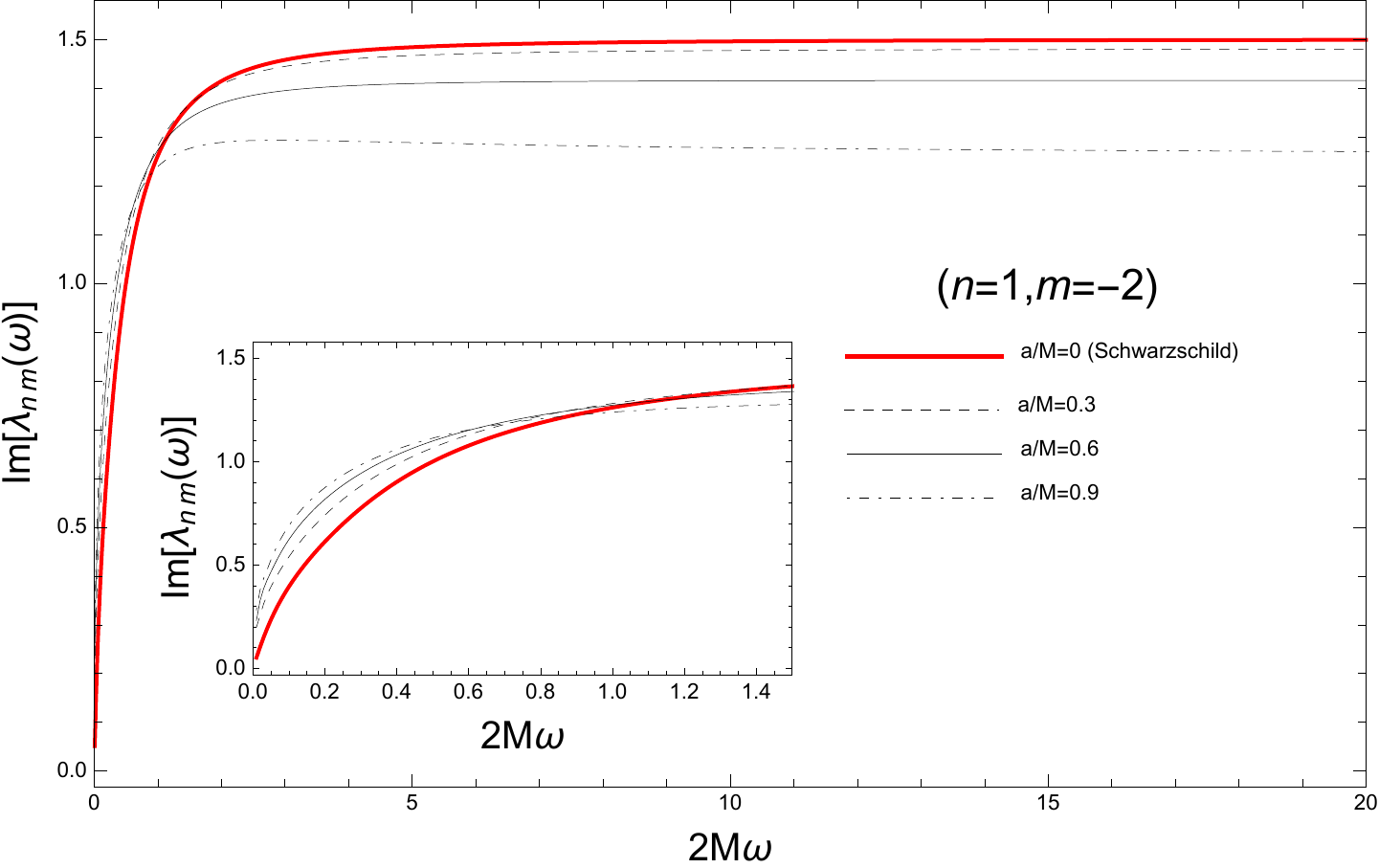}
\centering
\vspace{0.2cm}
\caption{\label{fig:RTPR2M0M2Mm2} Regge trajectories of the Regge poles $\lambda_{n=1,m=0}(\omega)$, $\lambda_{n=1,m=2}(\omega)$ and $\lambda_{n=1,m=-2}(\omega)$ of the Kerr BH for the rotation rates $a/M= 0, 0.30, 0.60, 0.90$. We consider the gravitational perturbations ($s=-2$) of the BH. The behavior of $\lambda_{n=1,m=0}(\omega)$ and $\lambda_{n=1,m=2}(\omega)$ is typical of the Kerr Regge poles $\lambda_{n=1, m}(\omega)$ with $m \ge 0$ while the behavior of $\lambda_{n=1,m=-2}(\omega)$ is typical of the Kerr Regge poles $\lambda_{n=1, m}(\omega)$ with $m < 0$. In particular, it should be that, for a given value of $a/M$, the Regge trajectories of $\mathrm{Re}[\lambda_{n=1,m=0}(\omega)]$ and $\mathrm{Re}[\lambda_{n=1,m=2}(\omega)]$ lie below that of the corresponding Schwarzschild Regge pole and those with lower rotation rates while the Regge trajectory of $\mathrm{Re}[\lambda_{n=1,m=-2}(\omega)]$ lies below that of the corresponding Schwarzschild Regge pole and those with lower rotation rates only for high frequencies. The results obtained for $m \ge 0$ are affected for low frequencies by numerical instabilities (see explanations and discussions in the text).}
\end{figure*}

It is interesting to note that, for a given value of $a/M$, the Regge trajectory of $\mathrm{Re}[\lambda_{n,m}(\omega)]$ with $m \ge 0$ lies below that of the corresponding Schwarzschild Regge pole and those with lower rotation rates while the Regge trajectory of $\mathrm{Re}[\lambda_{n,m}(\omega)]$ with $m < 0$ lies below that of the corresponding Schwarzschild Regge pole and those with lower rotation rates only for high frequencies. These facts can be observed in Figs.~\ref{fig:RTPR1M0M2Mm2} and \ref{fig:RTPR2M0M2Mm2} for $n=0, 1$ and $m=-2, 0, +2$. Such results combined with the semiclassical relations (\ref{sc1}) and (\ref{sc2}) explain the behavior of the quasinormal frequencies $\omega_{\ell m n}$ as $a/M$ increases, i.e., the fact that their real part always increases when $m \ge 0$ but decreases for low values of $\ell$ when $m<0$.

We can finally remark that Tables \ref{tab:tablePR1m0}-\ref{tab:tablePR2mm2} exhibit the breaking of the azimuthal degeneracy of the quasinormal frequencies of the Schwarzschild BH due to rotation.  Of course, this fact is well known (see, e.g., Ref~\cite{Detweiler:1980gk}) but it is interesting to realize that it can be here interpreted as a consequence, via the semiclassical relations (\ref{sc1}) and (\ref{sc2}), of the splitting of the Schwarzschild Regge poles and Regge trajectories discussed in Secs.~\ref{RTQNF_res1} and \ref{RTQNF_res2}.

\begingroup
\squeezetable
\begin{table}[H]
\caption{\label{tab:tablePR1m0} QNM frequencies $\omega_{\ell m n}$ for gravitational perturbations ($s=-2$) of the Kerr BH: semiclassical results obtained from the trajectory of the $(n=0,m=0)$-Regge pole versus ``exact" ones obtained using Leaver's method and relative errors.}
\smallskip
\centering
\begin{ruledtabular}
\begin{tabular}{cccc}
$\ell $ & $a/M$  & $2 M\omega_{\ell 0 0}$ (exact) & $2M\omega_{\ell 0 0}$ (semiclassical)\\
\hline

 & & & \\

$2$  & $0$ &$ 0.74734337-0.17792463 i $  & $ 0.75812003 - 0.17620450 i $   \\
     &     &                  &       ($-1.4 \times 10^{-2}$, $9.7 \times 10^{-3}$)                     \\
     & $0.3$ &$ 0.75397012 - 0.17670656 i $  &$ 0.7645048 - 0.17509216 i $   \\
          &     &          &   ($-1.4 \times 10^{-2}$, $9.1 \times 10^{-3}$) \\
     & $0.6$ &$  0.77610784 - 0.17198934 i $  &$ 0.78581861 - 0.17066938 i $   \\
          &     &          &   ($  -1.3\times 10^{-2}$, $  7.7\times 10^{-3}$)                                                 \\
\end{tabular}
\end{ruledtabular}
\end{table}
\endgroup

\begingroup
\squeezetable
\begin{table}[H]
\caption{\label{tab:tablePR1m1} QNM frequencies $\omega_{\ell m n}$ for gravitational perturbations ($s=-2$) of the Kerr BH: semiclassical results obtained from the trajectory of the $(n=0,m=1)$-Regge pole versus ``exact" ones obtained using Leaver's method and relative errors.}
\smallskip
\centering
\begin{ruledtabular}
\begin{tabular}{cccc}
$\ell $ & $a/M$  & $2 M\omega_{\ell 1 0}$ (exact) & $2M\omega_{\ell 1 0}$ (semiclassical)\\
\hline

 & & & \\

$2$  & $0$ &$ 0.74734337-0.17792463 i $  & $ 0.75812003 - 0.17620450 i $   \\
     &     &                  &       ($-1.4 \times 10^{-2}$, $9.7 \times 10^{-3}$)                     \\
     & $0.3$ &$ 0.79466079 - 0.17599034 i $  &$ 0.80482722 - 0.17482294 i $   \\
          &     &          &   ($ -1.3 \times 10^{-2}$, $  6.6\times 10^{-3}$)                                             \\
     & $0.6$ &$ 0.87193694 - 0.16912836 i $  &$ 0.88029339 - 0.16888048 i $   \\
          &     &          &   ($  -9.6\times 10^{-3}$, $  1.4\times 10^{-3}$)                                                 \\
\end{tabular}
\end{ruledtabular}
\end{table}
\endgroup

\begingroup
\squeezetable
\begin{table}[H]
\caption{\label{tab:tablePR1m2} QNM frequencies $\omega_{\ell m n}$ for gravitational perturbations ($s=-2$) of the Kerr BH: semiclassical results obtained from the trajectory of the $(n=0,m=2)$-Regge pole versus ``exact" ones obtained using Leaver's method and relative errors. }
\smallskip
\centering
\begin{ruledtabular}
\begin{tabular}{cccc}
$\ell $ & $a/M$  & $2 M\omega_{\ell 2 0}$ (exact) & $2M\omega_{\ell 2 0}$ (semiclassical)\\
\hline

 & & & \\

$2$  & $0$ &$ 0.74734337-0.17792463 i $  & $ 0.75812003 - 0.17620450 i $   \\
     &     &                  &       ($-1.4 \times 10^{-2}$, $9.7 \times 10^{-3}$)                     \\
     & $0.3$ &$ 0.83905336 - 0.17545854 i $  &$ 0.84858806 - 0.17476829 i $   \\
          &     &          &   ($-1.1 \times 10^{-2}$, $3.9 \times 10^{-3}$) \\
     & $0.6$ &$  0.98808956 - 0.16753040 i $  &$ 0.99400184 - 0.16816867 i $   \\
          &     &          &   ($  6.0\times 10^{-3}$, $ -3.8\times 10^{-3}$)                                                 \\
\end{tabular}
\end{ruledtabular}
\end{table}
\endgroup

\begingroup
\squeezetable
\begin{table}[H]
\caption{\label{tab:tablePR1m4} QNM frequencies $\omega_{\ell m n}$ for gravitational perturbations ($s=-2$) of the Kerr BH: semiclassical results obtained from the trajectory of the $(n=0,m=4)$-Regge pole versus ``exact" ones obtained using Leaver's method and relative errors.}
\smallskip
\centering
\begin{ruledtabular}
\begin{tabular}{cccc}
$\ell $ & $a/M$  & $2 M\omega_{\ell 4 0}$ (exact) & $2M\omega_{\ell 4 0}$ (semiclassical)\\
\hline

 & & & \\

$4$  & $0$ &$ 1.61835676 - 0.18832792 i $  & $ 1.62051635 - 0.18797143 i$   \\
     &     &                  &       ($-1.3\times 10^{-3}$, $1.9 \times 10^{-3}$)                                                                                    \\ & $0.3$ &$ 1.81848331 - 0.18572991 i $  &$ 1.82093859 - 0.18534858 i $   \\
          &     &          &   ($ -1.4\times 10^{-3}$, $ 2.0 \times 10^{-3}$)                                                 \\
     & $0.6$ &$ 2.12996276 - 0.17581884  i $  &$ 2.13221213 - 0.17552052 i $   \\
          &     &          &   ($ -1.1 \times 10^{-3}$, $ 1.7 \times 10^{-3}$)                                                 \\
\end{tabular}
\end{ruledtabular}
\end{table}
\endgroup

\begingroup
\squeezetable
\begin{table}[H]
\caption{\label{tab:tablePR1mm1} QNM frequencies $\omega_{\ell m n}$ for gravitational perturbations ($s=-2$) of the Kerr BH: semiclassical results obtained from the trajectory of the $(n=0,m=-1)$-Regge pole versus ``exact" ones obtained using Leaver's method and relative errors.}
\smallskip
\centering
\begin{ruledtabular}
\begin{tabular}{cccc}
$\ell $ & $a/M$  & $2 M\omega_{\ell, -1, 0}$ (exact) & $2M\omega_{\ell, -1, 0}$ (semiclassical)\\
\hline

 & & & \\

$2$  & $0$ &$ 0.74734337-0.17792463 i $  & $ 0.75812003 - 0.17620450 i $   \\
     &     &                  &       ($-1.4 \times 10^{-2}$, $9.7 \times 10^{-3}$)                     \\
     & $0.3$ &$ 0.71673262 - 0.17752552 i $  &$ 0.72740309 - 0.17556047 i $   \\
          &     &          &   ($ -1.5 \times 10^{-2}$, $ 1.1 \times 10^{-2}$)                                             \\
     & $0.6$ &$  0.69782136 - 0.17513239 i $  &$ 0.70776543 - 0.17308593 i $   \\
          &     &          &   ($ -1.4 \times 10^{-2}$, $ 1.2 \times 10^{-2}$)                                                 \\
\end{tabular}
\end{ruledtabular}
\end{table}
\endgroup

\begingroup
\squeezetable
\begin{table}[H]
\caption{\label{tab:tablePR1mm2} QNM frequencies $\omega_{\ell m n}$ for gravitational perturbations ($s=-2$) of the Kerr BH: semiclassical results obtained from the trajectory of the $(n=0,m=-2)$-Regge pole versus ``exact" ones obtained using Leaver's method and relative errors.}
\smallskip
\centering
\begin{ruledtabular}
\begin{tabular}{cccc}
$\ell $ & $a/M$  & $2 M\omega_{\ell, -2, 0}$ (exact) & $2M\omega_{\ell, -2, 0}$ (semiclassical)\\
\hline

 & & & \\

$2$  & $0$ &$ 0.74734337-0.17792463 i $  & $ 0.75812003 - 0.17620450 i $   \\
     &     &                  &       ($-1.4 \times 10^{-2}$, $9.7 \times 10^{-3}$)                     \\
     & $0.3$ &$ 0.68266585 - 0.17836910 i $  &$ 0.69329002 - 0.17612130 i $   \\
          &     &          &   ($ -1.6 \times 10^{-2}$, $  1.3 \times 10^{-2}$)                                             \\
     & $0.6$ &$ 0.63356799 - 0.17778349  i $  &$ 0.6431444 - 0.17523126 i $   \\
          &     &          &   ($ -1.5 \times 10^{-2}$, $  1.4\times 10^{-2}$)                                                 \\
\end{tabular}
\end{ruledtabular}
\end{table}
\endgroup

\begingroup
\squeezetable
\begin{table}[H]
\caption{\label{tab:tablePR2m0} QNM frequencies $\omega_{\ell m n}$ for gravitational perturbations ($s=-2$) of the Kerr BH: semiclassical results obtained from the trajectory of the $(n=1,m=0)$-Regge pole versus ``exact" ones obtained using Leaver's method and relative errors.}
\smallskip
\centering
\begin{ruledtabular}
\begin{tabular}{cccc}
$\ell $ & $a/M$  & $2 M\omega_{\ell 0 1}$ (exact) & $2M\omega_{\ell 0 1}$ (semiclassical)\\
\hline

 & & & \\

$2$  & $0$ &$ 0.69342199 - 0.54782975 i $  & $ 0.78133808 - 0.49982731 i $   \\
     &     &                  &       ($-1.3\times 10^{-1}$, $8.8 \times 10^{-2}$)                                        \\ & $0.3$ &$ 0.70223138 - 0.54364265 i $  &$ 0.78806072 - 0.49706781 i $   \\
          &     &          &   ($ -1.2 \times 10^{-1}$, $ 8.6 \times 10^{-2}$)                                                 \\
     & $0.6$ &$  0.73106708 - 0.52750718 i $  &$ 0.81021312 - 0.48585265 i $   \\
          &     &          &   ($ -1.1 \times 10^{-1}$, $ 7.9 \times 10^{-2}$)                                                 \\
\end{tabular}
\end{ruledtabular}
\end{table}
\endgroup

\begingroup
\squeezetable
\begin{table}[H]
\caption{\label{tab:tablePR2m1} QNM frequencies $\omega_{\ell m n}$ for gravitational perturbations ($s=-2$) of the Kerr BH: semiclassical results obtained from the trajectory of the $(n=1,m=1)$-Regge pole versus ``exact" ones obtained using Leaver's method and relative errors.}
\smallskip
\centering
\begin{ruledtabular}
\begin{tabular}{cccc}
$\ell $ & $a/M$  & $2 M\omega_{\ell 1 1}$ (exact) & $2M\omega_{\ell 1 1}$ (semiclassical)\\
\hline

 & & & \\

$2$  & $0$ &$ 0.69342199 - 0.54782975 i $  & $ 0.78133808 - 0.49982731 i $   \\
     &     &                  &       ($-1.3\times 10^{-1}$, $8.8 \times 10^{-2}$)                                        \\ & $0.3$ &$ 0.74802871 - 0.53950712 i $  &$ 0.83527633 - 0.49307644 i $   \\
          &     &          &   ($ -1.2\times 10^{-1}$, $ 8.6 \times 10^{-2}$)                                                 \\
     & $0.6$ &$  0.83667406 - 0.51509385 i $  &$ 0.91882854 - 0.4722063 i $   \\
          &     &          &   ($ -9.8 \times 10^{-2}$, $ 8.3 \times 10^{-2}$)                                                 \\
\end{tabular}
\end{ruledtabular}
\end{table}
\endgroup

\begingroup
\squeezetable
\begin{table}[H]
\caption{\label{tab:tablePR2m2} QNM frequencies $\omega_{\ell m n}$ for gravitational perturbations ($s=-2$) of the Kerr BH: semiclassical results obtained from the trajectory of the $(n=1,m=2)$-Regge pole versus ``exact" ones obtained using Leaver's method and relative errors.}
\smallskip
\centering
\begin{ruledtabular}
\begin{tabular}{cccc}
$\ell $ & $a/M$  & $2 M\omega_{\ell 2 1}$ (exact) & $2M\omega_{\ell 2 1}$ (semiclassical)\\
\hline

 & & & \\

$2$  & $0$ &$ 0.69342199 - 0.54782975 i $  & $ 0.78133808 - 0.49982731 i $   \\
     &     &                  &       ($-1.3\times 10^{-1}$, $8.8 \times 10^{-2}$)                                        \\ & $0.3$ &$ 0.79678065 - 0.53609714 i $  &$ 0.88570063 - 0.48827981 i $   \\
          &     &          &   ($  -1.1\times 10^{-1}$, $  8.9\times 10^{-2}$)                                                 \\
     & $0.6$ &$  0.95961333 - 0.50769373 i $  &$ 1.0430397 - 0.45712999 i $   \\
          &     &          &   ($ -8.7 \times 10^{-2}$, $  1.0\times 10^{-1}$)                                                 \\
\end{tabular}
\end{ruledtabular}
\end{table}
\endgroup

\begingroup
\squeezetable
\begin{table}[H]
\caption{\label{tab:tablePR2m4} QNM frequencies $\omega_{\ell m n}$ for gravitational perturbations ($s=-2$) of the Kerr BH: semiclassical results obtained from the trajectory of the $(n=1,m=4)$-Regge pole versus ``exact" ones obtained using Leaver's method and relative errors.}
\smallskip
\centering
\begin{ruledtabular}
\begin{tabular}{cccc}
$\ell $ & $a/M$  & $2 M\omega_{\ell 4 1}$ (exact) & $2M\omega_{\ell 4 1}$ (semiclassical)\\
\hline

 & & & \\

$4$  & $0$ &$ 1.59326306 - 0.56866870 i $  & $ 1.61119084 - 0.56028355 i$   \\
     &     &                  &       ($-1.1\times 10^{-2}$, $1.5 \times 10^{-2}$)                                                            \\ & $0.3$ &$ 1.79940517 - 0.55982668 i $  &$ 1.81910724 - 0.55096038 i $   \\
          &     &          &   ($ -1.1 \times 10^{-2}$, $  1.6\times 10^{-2}$)                                                 \\
     & $0.6$ &$ 2.11818846 - 0.52890413  i $  &$ 2.13674231 - 0.52066689 i $   \\
          &     &          &   ($ -8.8 \times 10^{-3}$, $  1.6\times 10^{-2}$)                                                 \\
\end{tabular}
\end{ruledtabular}
\end{table}
\endgroup

\begingroup
\squeezetable
\begin{table}[H]
\caption{\label{tab:tablePR2mm1} QNM frequencies $\omega_{\ell m n}$ for gravitational perturbations ($s=-2$) of the Kerr BH: semiclassical results obtained from the trajectory of the $(n=1,m=-1)$-Regge pole versus ``exact" ones obtained using Leaver's method and relative errors.}
\smallskip
\centering
\begin{ruledtabular}
\begin{tabular}{cccc}
$\ell $ & $a/M$  & $2 M\omega_{\ell, -1, 1}$ (exact) & $2M\omega_{\ell, -1, 1}$ (semiclassical)\\
\hline

 & & & \\

$2$  & $0$ &$ 0.69342199 - 0.54782975 i $  & $ 0.78133808 - 0.49982731 i $   \\
     &     &                  &       ($-1.3\times 10^{-1}$, $8.8 \times 10^{-2}$)                                        \\ & $0.3$ &$ 0.65916485 - 0.54822313 i $  &$ 0.74404009 - 0.50010244 i $   \\
          &     &          &   ($ -1.3\times 10^{-1}$, $ 8.8 \times 10^{-2}$)                                                 \\
     & $0.6$ &$  0.64012741 - 0.54156487 i $  &$ 0.71741656 - 0.49521128 i $   \\
          &     &          &   ($ -1.2 \times 10^{-1}$, $ 8.6 \times 10^{-2}$)                                                 \\
\end{tabular}
\end{ruledtabular}
\end{table}
\endgroup

\begingroup
\squeezetable
\begin{table}[H]
\caption{\label{tab:tablePR2mm2} QNM frequencies $\omega_{\ell m n}$ for gravitational perturbations ($s=-2$) of the Kerr BH: semiclassical results obtained from the trajectory of the $(n=1,m=-2)$-Regge pole versus ``exact" ones obtained using Leaver's method and relative errors.}
\smallskip
\centering
\begin{ruledtabular}
\begin{tabular}{cccc}
$\ell $ & $a/M$  & $2 M\omega_{\ell, -2, 1}$ (exact) & $2M\omega_{\ell, -2, 1}$ (semiclassical)\\
\hline

 & & & \\

$2$  & $0$ &$ 0.69342199 - 0.54782975 i $  & $ 0.78133808 - 0.49982731 i $   \\
     &     &                  &       ($-1.3\times 10^{-1}$, $8.8 \times 10^{-2}$)                                        \\ & $0.3$ &$ 0.61858926 - 0.55299074 i $  &$ 0.70315887 - 0.50205236 i $   \\
          &     &          &   ($ -1.4 \times 10^{-1}$, $  9.2\times 10^{-2}$)                                                 \\
     & $0.6$ &$  0.56068443 - 0.55465252 i $  &$ 0.63943331 - 0.49866793 i $   \\
          &     &          &   ($ -1.4 \times 10^{-1}$, $  1.0\times 10^{-1}$)                                                 \\
\end{tabular}
\end{ruledtabular}
\end{table}
\endgroup

\section{Conclusion and perspectives}
\label{CandP}

\subsection{Main results}
\label{MainR}

In the present paper, we have derived and subsequently utilized semiclassical relations permitting us to relate numerically the quasinormal frequencies of the Kerr BH with its Regge poles or, more precisely, to obtain the complex frequencies of the weakly damped QNMs from the Regge trajectories, i.e., from the curves traced out in the CAM plane by the (lowest) Regge poles as a function of the frequency $\omega \in \mathbb{R}$. It is important to keep in mind that the semiclassical formulas (\ref{sc1}) and (\ref{sc2}) used in our calculations have been obtained under assumptions [see Eqs.~(\ref{hyp1_scQNM}) and (\ref{hyp2_scQNM})] that do not hold for higher overtones, i.e, higher values of $n$ and are in fact formally valid for high frequencies. In the eikonal regime, i.e., for $\ell \gg 1$, the agreement between exact and semiclassical results is impressive and, despite the assumptions previously mentioned, this agreement remains good even for rather low frequencies. We have moreover shown that the splitting of each Regge pole of the Schwarzschild BH into an infinite number of Kerr Regge poles can explain semiclassically the breaking of the azimuthal degeneracy of the quasinormal frequencies of the Schwarzschild BH due to rotation.

\subsection{Existing challenges}
\label{ExistingC}
It is important to point out that the Regge pole analysis of the QNM frequencies we have performed in this article cannot be used, in its current form, to interpret several important aspects of Kerr BH physics and, in particular, the superradiance effect (see, e.g., Ref.~\cite{Brito:2015oca} and references therein) which occurs for low or very low frequencies, the existence of purely imaginary quasinormal frequencies \cite{Chandrashekhar1984,Leaver:1985ax} (see also Ref.~\cite{Cook:2016ngj} and references therein) and of quasinormal frequencies with very large imaginary part \cite{Leaver:1985ax}. Moreover, the instabilities we have encountered when $a/M \to 1$, which are due to our numerical method, prevent us from recovering the accumulation of the quasinormal frequencies $\omega_{\ell m n}$ near the undamped critical frequency $\omega_c =m$ \cite{Detweiler:1980gk} even if such a trend seems to be present in Tables \ref{tab:tablePR1m2} and \ref{tab:tablePR2m2}.

At the moment, we have not succeeded in interpreting geometrically the Kerr Regge poles in order to derive accurate analytical formulas for the associated Regge trajectories and, as a by-product, for the complex frequencies of the weakly damped QNMs. To tell the truth, that was our initial motivation and, by starting this work, we hoped to verify the analytical formulas obtained from our numerical results. Let us recall that, in the case of static spherically symmetric BHs, it was possible to achieve such a program \cite{Decanini:2009mu,Decanini:2010fz} by emphasizing the role played by the BH photon sphere, i.e., by the hypersurface on which massless particles can orbit a BH on unstable circular null geodesics. Indeed, by adapting the WKB methods \cite{Schutz:1985km,Iyer:1986np,Iyer:1986nq,Will:1988zz,Konoplya:2003ii} developed to
study the resonant behaviour of BHs, it was possible to derive high-frequency asymptotic expansions for the Regge poles taking into account the spin dependence of the field considered. This permitted us to describe very accurately the Regge trajectories and to recover from them well-known analytical expressions for the quasinormal frequencies. For the Kerr BH, we have encountered various difficulties mainly due to the rich structure of the set of periodic unstable null geodesics of this spacetime, the coupling of the Teukolsky angular and radial equations and the behaviour of the Teukolsky potential which have prevented us from carrying out an analogous program. For potential avenues that might circumvent these difficulties, see Sec.~\ref{furthav}.

In fact, similar difficulties have been encountered by the various authors who have directly considered the computation of the QNM frequencies of the Kerr BH from WKB methods (see, e.g., the pioneering article by Seidel and Iyer \cite{Seidel:1989bp} and the more recent articles by Yang {\it et al.} \cite{Yang:2012he,Yang:2012pj,Yang:2013uba}). In our opinion, we cannot be satisfied with their results which have been derived by neglecting the spin dependent terms and/or having high error bars for the astrophysically relevant $s=-2$ case even if, in some particular cases, appealing interpretation of the quasinormal frequencies $\omega_{\ell m n}$ and of the separation constants ${_{s}A}_{\ell m}(a \omega_{\ell m n})$ have been provided in terms of conserved quantities in geodesic motion \cite{Yang:2012he}. Anyway, the analytical results displayed in these articles are not as elegant as those obtained for static spherically symmetric BHs but, of course, it seems that this is the price to pay when working in Kerr spacetime.

It is interesting to recall that Dolan and Ottewill in Ref.~\cite{Dolan:2009nk} have considered the determination of the quasinormal frequencies or the Regge poles of static spherically symmetric BHs from an alternative point of view based on an ansatz for the QNMs or the Regge modes which relates the high-$\ell $ modes or the high-frequency modes to null geodesics starting at infinity and ending around the photon sphere (see also Ref.~\cite{Decanini:2011eh} for an extension of this method to the massive scalar field in Schwarzschild spacetime). This powerful method, which permits one to recover all the results obtained by using WKB methods seems more efficient at capturing higher-order terms in the asymptotic expansions of quasinormal frequencies and Regge poles. In Ref.~\cite{Dolan:2010wr}, Dolan has adapted his method to the Kerr BH which allowed him to obtain for the complex frequencies $\omega_{\ell m n}$ with $m = \pm \ell$ and $m=0$ of the weakly damped QNMs interesting expansions with spin-dependent corrections which are valid in the eikonal limit ($\ell \gg 1$). In his derivation, the unstable circular equatorial and polar null geodesics play a central role and the corresponding orbital frequencies and Lyapunov exponents are therefore present in the expansions obtained\footnote{Dolan's approach has permitted us to derive easily the leading-order terms for the Regge poles $\lambda_{n, m=0} (\omega) $. In the high-frequency regime, we can write $\lambda_{n, m=0} (\omega) = b_0  \, \omega + i \, (n-1/2) + {\cal O} (1/\omega)$ where $b_0 = \sqrt{ \frac{(3{r_0}^2 - a^2)({r_0}^2 + a^2)}{{r_0}^2 - a^2}}$ denotes the impact parameter associated with the unstable circular null orbit of radius $r_0 = M + 2\sqrt{M^2 - a^2/3} \, \cos \left[\frac{1}{3} \arccos \left(\frac{M(M^2-a^2)}{(M^2-a^2/3)^{3/2}}\right)\right]$ which corresponds to a photon with zero angular momentum \cite{Teo:2003}. Of course, from the semiclassical formulas (\ref{sc1}) and (\ref{sc2}), we can then obtain the leading-order terms for the complex frequencies $\omega_{\ell m n}$ when $m=0$. Unfortunately, taking into account for the Regge poles the dependence on the azimuthal number $m \not= 0$ and on the spin $s$ and capturing higher-order terms is not really obvious and leads to heavy formulas.}. However, even this appealing approach provides results which are not as elegant as those obtained for static spherically symmetric BHs and does not permit one to deal with the case where the azimuthal number $m$ is arbitrary.

\subsection{Further avenues to explore}
\label{furthav}
In our opinion, it seems very likely that some new analytical tools are required to establish the correspondence between Kerr QNMs and the spacetime dynamics in full generality. What works very well for static spherically symmetric BHs no longer works for the Kerr BH. In particular, to reason only on Kerr circular null geodesics oversimplifies the approach. Notwithstanding these difficulties, there are a number of potential avenues, all of which are active areas of research in their own right, whose synthesis might improve our understanding on the issue:

\begin{enumerate}[label=(\arabic*)]

   \item Utilization of Heun properties of the Teukolsky radial and angular equations: It has been established by Fiziev that for Kerr spacetime the Teukolsky radial and angular equations  (\ref{eq:TREhom}) and (\ref{eq:TAE}) are types of confluent Heun equations \cite{Fiziev:2009wn}. Very recently, the general Heun functions were used by Hatsuda for studying QNMs of Kerr-de~Sitter BHs \cite{Hatsuda:2020sbn}. However, if we wish to study Kerr QNMs using Regge pole theory in the Heun framework, we would have to circumvent the crucial difficulty we faced in imposition of boundary conditions. This indicates the need for developing better analytical tools with well-postured numerical abilities to handle coupled Heun differential equations. For recent work of one of the authors in this direction, see Ref.~\cite{Giscard:2020iqg}.

\item Utilization of the Sasaki-Nakamura form of the Teukolsky equation \cite{Sasaki:1981sx} to study the Regge pole behaviour: Recently, in a series of papers \cite{Nakamura:2016gri,Nakamura:2016yjl,Nakano:2016sgf}, Nakamura and collaborators have extended the conventional WKB method of Schutz, Will and Iyer used for obtaining QNM results for the Schwarzschild BH \cite{Schutz:1985km,Iyer:1986np,Iyer:1986nq} to deal with $s=-2$ Kerr QNMs by considering the ``potential term'' of the Sasaki-Nakamura equation (a short-ranged version of the Teukolsky potential). For a wide range of values of $a/M$, including the high spin $a/M = 0.99$ case, they have obtained interesting (but purely numerical) results by associating the maximum of the Sasaki-Nakamura potential with the dominant Kerr QNM. Therefore, exploring whether the Sasaki-Nakamura equation allows us to obtain analytical results for Regge poles seems like a plausible avenue for future research, although the cumbersome nature of the Sasaki-Nakamura equation is expected to make the analysis difficult to carry out, to say the least.

\item Lessons from the hydrogen molecular ion ${\mathrm{H}_2}^+$ and the helium atom $\mathrm{He}$: The interpretation problem of the Kerr QNMs has some analogies with that of the semiclassical quantization of the hydrogen molecular ion (let us recall that its study has inspired Leaver's method \cite{Leaver:1985ax,LeaverJMP1986}) and of the helium atom. In the early 20th century, there had been unsuccessful attempts to compute the energy levels of these structures using the Bohr-Sommerfeld quantization method. This was only later made possible by solving the Schr{\"o}dinger equation. The relevant fact for our study is that this semiclassical quantization was carried out rather recently for the hydrogen molecular ion (see, e.g., Refs.~\cite{StrandReinhardt1979,DuanYuanBao1995,DuanYuan1999}) and for the helium atom (see, e.g., Refs.~\cite{Ezra1991,WintgenRichterTanner1992}) from modern semiclassical techniques (trace formulas, spectral determinants, cycle expansions, EBK and uniform EBK rules,\dots) by exploiting the full classical dynamics of the electrons. It would be interesting to explore whether such semiclassical techniques can provide analytical tools necessary to circumvent the difficulties that we faced. For example, keeping in mind the integrability of geodesic motion in Kerr \cite{Carter:1968rr}, an approach based on the Berry-Tabor trace formula \cite{BerryTabor1976,BerryTabor1977}, or more precisely on its extension to open systems, by taking into account the whole set of periodic null geodesics, could be very interesting. The major caveat is that even if they can give us accurate results for the entire range of QNMs (not just for the weakly damped ones), these methods will neither provide simple analytical formulas linked with spacetime properties, nor provide a clear physical interpretation.

    \item Extension of the CAM method: In this article, we have derived a CAM representation and a Regge pole approximation of the retarded Green's function by complexifying the ``angular momentum number'' $\ell$ appearing in its spectral decomposition (\ref{eq:GreenFexp_def}). This is a rather natural approach if we want to understand the transition from the Schwarzschild to the Kerr BH and, in particular, the breaking of the azimuthal degeneracy due to rotation and its consequences as far as Regge poles and quasinormal frequencies are concerned. However, keeping in mind effective resummations of the partial wave expansion defining the retarded Green's function, it would be interesting to complexify the ``magnetic number'' $m$ or to complexify both $\ell$ and $m$. In that last case, it would be necessary to generalize the Sommerfeld-Watson transformation involved in Eq.~(\ref{eq:GreenFexp_SW}) and to use Cauchy's residue theorem for functions of two complex variables. Tools to work in this direction have been forged a long time ago by Poincar\'e \cite{Poincare1887} (see also, for interesting remarks on this topic and additional references, the recent article by Esposito \cite{Esposito:2020prr} where the author considers the possibility to analyze potential scattering in quantum mechanics by complexifying both the angular momentum number $\ell$ and the space dimension).
\end{enumerate}

\subsection{Epilogue}
 We hope that the results obtained in this paper and in previous ones \cite{Andersson:1994rk,Andersson:1994rm,Decanini:2002ha,Glampedakis:2003dn,
 Decanini:2009mu,Decanini:2009dn,Dolan:2009nk,Decanini:2010fz,Decanini:2011xi,Decanini:2011eh,Macedo:2013afa,Folacci:2019cmc,
Folacci:2019vtt,OuldElHadj:2019kji,Folacci:2018sef,Folacci:2020ekl} as well as the aforementioned avenues would stimulate research in incorporating Regge pole theory into modern gravitational physics and, in particular, in the fields of strong lensing and BH spectroscopy.

\section{Acknowledgements}
AF wishes to thank Yves Decanini for conversations concerning the CAM approach to BH physics during the last years. AT acknowledges discussions with Felix Finster in early stages of this work on various features of WKB analysis using the complex potential form of the Teukolsky equation.

\bibliography{QNMReggeT}

\begin{thebibliography}{123}%
\makeatletter
\providecommand \@ifxundefined [1]{%
 \@ifx{#1\undefined}
}%
\providecommand \@ifnum [1]{%
 \ifnum #1\expandafter \@firstoftwo
 \else \expandafter \@secondoftwo
 \fi
}%
\providecommand \@ifx [1]{%
 \ifx #1\expandafter \@firstoftwo
 \else \expandafter \@secondoftwo
 \fi
}%
\providecommand \natexlab [1]{#1}%
\providecommand \enquote  [1]{``#1''}%
\providecommand \bibnamefont  [1]{#1}%
\providecommand \bibfnamefont [1]{#1}%
\providecommand \citenamefont [1]{#1}%
\providecommand \href@noop [0]{\@secondoftwo}%
\providecommand \href [0]{\begingroup \@sanitize@url \@href}%
\providecommand \@href[1]{\@@startlink{#1}\@@href}%
\providecommand \@@href[1]{\endgroup#1\@@endlink}%
\providecommand \@sanitize@url [0]{\catcode `\\12\catcode `\$12\catcode
  `\&12\catcode `\#12\catcode `\^12\catcode `\_12\catcode `\%12\relax}%
\providecommand \@@startlink[1]{}%
\providecommand \@@endlink[0]{}%
\providecommand \url  [0]{\begingroup\@sanitize@url \@url }%
\providecommand \@url [1]{\endgroup\@href {#1}{\urlprefix }}%
\providecommand \urlprefix  [0]{URL }%
\providecommand \Eprint [0]{\href }%
\providecommand \doibase [0]{https://doi.org/}%
\providecommand \selectlanguage [0]{\@gobble}%
\providecommand \bibinfo  [0]{\@secondoftwo}%
\providecommand \bibfield  [0]{\@secondoftwo}%
\providecommand \translation [1]{[#1]}%
\providecommand \BibitemOpen [0]{}%
\providecommand \bibitemStop [0]{}%
\providecommand \bibitemNoStop [0]{.\EOS\space}%
\providecommand \EOS [0]{\spacefactor3000\relax}%
\providecommand \BibitemShut  [1]{\csname bibitem#1\endcsname}%
\let\auto@bib@innerbib\@empty
\bibitem [{\citenamefont {Kerr}(1963)}]{Kerr:1963ud}%
  \BibitemOpen
  \bibfield  {author} {\bibinfo {author} {\bibfnamefont {R.~P.}\ \bibnamefont
  {Kerr}},\ }\bibfield  {title} {\bibinfo {title} {{Gravitational field of a
  spinning mass as an example of algebraically special metrics}},\ }\href
  {https://doi.org/10.1103/PhysRevLett.11.237} {\bibfield  {journal} {\bibinfo
  {journal} {Phys.\ Rev.\ Lett.}\ }\textbf {\bibinfo {volume} {11}},\ \bibinfo
  {pages} {237} (\bibinfo {year} {1963})}\BibitemShut {NoStop}%
\bibitem [{\citenamefont {Chandrasekhar}(1983)}]{Chandrasekhar:1985kt}%
  \BibitemOpen
  \bibfield  {author} {\bibinfo {author} {\bibfnamefont {S.}~\bibnamefont
  {Chandrasekhar}},\ }\href@noop {} {\emph {\bibinfo {title} {{The Mathematical
  Theory of Black Holes}}}}\ (\bibinfo  {publisher} {Oxford University Press,
  Oxford},\ \bibinfo {year} {1983})\BibitemShut {NoStop}%
\bibitem [{\citenamefont {O’Niell}(1995)}]{BarrettONiell1995}%
  \BibitemOpen
  \bibfield  {author} {\bibinfo {author} {\bibfnamefont {B.}~\bibnamefont
  {O’Niell}},\ }\href@noop {} {\emph {\bibinfo {title} {{Geometry of Kerr
  Black Holes}}}}\ (\bibinfo  {publisher} {A K Peters, Wellesley},\ \bibinfo
  {year} {1995})\BibitemShut {NoStop}%
\bibitem [{\citenamefont {Teukolsky}(2015)}]{Teukolsky:2014vca}%
  \BibitemOpen
  \bibfield  {author} {\bibinfo {author} {\bibfnamefont {S.~A.}\ \bibnamefont
  {Teukolsky}},\ }\bibfield  {title} {\bibinfo {title} {{The Kerr metric}},\
  }\href {https://doi.org/10.1088/0264-9381/32/12/124006} {\bibfield  {journal}
  {\bibinfo  {journal} {Class.\ Quant.\ Grav.}\ }\textbf {\bibinfo {volume}
  {32}},\ \bibinfo {pages} {124006} (\bibinfo {year} {2015})},\ \Eprint
  {https://arxiv.org/abs/1410.2130} {arXiv:1410.2130 [gr-qc]} \BibitemShut
  {NoStop}%
\bibitem [{\citenamefont {Abbott}\ \emph {et~al.}(2016)\citenamefont {Abbott}
  \emph {et~al.}}]{Abbott:2016blz}%
  \BibitemOpen
  \bibfield  {author} {\bibinfo {author} {\bibfnamefont {B.~P.}\ \bibnamefont
  {Abbott}} \emph {et~al.} (\bibinfo {collaboration} {LIGO Scientific
  Collaboration, Virgo Collaboration}),\ }\bibfield  {title} {\bibinfo {title}
  {{Observation of gravitational waves from a binary black hole merger}},\
  }\href {https://doi.org/10.1103/PhysRevLett.116.061102} {\bibfield  {journal}
  {\bibinfo  {journal} {Phys.\ Rev.\ Lett.}\ }\textbf {\bibinfo {volume}
  {116}},\ \bibinfo {pages} {061102} (\bibinfo {year} {2016})},\ \Eprint
  {https://arxiv.org/abs/1602.03837} {arXiv:1602.03837 [gr-qc]} \BibitemShut
  {NoStop}%
\bibitem [{\citenamefont {Akiyama}\ \emph {et~al.}(2019)\citenamefont {Akiyama}
  \emph {et~al.}}]{Akiyama:2019cqa}%
  \BibitemOpen
  \bibfield  {author} {\bibinfo {author} {\bibfnamefont {K.}~\bibnamefont
  {Akiyama}} \emph {et~al.} (\bibinfo {collaboration} {Event Horizon Telescope
  Collaboration}),\ }\bibfield  {title} {\bibinfo {title} {{First M87 Event
  Horizon Telescope results. I. The shadow of the supermassive black hole}},\
  }\href {https://doi.org/10.3847/2041-8213/ab0ec7} {\bibfield  {journal}
  {\bibinfo  {journal} {Astrophys.\ J.\ Lett.}\ }\textbf {\bibinfo {volume}
  {875}},\ \bibinfo {pages} {L1} (\bibinfo {year} {2019})},\ \Eprint
  {https://arxiv.org/abs/1906.11238} {arXiv:1906.11238 [astro-ph.GA]}
  \BibitemShut {NoStop}%
\bibitem [{\citenamefont {Jusufi}(2020)}]{Jusufi:2020dhz}%
  \BibitemOpen
  \bibfield  {author} {\bibinfo {author} {\bibfnamefont {K.}~\bibnamefont
  {Jusufi}},\ }\bibfield  {title} {\bibinfo {title} {{Connection between the
  shadow radius and quasinormal modes in rotating spacetimes}},\ }\href
  {https://doi.org/10.1103/PhysRevD.101.124063} {\bibfield  {journal} {\bibinfo
   {journal} {Phys.\ Rev.\ D}\ }\textbf {\bibinfo {volume} {101}},\ \bibinfo
  {pages} {124063} (\bibinfo {year} {2020})},\ \Eprint
  {https://arxiv.org/abs/2004.04664} {arXiv:2004.04664 [gr-qc]} \BibitemShut
  {NoStop}%
\bibitem [{\citenamefont {Chesler}\ \emph {et~al.}(2020)\citenamefont
  {Chesler}, \citenamefont {Blackburn}, \citenamefont {Doeleman}, \citenamefont
  {Johnson}, \citenamefont {Moran}, \citenamefont {Narayan},\ and\
  \citenamefont {Wielgus}}]{Chesler:2020gtw}%
  \BibitemOpen
  \bibfield  {author} {\bibinfo {author} {\bibfnamefont {P.~M.}\ \bibnamefont
  {Chesler}}, \bibinfo {author} {\bibfnamefont {L.}~\bibnamefont {Blackburn}},
  \bibinfo {author} {\bibfnamefont {S.~S.}\ \bibnamefont {Doeleman}}, \bibinfo
  {author} {\bibfnamefont {M.~D.}\ \bibnamefont {Johnson}}, \bibinfo {author}
  {\bibfnamefont {J.~M.}\ \bibnamefont {Moran}}, \bibinfo {author}
  {\bibfnamefont {R.}~\bibnamefont {Narayan}},\ and\ \bibinfo {author}
  {\bibfnamefont {M.}~\bibnamefont {Wielgus}},\ }\bibfield  {title} {\bibinfo
  {title} {{Light echos and coherent autocorrelations in a black hole
  spacetime}},\ }\href@noop {} {\  (\bibinfo {year} {2020})},\ \Eprint
  {https://arxiv.org/abs/2012.11778} {arXiv:2012.11778 [gr-qc]} \BibitemShut
  {NoStop}%
\bibitem [{\citenamefont {Nakamura}\ \emph {et~al.}(2016)\citenamefont
  {Nakamura}, \citenamefont {Nakano},\ and\ \citenamefont
  {Tanaka}}]{Nakamura:2016gri}%
  \BibitemOpen
  \bibfield  {author} {\bibinfo {author} {\bibfnamefont {T.}~\bibnamefont
  {Nakamura}}, \bibinfo {author} {\bibfnamefont {H.}~\bibnamefont {Nakano}},\
  and\ \bibinfo {author} {\bibfnamefont {T.}~\bibnamefont {Tanaka}},\
  }\bibfield  {title} {\bibinfo {title} {{Detecting quasinormal modes of binary
  black hole mergers with second-generation gravitational-wave detectors}},\
  }\href {https://doi.org/10.1103/PhysRevD.93.044048} {\bibfield  {journal}
  {\bibinfo  {journal} {Phys.\ Rev.\ D}\ }\textbf {\bibinfo {volume} {93}},\
  \bibinfo {pages} {044048} (\bibinfo {year} {2016})},\ \Eprint
  {https://arxiv.org/abs/1601.00356} {arXiv:1601.00356 [astro-ph.HE]}
  \BibitemShut {NoStop}%
\bibitem [{\citenamefont {Chirenti}(2018)}]{Chirenti:2017mwe}%
  \BibitemOpen
  \bibfield  {author} {\bibinfo {author} {\bibfnamefont {C.}~\bibnamefont
  {Chirenti}},\ }\bibfield  {title} {\bibinfo {title} {{Black hole quasinormal
  modes in the era of LIGO}},\ }\href
  {https://doi.org/10.1007/s13538-017-0543-7} {\bibfield  {journal} {\bibinfo
  {journal} {Braz.\ J.\ Phys.}\ }\textbf {\bibinfo {volume} {48}},\ \bibinfo
  {pages} {102} (\bibinfo {year} {2018})},\ \Eprint
  {https://arxiv.org/abs/1708.04476} {arXiv:1708.04476 [gr-qc]} \BibitemShut
  {NoStop}%
\bibitem [{\citenamefont {Berti}\ \emph {et~al.}(2018)\citenamefont {Berti},
  \citenamefont {Yagi}, \citenamefont {Yang},\ and\ \citenamefont
  {Yunes}}]{Berti:2018vdi}%
  \BibitemOpen
  \bibfield  {author} {\bibinfo {author} {\bibfnamefont {E.}~\bibnamefont
  {Berti}}, \bibinfo {author} {\bibfnamefont {K.}~\bibnamefont {Yagi}},
  \bibinfo {author} {\bibfnamefont {H.}~\bibnamefont {Yang}},\ and\ \bibinfo
  {author} {\bibfnamefont {N.}~\bibnamefont {Yunes}},\ }\bibfield  {title}
  {\bibinfo {title} {{Extreme gravity tests with gravitational waves from
  compact binary coalescences: (II) Ringdown}},\ }\href
  {https://doi.org/10.1007/s10714-018-2372-6} {\bibfield  {journal} {\bibinfo
  {journal} {Gen.\ Rel.\ Grav.}\ }\textbf {\bibinfo {volume} {50}},\ \bibinfo
  {pages} {49} (\bibinfo {year} {2018})},\ \Eprint
  {https://arxiv.org/abs/1801.03587} {arXiv:1801.03587 [gr-qc]} \BibitemShut
  {NoStop}%
\bibitem [{\citenamefont {Barack}\ \emph {et~al.}(2019)\citenamefont {Barack}
  \emph {et~al.}}]{Barack:2018yly}%
  \BibitemOpen
  \bibfield  {author} {\bibinfo {author} {\bibfnamefont {L.}~\bibnamefont
  {Barack}} \emph {et~al.},\ }\bibfield  {title} {\bibinfo {title} {{Black
  holes, gravitational waves and fundamental physics: A roadmap}},\ }\href
  {https://doi.org/10.1088/1361-6382/ab0587} {\bibfield  {journal} {\bibinfo
  {journal} {Class.\ Quant.\ Grav.}\ }\textbf {\bibinfo {volume} {36}},\
  \bibinfo {pages} {143001} (\bibinfo {year} {2019})},\ \Eprint
  {https://arxiv.org/abs/1806.05195} {arXiv:1806.05195 [gr-qc]} \BibitemShut
  {NoStop}%
\bibitem [{\citenamefont {Baibhav}\ and\ \citenamefont
  {Berti}(2019)}]{Baibhav:2018rfk}%
  \BibitemOpen
  \bibfield  {author} {\bibinfo {author} {\bibfnamefont {V.}~\bibnamefont
  {Baibhav}}\ and\ \bibinfo {author} {\bibfnamefont {E.}~\bibnamefont
  {Berti}},\ }\bibfield  {title} {\bibinfo {title} {{Multimode black hole
  spectroscopy}},\ }\href {https://doi.org/10.1103/PhysRevD.99.024005}
  {\bibfield  {journal} {\bibinfo  {journal} {Phys.\ Rev.\ D}\ }\textbf
  {\bibinfo {volume} {99}},\ \bibinfo {pages} {024005} (\bibinfo {year}
  {2019})},\ \Eprint {https://arxiv.org/abs/1809.03500} {arXiv:1809.03500
  [gr-qc]} \BibitemShut {NoStop}%
\bibitem [{\citenamefont {Baibhav}\ \emph {et~al.}(2019)\citenamefont {Baibhav}
  \emph {et~al.}}]{Baibhav:2019rsa}%
  \BibitemOpen
  \bibfield  {author} {\bibinfo {author} {\bibfnamefont {V.}~\bibnamefont
  {Baibhav}} \emph {et~al.},\ }\bibfield  {title} {\bibinfo {title} {{Probing
  the nature of black holes: Deep in the mHz gravitational-wave sky}},\
  }\href@noop {} {\  (\bibinfo {year} {2019})},\ \Eprint
  {https://arxiv.org/abs/1908.11390} {arXiv:1908.11390 [astro-ph.HE]}
  \BibitemShut {NoStop}%
\bibitem [{\citenamefont {Giesler}\ \emph {et~al.}(2019)\citenamefont
  {Giesler}, \citenamefont {Isi}, \citenamefont {Scheel},\ and\ \citenamefont
  {Teukolsky}}]{Giesler:2019uxc}%
  \BibitemOpen
  \bibfield  {author} {\bibinfo {author} {\bibfnamefont {M.}~\bibnamefont
  {Giesler}}, \bibinfo {author} {\bibfnamefont {M.}~\bibnamefont {Isi}},
  \bibinfo {author} {\bibfnamefont {M.~A.}\ \bibnamefont {Scheel}},\ and\
  \bibinfo {author} {\bibfnamefont {S.~A.}\ \bibnamefont {Teukolsky}},\
  }\bibfield  {title} {\bibinfo {title} {{Black hole ringdown: The importance
  of overtones}},\ }\href {https://doi.org/10.1103/PhysRevX.9.041060}
  {\bibfield  {journal} {\bibinfo  {journal} {Phys.\ Rev.\ X}\ }\textbf
  {\bibinfo {volume} {9}},\ \bibinfo {pages} {041060} (\bibinfo {year}
  {2019})},\ \Eprint {https://arxiv.org/abs/1903.08284} {arXiv:1903.08284
  [gr-qc]} \BibitemShut {NoStop}%
\bibitem [{\citenamefont {Isi}\ \emph {et~al.}(2019)\citenamefont {Isi},
  \citenamefont {Giesler}, \citenamefont {Farr}, \citenamefont {Scheel},\ and\
  \citenamefont {Teukolsky}}]{Isi:2019aib}%
  \BibitemOpen
  \bibfield  {author} {\bibinfo {author} {\bibfnamefont {M.}~\bibnamefont
  {Isi}}, \bibinfo {author} {\bibfnamefont {M.}~\bibnamefont {Giesler}},
  \bibinfo {author} {\bibfnamefont {W.~M.}\ \bibnamefont {Farr}}, \bibinfo
  {author} {\bibfnamefont {M.~A.}\ \bibnamefont {Scheel}},\ and\ \bibinfo
  {author} {\bibfnamefont {S.~A.}\ \bibnamefont {Teukolsky}},\ }\bibfield
  {title} {\bibinfo {title} {{Testing the no-hair theorem with GW150914}},\
  }\href {https://doi.org/10.1103/PhysRevLett.123.111102} {\bibfield  {journal}
  {\bibinfo  {journal} {Phys.\ Rev.\ Lett.}\ }\textbf {\bibinfo {volume}
  {123}},\ \bibinfo {pages} {111102} (\bibinfo {year} {2019})},\ \Eprint
  {https://arxiv.org/abs/1905.00869} {arXiv:1905.00869 [gr-qc]} \BibitemShut
  {NoStop}%
\bibitem [{\citenamefont {Bhagwat}\ \emph {et~al.}(2020)\citenamefont
  {Bhagwat}, \citenamefont {Forteza}, \citenamefont {Pani},\ and\ \citenamefont
  {Ferrari}}]{Bhagwat:2019dtm}%
  \BibitemOpen
  \bibfield  {author} {\bibinfo {author} {\bibfnamefont {S.}~\bibnamefont
  {Bhagwat}}, \bibinfo {author} {\bibfnamefont {X.~J.}\ \bibnamefont
  {Forteza}}, \bibinfo {author} {\bibfnamefont {P.}~\bibnamefont {Pani}},\ and\
  \bibinfo {author} {\bibfnamefont {V.}~\bibnamefont {Ferrari}},\ }\bibfield
  {title} {\bibinfo {title} {{Ringdown overtones, black hole spectroscopy, and
  no-hair theorem tests}},\ }\href
  {https://doi.org/10.1103/PhysRevD.101.044033} {\bibfield  {journal} {\bibinfo
   {journal} {Phys.\ Rev.\ D}\ }\textbf {\bibinfo {volume} {101}},\ \bibinfo
  {pages} {044033} (\bibinfo {year} {2020})},\ \Eprint
  {https://arxiv.org/abs/1910.08708} {arXiv:1910.08708 [gr-qc]} \BibitemShut
  {NoStop}%
\bibitem [{\citenamefont {Barausse}\ \emph {et~al.}(2020)\citenamefont
  {Barausse} \emph {et~al.}}]{Barausse:2020rsu}%
  \BibitemOpen
  \bibfield  {author} {\bibinfo {author} {\bibfnamefont {E.}~\bibnamefont
  {Barausse}} \emph {et~al.},\ }\bibfield  {title} {\bibinfo {title}
  {{Prospects for fundamental physics with LISA}},\ }\href
  {https://doi.org/10.1007/s10714-020-02691-1} {\bibfield  {journal} {\bibinfo
  {journal} {Gen.\ Rel.\ Grav.}\ }\textbf {\bibinfo {volume} {52}},\ \bibinfo
  {pages} {81} (\bibinfo {year} {2020})},\ \Eprint
  {https://arxiv.org/abs/2001.09793} {arXiv:2001.09793 [gr-qc]} \BibitemShut
  {NoStop}%
\bibitem [{\citenamefont {Johnson}\ \emph {et~al.}(2020)\citenamefont {Johnson}
  \emph {et~al.}}]{Johnson:2019ljv}%
  \BibitemOpen
  \bibfield  {author} {\bibinfo {author} {\bibfnamefont {M.~D.}\ \bibnamefont
  {Johnson}} \emph {et~al.},\ }\bibfield  {title} {\bibinfo {title} {{Universal
  interferometric signatures of a black hole\textquoteright{}s photon ring}},\
  }\href {https://doi.org/10.1126/sciadv.aaz1310} {\bibfield  {journal}
  {\bibinfo  {journal} {Sci.\ Adv.}\ }\textbf {\bibinfo {volume} {6}},\
  \bibinfo {pages} {eaaz1310} (\bibinfo {year} {2020})},\ \Eprint
  {https://arxiv.org/abs/1907.04329} {arXiv:1907.04329 [astro-ph.IM]}
  \BibitemShut {NoStop}%
\bibitem [{\citenamefont {Gralla}(2021)}]{Gralla:2020pra}%
  \BibitemOpen
  \bibfield  {author} {\bibinfo {author} {\bibfnamefont {S.~E.}\ \bibnamefont
  {Gralla}},\ }\bibfield  {title} {\bibinfo {title} {{Can the EHT M87 results
  be used to test general relativity?}},\ }\href
  {https://doi.org/10.1103/PhysRevD.103.024023} {\bibfield  {journal} {\bibinfo
   {journal} {Phys.\ Rev.\ D}\ }\textbf {\bibinfo {volume} {103}},\ \bibinfo
  {pages} {024023} (\bibinfo {year} {2021})},\ \Eprint
  {https://arxiv.org/abs/2010.08557} {arXiv:2010.08557 [astro-ph.HE]}
  \BibitemShut {NoStop}%
\bibitem [{\citenamefont
  {Vishveshwara}(1970{\natexlab{a}})}]{Vishveshwara:1970zz}%
  \BibitemOpen
  \bibfield  {author} {\bibinfo {author} {\bibfnamefont {C.~V.}\ \bibnamefont
  {Vishveshwara}},\ }\bibfield  {title} {\bibinfo {title} {{Scattering of
  gravitational radiation by a Schwarzschild black hole}},\ }\href
  {https://doi.org/10.1038/227936a0} {\bibfield  {journal} {\bibinfo  {journal}
  {Nature}\ }\textbf {\bibinfo {volume} {227}},\ \bibinfo {pages} {936}
  (\bibinfo {year} {1970}{\natexlab{a}})}\BibitemShut {NoStop}%
\bibitem [{\citenamefont
  {Vishveshwara}(1970{\natexlab{b}})}]{Vishveshwara:1970cc}%
  \BibitemOpen
  \bibfield  {author} {\bibinfo {author} {\bibfnamefont {C.~V.}\ \bibnamefont
  {Vishveshwara}},\ }\bibfield  {title} {\bibinfo {title} {{Stability of the
  Schwarzschild metric}},\ }\href {https://doi.org/10.1103/PhysRevD.1.2870}
  {\bibfield  {journal} {\bibinfo  {journal} {Phys.\ Rev.\ D}\ }\textbf
  {\bibinfo {volume} {1}},\ \bibinfo {pages} {2870} (\bibinfo {year}
  {1970}{\natexlab{b}})}\BibitemShut {NoStop}%
\bibitem [{\citenamefont {Kokkotas}\ and\ \citenamefont
  {Schmidt}(1999)}]{Kokkotas:1999bd}%
  \BibitemOpen
  \bibfield  {author} {\bibinfo {author} {\bibfnamefont {K.~D.}\ \bibnamefont
  {Kokkotas}}\ and\ \bibinfo {author} {\bibfnamefont {B.~G.}\ \bibnamefont
  {Schmidt}},\ }\bibfield  {title} {\bibinfo {title} {{Quasinormal modes of
  stars and black holes}},\ }\href {https://doi.org/10.12942/lrr-1999-2}
  {\bibfield  {journal} {\bibinfo  {journal} {Living Rev.\ Rel.}\ }\textbf
  {\bibinfo {volume} {2}},\ \bibinfo {pages} {2} (\bibinfo {year} {1999})},\
  \Eprint {https://arxiv.org/abs/gr-qc/9909058} {arXiv:gr-qc/9909058}
  \BibitemShut {NoStop}%
\bibitem [{\citenamefont {Nollert}(1999)}]{Nollert:1999ji}%
  \BibitemOpen
  \bibfield  {author} {\bibinfo {author} {\bibfnamefont {H.-P.}\ \bibnamefont
  {Nollert}},\ }\bibfield  {title} {\bibinfo {title} {{Quasinormal modes: The
  characteristic `sound' of black holes and neutron stars}},\ }\href
  {https://doi.org/10.1088/0264-9381/16/12/201} {\bibfield  {journal} {\bibinfo
   {journal} {Class.\ Quant.\ Grav.}\ }\textbf {\bibinfo {volume} {16}},\
  \bibinfo {pages} {R159} (\bibinfo {year} {1999})}\BibitemShut {NoStop}%
\bibitem [{\citenamefont {Berti}\ \emph {et~al.}(2009)\citenamefont {Berti},
  \citenamefont {Cardoso},\ and\ \citenamefont {Starinets}}]{Berti:2009kk}%
  \BibitemOpen
  \bibfield  {author} {\bibinfo {author} {\bibfnamefont {E.}~\bibnamefont
  {Berti}}, \bibinfo {author} {\bibfnamefont {V.}~\bibnamefont {Cardoso}},\
  and\ \bibinfo {author} {\bibfnamefont {A.~O.}\ \bibnamefont {Starinets}},\
  }\bibfield  {title} {\bibinfo {title} {{Quasinormal modes of black holes and
  black branes}},\ }\href {https://doi.org/10.1088/0264-9381/26/16/163001}
  {\bibfield  {journal} {\bibinfo  {journal} {Class.\ Quant.\ Grav.}\ }\textbf
  {\bibinfo {volume} {26}},\ \bibinfo {pages} {163001} (\bibinfo {year}
  {2009})},\ \Eprint {https://arxiv.org/abs/0905.2975} {arXiv:0905.2975
  [gr-qc]} \BibitemShut {NoStop}%
\bibitem [{\citenamefont {Konoplya}\ and\ \citenamefont
  {Zhidenko}(2011)}]{Konoplya:2011qq}%
  \BibitemOpen
  \bibfield  {author} {\bibinfo {author} {\bibfnamefont {R.~A.}\ \bibnamefont
  {Konoplya}}\ and\ \bibinfo {author} {\bibfnamefont {A.}~\bibnamefont
  {Zhidenko}},\ }\bibfield  {title} {\bibinfo {title} {{Quasinormal modes of
  black holes: From astrophysics to string theory}},\ }\href
  {https://doi.org/10.1103/RevModPhys.83.793} {\bibfield  {journal} {\bibinfo
  {journal} {Rev.\ Mod.\ Phys.}\ }\textbf {\bibinfo {volume} {83}},\ \bibinfo
  {pages} {793} (\bibinfo {year} {2011})},\ \Eprint
  {https://arxiv.org/abs/1102.4014} {arXiv:1102.4014 [gr-qc]} \BibitemShut
  {NoStop}%
\bibitem [{\citenamefont {Leaver}(1985)}]{Leaver:1985ax}%
  \BibitemOpen
  \bibfield  {author} {\bibinfo {author} {\bibfnamefont {E.~W.}\ \bibnamefont
  {Leaver}},\ }\bibfield  {title} {\bibinfo {title} {{An Analytic
  representation for the quasi-normal modes of Kerr black holes}},\ }\href
  {https://doi.org/10.1098/rspa.1985.0119} {\bibfield  {journal} {\bibinfo
  {journal} {Proc.\ Roy.\ Soc.\ Lond.\ A}\ }\textbf {\bibinfo {volume} {402}},\
  \bibinfo {pages} {285} (\bibinfo {year} {1985})}\BibitemShut {NoStop}%
\bibitem [{\citenamefont {Leaver}(1986{\natexlab{a}})}]{LeaverJMP1986}%
  \BibitemOpen
  \bibfield  {author} {\bibinfo {author} {\bibfnamefont {E.~W.}\ \bibnamefont
  {Leaver}},\ }\bibfield  {title} {\bibinfo {title} {Solutions to a generalized
  spheroidal wave equation: Teukolsky’s equations in general relativity, and
  the two-center problem in molecular quantum mechanics},\ }\href
  {https://doi.org/10.1063/1.527130} {\bibfield  {journal} {\bibinfo  {journal}
  {J.\ Math.\ Phys.}\ }\textbf {\bibinfo {volume} {27}},\ \bibinfo {pages}
  {1238} (\bibinfo {year} {1986}{\natexlab{a}})}\BibitemShut {NoStop}%
\bibitem [{\citenamefont {Nollert}(1993)}]{Nollert:1993zz}%
  \BibitemOpen
  \bibfield  {author} {\bibinfo {author} {\bibfnamefont {H.-P.}\ \bibnamefont
  {Nollert}},\ }\bibfield  {title} {\bibinfo {title} {Quasinormal modes of
  schwarzschild black holes: The determination of quasinormal frequencies with
  very large imaginary parts},\ }\href
  {https://doi.org/10.1103/PhysRevD.47.5253} {\bibfield  {journal} {\bibinfo
  {journal} {Phys.\ Rev.\ D}\ }\textbf {\bibinfo {volume} {47}},\ \bibinfo
  {pages} {5253} (\bibinfo {year} {1993})}\BibitemShut {NoStop}%
\bibitem [{\citenamefont {Onozawa}\ \emph {et~al.}(1996)\citenamefont
  {Onozawa}, \citenamefont {Mishima}, \citenamefont {Okamura},\ and\
  \citenamefont {Ishihara}}]{Onozawa:1995vu}%
  \BibitemOpen
  \bibfield  {author} {\bibinfo {author} {\bibfnamefont {H.}~\bibnamefont
  {Onozawa}}, \bibinfo {author} {\bibfnamefont {T.}~\bibnamefont {Mishima}},
  \bibinfo {author} {\bibfnamefont {T.}~\bibnamefont {Okamura}},\ and\ \bibinfo
  {author} {\bibfnamefont {H.}~\bibnamefont {Ishihara}},\ }\bibfield  {title}
  {\bibinfo {title} {{Quasinormal modes of maximally charged black holes}},\
  }\href {https://doi.org/10.1103/PhysRevD.53.7033} {\bibfield  {journal}
  {\bibinfo  {journal} {Phys. Rev. D}\ }\textbf {\bibinfo {volume} {53}},\
  \bibinfo {pages} {7033} (\bibinfo {year} {1996})},\ \Eprint
  {https://arxiv.org/abs/gr-qc/9603021} {arXiv:gr-qc/9603021} \BibitemShut
  {NoStop}%
\bibitem [{\citenamefont {Onozawa}(1997)}]{Onozawa:1996ux}%
  \BibitemOpen
  \bibfield  {author} {\bibinfo {author} {\bibfnamefont {H.}~\bibnamefont
  {Onozawa}},\ }\bibfield  {title} {\bibinfo {title} {{A detailed study of
  quasinormal frequencies of the Kerr black hole}},\ }\href
  {https://doi.org/10.1103/PhysRevD.55.3593} {\bibfield  {journal} {\bibinfo
  {journal} {Phys.\ Rev.\ D}\ }\textbf {\bibinfo {volume} {55}},\ \bibinfo
  {pages} {3593} (\bibinfo {year} {1997})},\ \Eprint
  {https://arxiv.org/abs/gr-qc/9610048} {arXiv:gr-qc/9610048} \BibitemShut
  {NoStop}%
\bibitem [{Rin()}]{RingdownBerti}%
  \BibitemOpen
  \href@noop {} {}\bibinfo {howpublished}
  {\url{{https://pages.jh.edu/eberti2/ringdown/}}}\BibitemShut {NoStop}%
\bibitem [{\citenamefont {Goebel}(1972)}]{Goebel1972}%
  \BibitemOpen
  \bibfield  {author} {\bibinfo {author} {\bibfnamefont {C.~J.}\ \bibnamefont
  {Goebel}},\ }\bibfield  {title} {\bibinfo {title} {Comments on the vibrations
  of a black hole},\ }\href {https://doi.org/10.1086/180898} {\bibfield
  {journal} {\bibinfo  {journal} {Astrophys.\ J.}\ }\textbf {\bibinfo {volume}
  {172}},\ \bibinfo {pages} {L95} (\bibinfo {year} {1972})}\BibitemShut
  {NoStop}%
\bibitem [{\citenamefont {Ferrari}\ and\ \citenamefont
  {Mashhoon}(1984{\natexlab{a}})}]{Ferrari:1984ozr}%
  \BibitemOpen
  \bibfield  {author} {\bibinfo {author} {\bibfnamefont {V.}~\bibnamefont
  {Ferrari}}\ and\ \bibinfo {author} {\bibfnamefont {B.}~\bibnamefont
  {Mashhoon}},\ }\bibfield  {title} {\bibinfo {title} {{Oscillations of a black
  hole}},\ }\href {https://doi.org/10.1103/PhysRevLett.52.1361} {\bibfield
  {journal} {\bibinfo  {journal} {Phys.\ Rev.\ Lett.}\ }\textbf {\bibinfo
  {volume} {52}},\ \bibinfo {pages} {1361} (\bibinfo {year}
  {1984}{\natexlab{a}})}\BibitemShut {NoStop}%
\bibitem [{\citenamefont {Ferrari}\ and\ \citenamefont
  {Mashhoon}(1984{\natexlab{b}})}]{Ferrari:1984zz}%
  \BibitemOpen
  \bibfield  {author} {\bibinfo {author} {\bibfnamefont {V.}~\bibnamefont
  {Ferrari}}\ and\ \bibinfo {author} {\bibfnamefont {B.}~\bibnamefont
  {Mashhoon}},\ }\bibfield  {title} {\bibinfo {title} {{New approach to the
  quasinormal modes of a black hole}},\ }\href
  {https://doi.org/10.1103/PhysRevD.30.295} {\bibfield  {journal} {\bibinfo
  {journal} {Phys.\ Rev.\ D}\ }\textbf {\bibinfo {volume} {30}},\ \bibinfo
  {pages} {295} (\bibinfo {year} {1984}{\natexlab{b}})}\BibitemShut {NoStop}%
\bibitem [{\citenamefont {Mashhoon}(1985)}]{Mashhoon:1985cya}%
  \BibitemOpen
  \bibfield  {author} {\bibinfo {author} {\bibfnamefont {B.}~\bibnamefont
  {Mashhoon}},\ }\bibfield  {title} {\bibinfo {title} {{Stability of charged
  rotating black holes in the eikonal approximation}},\ }\href
  {https://doi.org/10.1103/PhysRevD.31.290} {\bibfield  {journal} {\bibinfo
  {journal} {Phys.\ Rev.\ D}\ }\textbf {\bibinfo {volume} {31}},\ \bibinfo
  {pages} {290} (\bibinfo {year} {1985})}\BibitemShut {NoStop}%
\bibitem [{\citenamefont {Stewart}(1989)}]{Stewart89}%
  \BibitemOpen
  \bibfield  {author} {\bibinfo {author} {\bibfnamefont {J.~M.}\ \bibnamefont
  {Stewart}},\ }\bibfield  {title} {\bibinfo {title} {Solutions of the wave
  equation on a schwarzschild space-time with localized energy},\ }\href
  {https://doi.org/10.1098/rspa.1989.0078} {\bibfield  {journal} {\bibinfo
  {journal} {Proc.\ Roy.\ Soc.\ Lond.\ A.}\ }\textbf {\bibinfo {volume}
  {424}},\ \bibinfo {pages} {239} (\bibinfo {year} {1989})}\BibitemShut
  {NoStop}%
\bibitem [{\citenamefont {Andersson}\ and\ \citenamefont
  {Onozawa}(1996)}]{Andersson:1996xw}%
  \BibitemOpen
  \bibfield  {author} {\bibinfo {author} {\bibfnamefont {N.}~\bibnamefont
  {Andersson}}\ and\ \bibinfo {author} {\bibfnamefont {H.}~\bibnamefont
  {Onozawa}},\ }\bibfield  {title} {\bibinfo {title} {{Quasinormal modes of
  nearly extreme Reissner-Nordstrom black holes}},\ }\href
  {https://doi.org/10.1103/PhysRevD.54.7470} {\bibfield  {journal} {\bibinfo
  {journal} {Phys.\ Rev.\ D}\ }\textbf {\bibinfo {volume} {54}},\ \bibinfo
  {pages} {7470} (\bibinfo {year} {1996})},\ \Eprint
  {https://arxiv.org/abs/gr-qc/9607054} {arXiv:gr-qc/9607054} \BibitemShut
  {NoStop}%
\bibitem [{\citenamefont {Vanzo}\ and\ \citenamefont
  {Zerbini}(2004)}]{Vanzo:2004fy}%
  \BibitemOpen
  \bibfield  {author} {\bibinfo {author} {\bibfnamefont {L.}~\bibnamefont
  {Vanzo}}\ and\ \bibinfo {author} {\bibfnamefont {S.}~\bibnamefont
  {Zerbini}},\ }\bibfield  {title} {\bibinfo {title} {{Asymptotics of
  quasinormal modes for multihorizon black holes}},\ }\href
  {https://doi.org/10.1103/PhysRevD.70.044030} {\bibfield  {journal} {\bibinfo
  {journal} {Phys.\ Rev.\ D}\ }\textbf {\bibinfo {volume} {70}},\ \bibinfo
  {pages} {044030} (\bibinfo {year} {2004})},\ \Eprint
  {https://arxiv.org/abs/hep-th/0402103} {arXiv:hep-th/0402103} \BibitemShut
  {NoStop}%
\bibitem [{\citenamefont {Cardoso}\ \emph {et~al.}(2009)\citenamefont
  {Cardoso}, \citenamefont {Miranda}, \citenamefont {Berti}, \citenamefont
  {Witek},\ and\ \citenamefont {Zanchin}}]{Cardoso:2008bp}%
  \BibitemOpen
  \bibfield  {author} {\bibinfo {author} {\bibfnamefont {V.}~\bibnamefont
  {Cardoso}}, \bibinfo {author} {\bibfnamefont {A.~S.}\ \bibnamefont
  {Miranda}}, \bibinfo {author} {\bibfnamefont {E.}~\bibnamefont {Berti}},
  \bibinfo {author} {\bibfnamefont {H.}~\bibnamefont {Witek}},\ and\ \bibinfo
  {author} {\bibfnamefont {V.~T.}\ \bibnamefont {Zanchin}},\ }\bibfield
  {title} {\bibinfo {title} {{Geodesic stability, Lyapunov exponents and
  quasinormal modes}},\ }\href {https://doi.org/10.1103/PhysRevD.79.064016}
  {\bibfield  {journal} {\bibinfo  {journal} {Phys.\ Rev.\ D}\ }\textbf
  {\bibinfo {volume} {79}},\ \bibinfo {pages} {064016} (\bibinfo {year}
  {2009})},\ \Eprint {https://arxiv.org/abs/0812.1806} {arXiv:0812.1806
  [hep-th]} \BibitemShut {NoStop}%
\bibitem [{\citenamefont {Dolan}\ and\ \citenamefont
  {Ottewill}(2009)}]{Dolan:2009nk}%
  \BibitemOpen
  \bibfield  {author} {\bibinfo {author} {\bibfnamefont {S.~R.}\ \bibnamefont
  {Dolan}}\ and\ \bibinfo {author} {\bibfnamefont {A.~C.}\ \bibnamefont
  {Ottewill}},\ }\bibfield  {title} {\bibinfo {title} {{On an expansion method
  for black hole quasinormal modes and Regge poles}},\ }\href
  {https://doi.org/10.1088/0264-9381/26/22/225003} {\bibfield  {journal}
  {\bibinfo  {journal} {Class.\ Quant.\ Grav.}\ }\textbf {\bibinfo {volume}
  {26}},\ \bibinfo {pages} {225003} (\bibinfo {year} {2009})},\ \Eprint
  {https://arxiv.org/abs/0908.0329} {arXiv:0908.0329 [gr-qc]} \BibitemShut
  {NoStop}%
\bibitem [{\citenamefont {Hod}(2009)}]{Hod:2009td}%
  \BibitemOpen
  \bibfield  {author} {\bibinfo {author} {\bibfnamefont {S.}~\bibnamefont
  {Hod}},\ }\bibfield  {title} {\bibinfo {title} {{Black-hole quasinormal
  resonances: Wave analysis versus a geometric-optics approximation}},\ }\href
  {https://doi.org/10.1103/PhysRevD.80.064004} {\bibfield  {journal} {\bibinfo
  {journal} {Phys.\ Rev.\ D}\ }\textbf {\bibinfo {volume} {80}},\ \bibinfo
  {pages} {064004} (\bibinfo {year} {2009})},\ \Eprint
  {https://arxiv.org/abs/0909.0314} {arXiv:0909.0314 [gr-qc]} \BibitemShut
  {NoStop}%
\bibitem [{\citenamefont {Yang}\ \emph {et~al.}(2012)\citenamefont {Yang},
  \citenamefont {Nichols}, \citenamefont {Zhang}, \citenamefont {Zimmerman},
  \citenamefont {Zhang},\ and\ \citenamefont {Chen}}]{Yang:2012he}%
  \BibitemOpen
  \bibfield  {author} {\bibinfo {author} {\bibfnamefont {H.}~\bibnamefont
  {Yang}}, \bibinfo {author} {\bibfnamefont {D.~A.}\ \bibnamefont {Nichols}},
  \bibinfo {author} {\bibfnamefont {F.}~\bibnamefont {Zhang}}, \bibinfo
  {author} {\bibfnamefont {A.}~\bibnamefont {Zimmerman}}, \bibinfo {author}
  {\bibfnamefont {Z.}~\bibnamefont {Zhang}},\ and\ \bibinfo {author}
  {\bibfnamefont {Y.}~\bibnamefont {Chen}},\ }\bibfield  {title} {\bibinfo
  {title} {{Quasinormal-mode spectrum of Kerr black holes and its geometric
  interpretation}},\ }\href {https://doi.org/10.1103/PhysRevD.86.104006}
  {\bibfield  {journal} {\bibinfo  {journal} {Phys.\ Rev.\ D}\ }\textbf
  {\bibinfo {volume} {86}},\ \bibinfo {pages} {104006} (\bibinfo {year}
  {2012})},\ \Eprint {https://arxiv.org/abs/1207.4253} {arXiv:1207.4253
  [gr-qc]} \BibitemShut {NoStop}%
\bibitem [{\citenamefont {Glampedakis}\ \emph {et~al.}(2017)\citenamefont
  {Glampedakis}, \citenamefont {Pappas}, \citenamefont {Silva},\ and\
  \citenamefont {Berti}}]{Glampedakis:2017dvb}%
  \BibitemOpen
  \bibfield  {author} {\bibinfo {author} {\bibfnamefont {K.}~\bibnamefont
  {Glampedakis}}, \bibinfo {author} {\bibfnamefont {G.}~\bibnamefont {Pappas}},
  \bibinfo {author} {\bibfnamefont {H.~O.}\ \bibnamefont {Silva}},\ and\
  \bibinfo {author} {\bibfnamefont {E.}~\bibnamefont {Berti}},\ }\bibfield
  {title} {\bibinfo {title} {{Post-Kerr black hole spectroscopy}},\ }\href
  {https://doi.org/10.1103/PhysRevD.96.064054} {\bibfield  {journal} {\bibinfo
  {journal} {Phys.\ Rev.\ D}\ }\textbf {\bibinfo {volume} {96}},\ \bibinfo
  {pages} {064054} (\bibinfo {year} {2017})},\ \Eprint
  {https://arxiv.org/abs/1706.07658} {arXiv:1706.07658 [gr-qc]} \BibitemShut
  {NoStop}%
\bibitem [{\citenamefont {Dolan}(2010)}]{Dolan:2010wr}%
  \BibitemOpen
  \bibfield  {author} {\bibinfo {author} {\bibfnamefont {S.~R.}\ \bibnamefont
  {Dolan}},\ }\bibfield  {title} {\bibinfo {title} {{The quasinormal mode
  spectrum of a Kerr black kole in the eikonal limit}},\ }\href
  {https://doi.org/10.1103/PhysRevD.82.104003} {\bibfield  {journal} {\bibinfo
  {journal} {Phys.\ Rev.\ D}\ }\textbf {\bibinfo {volume} {82}},\ \bibinfo
  {pages} {104003} (\bibinfo {year} {2010})},\ \Eprint
  {https://arxiv.org/abs/1007.5097} {arXiv:1007.5097 [gr-qc]} \BibitemShut
  {NoStop}%
\bibitem [{\citenamefont {Decanini}\ \emph {et~al.}(2003)\citenamefont
  {Decanini}, \citenamefont {Folacci},\ and\ \citenamefont
  {Jensen}}]{Decanini:2002ha}%
  \BibitemOpen
  \bibfield  {author} {\bibinfo {author} {\bibfnamefont {Y.}~\bibnamefont
  {Decanini}}, \bibinfo {author} {\bibfnamefont {A.}~\bibnamefont {Folacci}},\
  and\ \bibinfo {author} {\bibfnamefont {B.}~\bibnamefont {Jensen}},\
  }\bibfield  {title} {\bibinfo {title} {{Complex angular momentum in black
  hole physics and the quasinormal modes}},\ }\href
  {https://doi.org/10.1103/PhysRevD.67.124017} {\bibfield  {journal} {\bibinfo
  {journal} {Phys.\ Rev.\ D}\ }\textbf {\bibinfo {volume} {67}},\ \bibinfo
  {pages} {124017} (\bibinfo {year} {2003})},\ \Eprint
  {https://arxiv.org/abs/gr-qc/0212093} {arXiv:gr-qc/0212093} \BibitemShut
  {NoStop}%
\bibitem [{\citenamefont {Decanini}\ and\ \citenamefont
  {Folacci}(2009)}]{Decanini:2009dn}%
  \BibitemOpen
  \bibfield  {author} {\bibinfo {author} {\bibfnamefont {Y.}~\bibnamefont
  {Decanini}}\ and\ \bibinfo {author} {\bibfnamefont {A.}~\bibnamefont
  {Folacci}},\ }\bibfield  {title} {\bibinfo {title} {{Quasinormal modes of the
  BTZ black hole are generated by surface waves supported by its boundary at
  infinity}},\ }\href {https://doi.org/10.1103/PhysRevD.79.044021} {\bibfield
  {journal} {\bibinfo  {journal} {Phys.\ Rev.\ D}\ }\textbf {\bibinfo {volume}
  {79}},\ \bibinfo {pages} {044021} (\bibinfo {year} {2009})},\ \Eprint
  {https://arxiv.org/abs/0901.1642} {arXiv:0901.1642 [hep-th]} \BibitemShut
  {NoStop}%
\bibitem [{\citenamefont {Decanini}\ and\ \citenamefont
  {Folacci}(2010)}]{Decanini:2009mu}%
  \BibitemOpen
  \bibfield  {author} {\bibinfo {author} {\bibfnamefont {Y.}~\bibnamefont
  {Decanini}}\ and\ \bibinfo {author} {\bibfnamefont {A.}~\bibnamefont
  {Folacci}},\ }\bibfield  {title} {\bibinfo {title} {{Regge poles of the
  Schwarzschild black hole: A WKB approach}},\ }\href
  {https://doi.org/10.1103/PhysRevD.81.024031} {\bibfield  {journal} {\bibinfo
  {journal} {Phys.\ Rev.\ D}\ }\textbf {\bibinfo {volume} {81}},\ \bibinfo
  {pages} {024031} (\bibinfo {year} {2010})},\ \Eprint
  {https://arxiv.org/abs/0906.2601} {arXiv:0906.2601 [gr-qc]} \BibitemShut
  {NoStop}%
\bibitem [{\citenamefont {Decanini}\ \emph {et~al.}(2010)\citenamefont
  {Decanini}, \citenamefont {Folacci},\ and\ \citenamefont
  {Raffaelli}}]{Decanini:2010fz}%
  \BibitemOpen
  \bibfield  {author} {\bibinfo {author} {\bibfnamefont {Y.}~\bibnamefont
  {Decanini}}, \bibinfo {author} {\bibfnamefont {A.}~\bibnamefont {Folacci}},\
  and\ \bibinfo {author} {\bibfnamefont {B.}~\bibnamefont {Raffaelli}},\
  }\bibfield  {title} {\bibinfo {title} {{Unstable circular null geodesics of
  static spherically symmetric black holes, Regge poles and quasinormal
  frequencies}},\ }\href {https://doi.org/10.1103/PhysRevD.81.104039}
  {\bibfield  {journal} {\bibinfo  {journal} {Phys.\ Rev.\ D}\ }\textbf
  {\bibinfo {volume} {81}},\ \bibinfo {pages} {104039} (\bibinfo {year}
  {2010})},\ \Eprint {https://arxiv.org/abs/1002.0121} {arXiv:1002.0121
  [gr-qc]} \BibitemShut {NoStop}%
\bibitem [{\citenamefont {Decanini}\ \emph
  {et~al.}(2011{\natexlab{a}})\citenamefont {Decanini}, \citenamefont
  {Folacci},\ and\ \citenamefont {Raffaelli}}]{Decanini:2011eh}%
  \BibitemOpen
  \bibfield  {author} {\bibinfo {author} {\bibfnamefont {Y.}~\bibnamefont
  {Decanini}}, \bibinfo {author} {\bibfnamefont {A.}~\bibnamefont {Folacci}},\
  and\ \bibinfo {author} {\bibfnamefont {B.}~\bibnamefont {Raffaelli}},\
  }\bibfield  {title} {\bibinfo {title} {{Resonance and absorption spectra of
  the Schwarzschild black hole for massive scalar perturbations: a complex
  angular momentum analysis}},\ }\href
  {https://doi.org/10.1103/PhysRevD.84.084035} {\bibfield  {journal} {\bibinfo
  {journal} {Phys.\ Rev.\ D}\ }\textbf {\bibinfo {volume} {84}},\ \bibinfo
  {pages} {084035} (\bibinfo {year} {2011}{\natexlab{a}})},\ \Eprint
  {https://arxiv.org/abs/1108.5076} {arXiv:1108.5076 [gr-qc]} \BibitemShut
  {NoStop}%
\bibitem [{\citenamefont {Poincar\'e}(1910)}]{Poincare1910}%
  \BibitemOpen
  \bibfield  {author} {\bibinfo {author} {\bibfnamefont {H.}~\bibnamefont
  {Poincar\'e}},\ }\bibfield  {title} {\bibinfo {title} {Sur la diffraction des
  ondes hertziennes},\ }\href {https://doi.org/10.1007/BF03014067} {\bibfield
  {journal} {\bibinfo  {journal} {Rend.\ Circ.\ Matem.\ Palermo}\ }\textbf
  {\bibinfo {volume} {29}},\ \bibinfo {pages} {169} (\bibinfo {year}
  {1910})}\BibitemShut {NoStop}%
\bibitem [{\citenamefont {Watson}(1918)}]{Watson1918}%
  \BibitemOpen
  \bibfield  {author} {\bibinfo {author} {\bibfnamefont {G.~N.}\ \bibnamefont
  {Watson}},\ }\bibfield  {title} {\bibinfo {title} {The diffraction of
  electric waves by the earth},\ }\href
  {https://doi.org/10.1098/rspa.1918.0050} {\bibfield  {journal} {\bibinfo
  {journal} {Proc.\ Roy.\ Soc.\ Lond.\ A.}\ }\textbf {\bibinfo {volume} {95}},\
  \bibinfo {pages} {83} (\bibinfo {year} {1918})}\BibitemShut {NoStop}%
\bibitem [{\citenamefont {Sommerfeld}(1949)}]{Sommerfeld1949}%
  \BibitemOpen
  \bibfield  {author} {\bibinfo {author} {\bibfnamefont {A.}~\bibnamefont
  {Sommerfeld}},\ }\href@noop {} {\emph {\bibinfo {title} {Partial Differential
  Equations in Physics}}}\ (\bibinfo  {publisher} {Academic Press, New York},\
  \bibinfo {year} {1949})\BibitemShut {NoStop}%
\bibitem [{\citenamefont {de~Alfaro}\ and\ \citenamefont
  {Regge}(1965)}]{deAlfaro:1965zz}%
  \BibitemOpen
  \bibfield  {author} {\bibinfo {author} {\bibfnamefont {V.}~\bibnamefont
  {de~Alfaro}}\ and\ \bibinfo {author} {\bibfnamefont {T.}~\bibnamefont
  {Regge}},\ }\href@noop {} {\emph {\bibinfo {title} {{Potential
  Scattering}}}}\ (\bibinfo  {publisher} {North-Holland Publishing Company,
  Amsterdam},\ \bibinfo {year} {1965})\BibitemShut {NoStop}%
\bibitem [{\citenamefont {Newton}(1982)}]{Newton:1982qc}%
  \BibitemOpen
  \bibfield  {author} {\bibinfo {author} {\bibfnamefont {R.~G.}\ \bibnamefont
  {Newton}},\ }\href@noop {} {\emph {\bibinfo {title} {{Scattering Theory of
  Waves and Particles}}}},\ \bibinfo {edition} {2nd}\ ed.\ (\bibinfo
  {publisher} {Springer-Verlag, New York},\ \bibinfo {year} {1982})\BibitemShut
  {NoStop}%
\bibitem [{\citenamefont {Nussenzveig}(1992)}]{Nussen1992}%
  \BibitemOpen
  \bibfield  {author} {\bibinfo {author} {\bibfnamefont {H.~M.}\ \bibnamefont
  {Nussenzveig}},\ }\href@noop {} {\emph {\bibinfo {title} {Diffraction Effects
  in Semiclassical Scattering}}}\ (\bibinfo  {publisher} {Cambridge University
  Press, Cambridge},\ \bibinfo {year} {1992})\BibitemShut {NoStop}%
\bibitem [{\citenamefont {Grandy}(2000)}]{Grandy2000}%
  \BibitemOpen
  \bibfield  {author} {\bibinfo {author} {\bibfnamefont {J.~W.~T.}\
  \bibnamefont {Grandy}},\ }\href@noop {} {\emph {\bibinfo {title} {Scattering
  of Waves from Large Spheres}}}\ (\bibinfo  {publisher} {Cambridge University
  Press, Cambridge},\ \bibinfo {year} {2000})\BibitemShut {NoStop}%
\bibitem [{\citenamefont {Collins}(1977)}]{Collins:1977jy}%
  \BibitemOpen
  \bibfield  {author} {\bibinfo {author} {\bibfnamefont {P.~D.~B.}\
  \bibnamefont {Collins}},\ }\href@noop {} {\emph {\bibinfo {title} {{An
  Introduction to Regge Theory and High-Energy Physics}}}}\ (\bibinfo
  {publisher} {Cambridge University Press, Cambridge},\ \bibinfo {year}
  {1977})\BibitemShut {NoStop}%
\bibitem [{\citenamefont {Barone}\ and\ \citenamefont
  {Predazzi}(2002)}]{Barone:2002cv}%
  \BibitemOpen
  \bibfield  {author} {\bibinfo {author} {\bibfnamefont {V.}~\bibnamefont
  {Barone}}\ and\ \bibinfo {author} {\bibfnamefont {E.}~\bibnamefont
  {Predazzi}},\ }\href@noop {} {\emph {\bibinfo {title} {{High-Energy Particle
  Diffraction}}}}\ (\bibinfo  {publisher} {Springer-Verlag, Berlin},\ \bibinfo
  {year} {2002})\BibitemShut {NoStop}%
\bibitem [{\citenamefont {Donnachie}\ \emph {et~al.}(2004)\citenamefont
  {Donnachie}, \citenamefont {Dosch}, \citenamefont {Nachtmann},\ and\
  \citenamefont {Landshoff}}]{Donnachie:2002en}%
  \BibitemOpen
  \bibfield  {author} {\bibinfo {author} {\bibfnamefont {S.}~\bibnamefont
  {Donnachie}}, \bibinfo {author} {\bibfnamefont {H.~G.}\ \bibnamefont
  {Dosch}}, \bibinfo {author} {\bibfnamefont {O.}~\bibnamefont {Nachtmann}},\
  and\ \bibinfo {author} {\bibfnamefont {P.}~\bibnamefont {Landshoff}},\
  }\href@noop {} {\emph {\bibinfo {title} {{Pomeron Physics and QCD}}}}\
  (\bibinfo  {publisher} {Cambridge University Press, Cambridge},\ \bibinfo
  {year} {2004})\BibitemShut {NoStop}%
\bibitem [{\citenamefont {Gribov}(2007)}]{Gribov:2003nw}%
  \BibitemOpen
  \bibfield  {author} {\bibinfo {author} {\bibfnamefont {V.~N.}\ \bibnamefont
  {Gribov}},\ }\href@noop {} {\emph {\bibinfo {title} {{The Theory of Complex
  Angular Momenta: Gribov Lectures on Theoretical Physics}}}}\ (\bibinfo
  {publisher} {Cambridge University Press, Cambridge},\ \bibinfo {year}
  {2007})\BibitemShut {NoStop}%
\bibitem [{\citenamefont {Andersson}\ and\ \citenamefont
  {Thylwe}(1994)}]{Andersson:1994rk}%
  \BibitemOpen
  \bibfield  {author} {\bibinfo {author} {\bibfnamefont {N.}~\bibnamefont
  {Andersson}}\ and\ \bibinfo {author} {\bibfnamefont {K.~E.}\ \bibnamefont
  {Thylwe}},\ }\bibfield  {title} {\bibinfo {title} {{Complex angular momentum
  approach to black hole scattering}},\ }\href
  {https://doi.org/10.1088/0264-9381/11/12/013} {\bibfield  {journal} {\bibinfo
   {journal} {Class.\ Quant.\ Grav.}\ }\textbf {\bibinfo {volume} {11}},\
  \bibinfo {pages} {2991} (\bibinfo {year} {1994})}\BibitemShut {NoStop}%
\bibitem [{\citenamefont {Andersson}(1994)}]{Andersson:1994rm}%
  \BibitemOpen
  \bibfield  {author} {\bibinfo {author} {\bibfnamefont {N.}~\bibnamefont
  {Andersson}},\ }\bibfield  {title} {\bibinfo {title} {{Complex angular
  momenta and the black hole glory}},\ }\href
  {https://doi.org/10.1088/0264-9381/11/12/014} {\bibfield  {journal} {\bibinfo
   {journal} {Class.\ Quant.\ Grav.}\ }\textbf {\bibinfo {volume} {11}},\
  \bibinfo {pages} {3003} (\bibinfo {year} {1994})}\BibitemShut {NoStop}%
\bibitem [{\citenamefont {Decanini}\ \emph
  {et~al.}(2011{\natexlab{b}})\citenamefont {Decanini}, \citenamefont
  {Esposito-Farese},\ and\ \citenamefont {Folacci}}]{Decanini:2011xi}%
  \BibitemOpen
  \bibfield  {author} {\bibinfo {author} {\bibfnamefont {Y.}~\bibnamefont
  {Decanini}}, \bibinfo {author} {\bibfnamefont {G.}~\bibnamefont
  {Esposito-Farese}},\ and\ \bibinfo {author} {\bibfnamefont {A.}~\bibnamefont
  {Folacci}},\ }\bibfield  {title} {\bibinfo {title} {{Universality of
  high-energy absorption cross sections for black holes}},\ }\href
  {https://doi.org/10.1103/PhysRevD.83.044032} {\bibfield  {journal} {\bibinfo
  {journal} {Phys.\ Rev.\ D}\ }\textbf {\bibinfo {volume} {83}},\ \bibinfo
  {pages} {044032} (\bibinfo {year} {2011}{\natexlab{b}})},\ \Eprint
  {https://arxiv.org/abs/1101.0781} {arXiv:1101.0781 [gr-qc]} \BibitemShut
  {NoStop}%
\bibitem [{\citenamefont {Macedo}\ \emph {et~al.}(2013)\citenamefont {Macedo},
  \citenamefont {Leite}, \citenamefont {Oliveira}, \citenamefont {Dolan},\ and\
  \citenamefont {Crispino}}]{Macedo:2013afa}%
  \BibitemOpen
  \bibfield  {author} {\bibinfo {author} {\bibfnamefont {C.~F.~B.}\
  \bibnamefont {Macedo}}, \bibinfo {author} {\bibfnamefont {L.~C.~S.}\
  \bibnamefont {Leite}}, \bibinfo {author} {\bibfnamefont {E.~S.}\ \bibnamefont
  {Oliveira}}, \bibinfo {author} {\bibfnamefont {S.~R.}\ \bibnamefont
  {Dolan}},\ and\ \bibinfo {author} {\bibfnamefont {L.~C.~B.}\ \bibnamefont
  {Crispino}},\ }\bibfield  {title} {\bibinfo {title} {{Absorption of planar
  massless scalar waves by Kerr black holes}},\ }\href
  {https://doi.org/10.1103/PhysRevD.88.064033} {\bibfield  {journal} {\bibinfo
  {journal} {Phys.\ Rev.\ D}\ }\textbf {\bibinfo {volume} {88}},\ \bibinfo
  {pages} {064033} (\bibinfo {year} {2013})},\ \Eprint
  {https://arxiv.org/abs/1308.0018} {arXiv:1308.0018 [gr-qc]} \BibitemShut
  {NoStop}%
\bibitem [{\citenamefont {Folacci}\ and\ \citenamefont {Ould
  El~Hadj}(2019{\natexlab{a}})}]{Folacci:2019cmc}%
  \BibitemOpen
  \bibfield  {author} {\bibinfo {author} {\bibfnamefont {A.}~\bibnamefont
  {Folacci}}\ and\ \bibinfo {author} {\bibfnamefont {M.}~\bibnamefont {Ould
  El~Hadj}},\ }\bibfield  {title} {\bibinfo {title} {{Regge pole description of
  scattering of scalar and electromagnetic waves by a Schwarzschild black
  hole}},\ }\href {https://doi.org/10.1103/PhysRevD.99.104079} {\bibfield
  {journal} {\bibinfo  {journal} {Phys.\ Rev.\ D}\ }\textbf {\bibinfo {volume}
  {99}},\ \bibinfo {pages} {104079} (\bibinfo {year} {2019}{\natexlab{a}})},\
  \Eprint {https://arxiv.org/abs/1901.03965} {arXiv:1901.03965 [gr-qc]}
  \BibitemShut {NoStop}%
\bibitem [{\citenamefont {Folacci}\ and\ \citenamefont {Ould
  El~Hadj}(2019{\natexlab{b}})}]{Folacci:2019vtt}%
  \BibitemOpen
  \bibfield  {author} {\bibinfo {author} {\bibfnamefont {A.}~\bibnamefont
  {Folacci}}\ and\ \bibinfo {author} {\bibfnamefont {M.}~\bibnamefont {Ould
  El~Hadj}},\ }\bibfield  {title} {\bibinfo {title} {{Regge pole description of
  scattering of gravitational waves by a Schwarzschild black hole}},\ }\href
  {https://doi.org/10.1103/PhysRevD.100.064009} {\bibfield  {journal} {\bibinfo
   {journal} {Phys.\ Rev.\ D}\ }\textbf {\bibinfo {volume} {100}},\ \bibinfo
  {pages} {064009} (\bibinfo {year} {2019}{\natexlab{b}})},\ \Eprint
  {https://arxiv.org/abs/1906.01441} {arXiv:1906.01441 [gr-qc]} \BibitemShut
  {NoStop}%
\bibitem [{\citenamefont {Ould El~Hadj}\ \emph {et~al.}(2020)\citenamefont
  {Ould El~Hadj}, \citenamefont {Stratton},\ and\ \citenamefont
  {Dolan}}]{OuldElHadj:2019kji}%
  \BibitemOpen
  \bibfield  {author} {\bibinfo {author} {\bibfnamefont {M.}~\bibnamefont {Ould
  El~Hadj}}, \bibinfo {author} {\bibfnamefont {T.}~\bibnamefont {Stratton}},\
  and\ \bibinfo {author} {\bibfnamefont {S.~R.}\ \bibnamefont {Dolan}},\
  }\bibfield  {title} {\bibinfo {title} {{Scattering from compact objects:
  Regge poles and the complex angular momentum method}},\ }\href
  {https://doi.org/10.1103/PhysRevD.101.104035} {\bibfield  {journal} {\bibinfo
   {journal} {Phys.\ Rev.\ D}\ }\textbf {\bibinfo {volume} {101}},\ \bibinfo
  {pages} {104035} (\bibinfo {year} {2020})},\ \Eprint
  {https://arxiv.org/abs/1912.11348} {arXiv:1912.11348 [gr-qc]} \BibitemShut
  {NoStop}%
\bibitem [{\citenamefont {Folacci}\ and\ \citenamefont {Ould
  El~Hadj}(2018)}]{Folacci:2018sef}%
  \BibitemOpen
  \bibfield  {author} {\bibinfo {author} {\bibfnamefont {A.}~\bibnamefont
  {Folacci}}\ and\ \bibinfo {author} {\bibfnamefont {M.}~\bibnamefont {Ould
  El~Hadj}},\ }\bibfield  {title} {\bibinfo {title} {{Alternative description
  of gravitational radiation from black holes based on the Regge poles of the
  ${\cal S}$-matrix and the associated residues}},\ }\href
  {https://doi.org/10.1103/PhysRevD.98.064052} {\bibfield  {journal} {\bibinfo
  {journal} {Phys.\ Rev.\ D}\ }\textbf {\bibinfo {volume} {98}},\ \bibinfo
  {pages} {064052} (\bibinfo {year} {2018})},\ \Eprint
  {https://arxiv.org/abs/1807.09056} {arXiv:1807.09056 [gr-qc]} \BibitemShut
  {NoStop}%
\bibitem [{\citenamefont {Folacci}\ and\ \citenamefont {Ould
  El~Hadj}(2020)}]{Folacci:2020ekl}%
  \BibitemOpen
  \bibfield  {author} {\bibinfo {author} {\bibfnamefont {A.}~\bibnamefont
  {Folacci}}\ and\ \bibinfo {author} {\bibfnamefont {M.}~\bibnamefont {Ould
  El~Hadj}},\ }\bibfield  {title} {\bibinfo {title} {{Electromagnetic radiation
  generated by a charged particle falling radially into a Schwarzschild black
  hole: A complex angular momentum description}},\ }\href
  {https://doi.org/10.1103/PhysRevD.102.024026} {\bibfield  {journal} {\bibinfo
   {journal} {Phys.\ Rev.\ D}\ }\textbf {\bibinfo {volume} {102}},\ \bibinfo
  {pages} {024026} (\bibinfo {year} {2020})},\ \Eprint
  {https://arxiv.org/abs/2004.07813} {arXiv:2004.07813 [gr-qc]} \BibitemShut
  {NoStop}%
\bibitem [{\citenamefont {Glampedakis}\ and\ \citenamefont
  {Andersson}(2003)}]{Glampedakis:2003dn}%
  \BibitemOpen
  \bibfield  {author} {\bibinfo {author} {\bibfnamefont {K.}~\bibnamefont
  {Glampedakis}}\ and\ \bibinfo {author} {\bibfnamefont {N.}~\bibnamefont
  {Andersson}},\ }\bibfield  {title} {\bibinfo {title} {{Quick and dirty
  methods for studying black hole resonances}},\ }\href
  {https://doi.org/10.1088/0264-9381/20/15/312} {\bibfield  {journal} {\bibinfo
   {journal} {Class.\ Quant.\ Grav.}\ }\textbf {\bibinfo {volume} {20}},\
  \bibinfo {pages} {3441} (\bibinfo {year} {2003})},\ \Eprint
  {https://arxiv.org/abs/gr-qc/0304030} {arXiv:gr-qc/0304030} \BibitemShut
  {NoStop}%
\bibitem [{\citenamefont {Schutz}\ and\ \citenamefont
  {Will}(1985)}]{Schutz:1985km}%
  \BibitemOpen
  \bibfield  {author} {\bibinfo {author} {\bibfnamefont {B.~F.}\ \bibnamefont
  {Schutz}}\ and\ \bibinfo {author} {\bibfnamefont {C.~M.}\ \bibnamefont
  {Will}},\ }\bibfield  {title} {\bibinfo {title} {{Black hole normal modes: A
  semianalytic approach}},\ }\href {https://doi.org/10.1086/184453} {\bibfield
  {journal} {\bibinfo  {journal} {Astrophys.\ J.\ Lett.}\ }\textbf {\bibinfo
  {volume} {291}},\ \bibinfo {pages} {L33} (\bibinfo {year}
  {1985})}\BibitemShut {NoStop}%
\bibitem [{\citenamefont {Iyer}\ and\ \citenamefont
  {Will}(1987)}]{Iyer:1986np}%
  \BibitemOpen
  \bibfield  {author} {\bibinfo {author} {\bibfnamefont {S.}~\bibnamefont
  {Iyer}}\ and\ \bibinfo {author} {\bibfnamefont {C.~M.}\ \bibnamefont
  {Will}},\ }\bibfield  {title} {\bibinfo {title} {{Black hole normal modes: A
  WKB approach. I. Foundations and application of a higher order WKB analysis
  of potential barrier scattering}},\ }\href
  {https://doi.org/10.1103/PhysRevD.35.3621} {\bibfield  {journal} {\bibinfo
  {journal} {Phys.\ Rev.\ D}\ }\textbf {\bibinfo {volume} {35}},\ \bibinfo
  {pages} {3621} (\bibinfo {year} {1987})}\BibitemShut {NoStop}%
\bibitem [{\citenamefont {Iyer}(1987)}]{Iyer:1986nq}%
  \BibitemOpen
  \bibfield  {author} {\bibinfo {author} {\bibfnamefont {S.}~\bibnamefont
  {Iyer}},\ }\bibfield  {title} {\bibinfo {title} {{Black hole normal modes: A
  WKB approach. II. Schwarzschild black holes}},\ }\href
  {https://doi.org/10.1103/PhysRevD.35.3632} {\bibfield  {journal} {\bibinfo
  {journal} {Phys.\ Rev.\ D}\ }\textbf {\bibinfo {volume} {35}},\ \bibinfo
  {pages} {3632} (\bibinfo {year} {1987})}\BibitemShut {NoStop}%
\bibitem [{\citenamefont {Will}\ and\ \citenamefont
  {Guinn}(1988)}]{Will:1988zz}%
  \BibitemOpen
  \bibfield  {author} {\bibinfo {author} {\bibfnamefont {C.~M.}\ \bibnamefont
  {Will}}\ and\ \bibinfo {author} {\bibfnamefont {J.~W.}\ \bibnamefont
  {Guinn}},\ }\bibfield  {title} {\bibinfo {title} {{Tunneling near the peaks
  of potential barriers: Consequences of higher-order Wentzel-Kramers-Brillouin
  corrections}},\ }\href {https://doi.org/10.1103/PhysRevA.37.3674} {\bibfield
  {journal} {\bibinfo  {journal} {Phys.\ Rev.\ A}\ }\textbf {\bibinfo {volume}
  {37}},\ \bibinfo {pages} {3674} (\bibinfo {year} {1988})}\BibitemShut
  {NoStop}%
\bibitem [{\citenamefont {Seidel}\ and\ \citenamefont
  {Iyer}(1990)}]{Seidel:1989bp}%
  \BibitemOpen
  \bibfield  {author} {\bibinfo {author} {\bibfnamefont {E.}~\bibnamefont
  {Seidel}}\ and\ \bibinfo {author} {\bibfnamefont {S.}~\bibnamefont {Iyer}},\
  }\bibfield  {title} {\bibinfo {title} {{Black-hole normal modes: A WKB
  approach. IV. Kerr black holes}},\ }\href
  {https://doi.org/10.1103/PhysRevD.41.374} {\bibfield  {journal} {\bibinfo
  {journal} {Phys.\ Rev.\ D}\ }\textbf {\bibinfo {volume} {41}},\ \bibinfo
  {pages} {374} (\bibinfo {year} {1990})}\BibitemShut {NoStop}%
\bibitem [{\citenamefont {Konoplya}(2003)}]{Konoplya:2003ii}%
  \BibitemOpen
  \bibfield  {author} {\bibinfo {author} {\bibfnamefont {R.~A.}\ \bibnamefont
  {Konoplya}},\ }\bibfield  {title} {\bibinfo {title} {{Quasinormal behavior of
  the d-dimensional Schwarzschild black hole and higher order WKB approach}},\
  }\href {https://doi.org/10.1103/PhysRevD.68.024018} {\bibfield  {journal}
  {\bibinfo  {journal} {Phys.\ Rev.\ D}\ }\textbf {\bibinfo {volume} {68}},\
  \bibinfo {pages} {024018} (\bibinfo {year} {2003})},\ \Eprint
  {https://arxiv.org/abs/gr-qc/0303052} {arXiv:gr-qc/0303052} \BibitemShut
  {NoStop}%
\bibitem [{\citenamefont {Teukolsky}(1972)}]{Teukolsky:1972my}%
  \BibitemOpen
  \bibfield  {author} {\bibinfo {author} {\bibfnamefont {S.~A.}\ \bibnamefont
  {Teukolsky}},\ }\bibfield  {title} {\bibinfo {title} {{Rotating black holes:
  Separable wave equations for gravitational and electromagnetic
  perturbations}},\ }\href {https://doi.org/10.1103/PhysRevLett.29.1114}
  {\bibfield  {journal} {\bibinfo  {journal} {Phys.\ Rev.\ Lett.}\ }\textbf
  {\bibinfo {volume} {29}},\ \bibinfo {pages} {1114} (\bibinfo {year}
  {1972})}\BibitemShut {NoStop}%
\bibitem [{\citenamefont {Teukolsky}(1973)}]{Teukolsky:1973ha}%
  \BibitemOpen
  \bibfield  {author} {\bibinfo {author} {\bibfnamefont {S.~A.}\ \bibnamefont
  {Teukolsky}},\ }\bibfield  {title} {\bibinfo {title} {{Perturbations of a
  rotating black hole. I. Fundamental equations for gravitational
  electromagnetic and neutrino field perturbations}},\ }\href
  {https://doi.org/10.1086/152444} {\bibfield  {journal} {\bibinfo  {journal}
  {Astrophys.\ J.}\ }\textbf {\bibinfo {volume} {185}},\ \bibinfo {pages} {635}
  (\bibinfo {year} {1973})}\BibitemShut {NoStop}%
\bibitem [{\citenamefont {Boyer}\ and\ \citenamefont
  {Lindquist}(1967)}]{Boyer:1966qh}%
  \BibitemOpen
  \bibfield  {author} {\bibinfo {author} {\bibfnamefont {R.~H.}\ \bibnamefont
  {Boyer}}\ and\ \bibinfo {author} {\bibfnamefont {R.~W.}\ \bibnamefont
  {Lindquist}},\ }\bibfield  {title} {\bibinfo {title} {{Maximal analytic
  extension of the Kerr metric}},\ }\href {https://doi.org/10.1063/1.1705193}
  {\bibfield  {journal} {\bibinfo  {journal} {J.\ Math.\ Phys.}\ }\textbf
  {\bibinfo {volume} {8}},\ \bibinfo {pages} {265} (\bibinfo {year}
  {1967})}\BibitemShut {NoStop}%
\bibitem [{\citenamefont {Misner}\ \emph {et~al.}(1973)\citenamefont {Misner},
  \citenamefont {Thorne},\ and\ \citenamefont {Wheeler}}]{Misner:1974qy}%
  \BibitemOpen
  \bibfield  {author} {\bibinfo {author} {\bibfnamefont {C.~W.}\ \bibnamefont
  {Misner}}, \bibinfo {author} {\bibfnamefont {K.~S.}\ \bibnamefont {Thorne}},\
  and\ \bibinfo {author} {\bibfnamefont {J.~A.}\ \bibnamefont {Wheeler}},\
  }\href@noop {} {\emph {\bibinfo {title} {{Gravitation}}}}\ (\bibinfo
  {publisher} {W. H. Freeman, San Francisco},\ \bibinfo {year}
  {1973})\BibitemShut {NoStop}%
\bibitem [{\citenamefont {Fernandes}\ and\ \citenamefont
  {Lun}(1997)}]{GaugeInv}%
  \BibitemOpen
  \bibfield  {author} {\bibinfo {author} {\bibfnamefont {J.~F.~Q.}\
  \bibnamefont {Fernandes}}\ and\ \bibinfo {author} {\bibfnamefont {A.~W.~C.}\
  \bibnamefont {Lun}},\ }\bibfield  {title} {\bibinfo {title} {{Gauge invariant
  perturbations of black holes. II: Kerr space-time}},\ }\href
  {https://doi.org/10.1063/1.531852} {\bibfield  {journal} {\bibinfo  {journal}
  {J.\ Math.\ Phys.}\ }\textbf {\bibinfo {volume} {38}},\ \bibinfo {pages}
  {330} (\bibinfo {year} {1997})}\BibitemShut {NoStop}%
\bibitem [{\citenamefont {Newman}\ and\ \citenamefont
  {Penrose}(1962)}]{Newman:1961qr}%
  \BibitemOpen
  \bibfield  {author} {\bibinfo {author} {\bibfnamefont {E.}~\bibnamefont
  {Newman}}\ and\ \bibinfo {author} {\bibfnamefont {R.}~\bibnamefont
  {Penrose}},\ }\bibfield  {title} {\bibinfo {title} {{An approach to
  gravitational radiation by a method of spin coefficients}},\ }\href
  {https://doi.org/10.1063/1.1724257} {\bibfield  {journal} {\bibinfo
  {journal} {J.\ Math.\ Phys.}\ }\textbf {\bibinfo {volume} {3}},\ \bibinfo
  {pages} {566} (\bibinfo {year} {1962})}\BibitemShut {NoStop}%
\bibitem [{\citenamefont {Kinnersley}(1969)}]{Kinnersley:1969zza}%
  \BibitemOpen
  \bibfield  {author} {\bibinfo {author} {\bibfnamefont {W.}~\bibnamefont
  {Kinnersley}},\ }\bibfield  {title} {\bibinfo {title} {{Type D vacuum
  metrics}},\ }\href {https://doi.org/10.1063/1.1664958} {\bibfield  {journal}
  {\bibinfo  {journal} {J.\ Math.\ Phys.}\ }\textbf {\bibinfo {volume} {10}},\
  \bibinfo {pages} {1195} (\bibinfo {year} {1969})}\BibitemShut {NoStop}%
\bibitem [{\citenamefont {Bardeen}\ and\ \citenamefont
  {Press}(1973)}]{Bardeen:1973xb}%
  \BibitemOpen
  \bibfield  {author} {\bibinfo {author} {\bibfnamefont {J.~M.}\ \bibnamefont
  {Bardeen}}\ and\ \bibinfo {author} {\bibfnamefont {W.~H.}\ \bibnamefont
  {Press}},\ }\bibfield  {title} {\bibinfo {title} {{Radiation fields in the
  Schwarzschild background}},\ }\href {https://doi.org/10.1063/1.1666175}
  {\bibfield  {journal} {\bibinfo  {journal} {J.\ Math.\ Phys.}\ }\textbf
  {\bibinfo {volume} {14}},\ \bibinfo {pages} {7} (\bibinfo {year}
  {1973})}\BibitemShut {NoStop}%
\bibitem [{\citenamefont {Chrzanowski}\ and\ \citenamefont
  {Misner}(1974)}]{Chrzanowski:1974nr}%
  \BibitemOpen
  \bibfield  {author} {\bibinfo {author} {\bibfnamefont {P.~L.}\ \bibnamefont
  {Chrzanowski}}\ and\ \bibinfo {author} {\bibfnamefont {C.~W.}\ \bibnamefont
  {Misner}},\ }\bibfield  {title} {\bibinfo {title} {{Geodesic synchrotron
  radiation in the Kerr geometry by the method of asymptotically factorized
  Green's functions}},\ }\href {https://doi.org/10.1103/PhysRevD.10.1701}
  {\bibfield  {journal} {\bibinfo  {journal} {Phys.\ Rev.\ D}\ }\textbf
  {\bibinfo {volume} {10}},\ \bibinfo {pages} {1701} (\bibinfo {year}
  {1974})}\BibitemShut {NoStop}%
\bibitem [{\citenamefont {Chrzanowski}(1975)}]{Chrzanowski:1975wv}%
  \BibitemOpen
  \bibfield  {author} {\bibinfo {author} {\bibfnamefont {P.~L.}\ \bibnamefont
  {Chrzanowski}},\ }\bibfield  {title} {\bibinfo {title} {{Vector potential and
  metric perturbations of a rotating black hole}},\ }\href
  {https://doi.org/10.1103/PhysRevD.11.2042} {\bibfield  {journal} {\bibinfo
  {journal} {Phys.\ Rev.\ D}\ }\textbf {\bibinfo {volume} {11}},\ \bibinfo
  {pages} {2042} (\bibinfo {year} {1975})}\BibitemShut {NoStop}%
\bibitem [{\citenamefont {Casals}\ \emph {et~al.}(2016)\citenamefont {Casals},
  \citenamefont {Kavanagh},\ and\ \citenamefont {Ottewill}}]{Casals:2016soq}%
  \BibitemOpen
  \bibfield  {author} {\bibinfo {author} {\bibfnamefont {M.}~\bibnamefont
  {Casals}}, \bibinfo {author} {\bibfnamefont {C.}~\bibnamefont {Kavanagh}},\
  and\ \bibinfo {author} {\bibfnamefont {A.~C.}\ \bibnamefont {Ottewill}},\
  }\bibfield  {title} {\bibinfo {title} {{High-order late-time tail in a Kerr
  spacetime}},\ }\href {https://doi.org/10.1103/PhysRevD.94.124053} {\bibfield
  {journal} {\bibinfo  {journal} {Phys.\ Rev.\ D}\ }\textbf {\bibinfo {volume}
  {94}},\ \bibinfo {pages} {124053} (\bibinfo {year} {2016})},\ \Eprint
  {https://arxiv.org/abs/1608.05392} {arXiv:1608.05392 [gr-qc]} \BibitemShut
  {NoStop}%
\bibitem [{\citenamefont {Casals}\ and\ \citenamefont
  {Longo~Micchi}(2019)}]{Casals:2019vdb}%
  \BibitemOpen
  \bibfield  {author} {\bibinfo {author} {\bibfnamefont {M.}~\bibnamefont
  {Casals}}\ and\ \bibinfo {author} {\bibfnamefont {L.~F.}\ \bibnamefont
  {Longo~Micchi}},\ }\bibfield  {title} {\bibinfo {title} {{Spectroscopy of
  extremal and near-extremal Kerr black holes}},\ }\href
  {https://doi.org/10.1103/PhysRevD.99.084047} {\bibfield  {journal} {\bibinfo
  {journal} {Phys.\ Rev.\ D}\ }\textbf {\bibinfo {volume} {99}},\ \bibinfo
  {pages} {084047} (\bibinfo {year} {2019})},\ \Eprint
  {https://arxiv.org/abs/1901.04586} {arXiv:1901.04586 [gr-qc]} \BibitemShut
  {NoStop}%
\bibitem [{\citenamefont {Fackerell}\ and\ \citenamefont
  {Crossman}(1977)}]{Fackerell1977}%
  \BibitemOpen
  \bibfield  {author} {\bibinfo {author} {\bibfnamefont {E.~D.}\ \bibnamefont
  {Fackerell}}\ and\ \bibinfo {author} {\bibfnamefont {R.~G.}\ \bibnamefont
  {Crossman}},\ }\bibfield  {title} {\bibinfo {title} {Spin-weighted angular
  spheroidal functions},\ }\href {https://doi.org/10.1063/1.523499} {\bibfield
  {journal} {\bibinfo  {journal} {J.\ Math.\ Phys.}\ }\textbf {\bibinfo
  {volume} {18}},\ \bibinfo {pages} {1849} (\bibinfo {year}
  {1977})}\BibitemShut {NoStop}%
\bibitem [{\citenamefont {Breuer}\ \emph {et~al.}(1977)\citenamefont {Breuer},
  \citenamefont {Ryan~Jr},\ and\ \citenamefont {Waller}}]{Breuer1977}%
  \BibitemOpen
  \bibfield  {author} {\bibinfo {author} {\bibfnamefont {R.~A.}\ \bibnamefont
  {Breuer}}, \bibinfo {author} {\bibfnamefont {M.~P.}\ \bibnamefont
  {Ryan~Jr}},\ and\ \bibinfo {author} {\bibfnamefont {S.}~\bibnamefont
  {Waller}},\ }\bibfield  {title} {\bibinfo {title} {{Some properties of
  spin-weighted spheroidal harmonics}},\ }\href
  {https://doi.org/doi.org/10.1098/rspa.1977.0187} {\bibfield  {journal}
  {\bibinfo  {journal} {Proc.\ Roy.\ Soc.\ Lond.\ A}\ }\textbf {\bibinfo
  {volume} {358}},\ \bibinfo {pages} {71} (\bibinfo {year} {1977})}\BibitemShut
  {NoStop}%
\bibitem [{\citenamefont {Seidel}(1989)}]{Seidel:1988ue}%
  \BibitemOpen
  \bibfield  {author} {\bibinfo {author} {\bibfnamefont {E.}~\bibnamefont
  {Seidel}},\ }\bibfield  {title} {\bibinfo {title} {{A comment on the
  eigenvalues of spin-weighted spheroidal functions}},\ }\href
  {https://doi.org/10.1088/0264-9381/6/7/012} {\bibfield  {journal} {\bibinfo
  {journal} {Class.\ Quant.\ Grav.}\ }\textbf {\bibinfo {volume} {6}},\
  \bibinfo {pages} {1057} (\bibinfo {year} {1989})}\BibitemShut {NoStop}%
\bibitem [{\citenamefont {Leaver}(1986{\natexlab{b}})}]{Leaver:1986gd}%
  \BibitemOpen
  \bibfield  {author} {\bibinfo {author} {\bibfnamefont {E.~W.}\ \bibnamefont
  {Leaver}},\ }\bibfield  {title} {\bibinfo {title} {{Spectral decomposition of
  the perturbation response of the Schwarzschild geometry}},\ }\href
  {https://doi.org/10.1103/PhysRevD.34.384} {\bibfield  {journal} {\bibinfo
  {journal} {Phys.\ Rev.\ D}\ }\textbf {\bibinfo {volume} {34}},\ \bibinfo
  {pages} {384} (\bibinfo {year} {1986}{\natexlab{b}})}\BibitemShut {NoStop}%
\bibitem [{\citenamefont {Casals}\ and\ \citenamefont
  {Zimmerman}(2019)}]{Casals:2018eev}%
  \BibitemOpen
  \bibfield  {author} {\bibinfo {author} {\bibfnamefont {M.}~\bibnamefont
  {Casals}}\ and\ \bibinfo {author} {\bibfnamefont {P.}~\bibnamefont
  {Zimmerman}},\ }\bibfield  {title} {\bibinfo {title} {{Perturbations of an
  extremal Kerr spacetime: Analytic framework and late-time tails}},\ }\href
  {https://doi.org/10.1103/PhysRevD.100.124027} {\bibfield  {journal} {\bibinfo
   {journal} {Phys.\ Rev.\ D}\ }\textbf {\bibinfo {volume} {100}},\ \bibinfo
  {pages} {124027} (\bibinfo {year} {2019})},\ \Eprint
  {https://arxiv.org/abs/1801.05830} {arXiv:1801.05830 [gr-qc]} \BibitemShut
  {NoStop}%
\bibitem [{\citenamefont {Berti}\ \emph
  {et~al.}(2006{\natexlab{a}})\citenamefont {Berti}, \citenamefont {Cardoso},\
  and\ \citenamefont {Will}}]{Berti:2005ys}%
  \BibitemOpen
  \bibfield  {author} {\bibinfo {author} {\bibfnamefont {E.}~\bibnamefont
  {Berti}}, \bibinfo {author} {\bibfnamefont {V.}~\bibnamefont {Cardoso}},\
  and\ \bibinfo {author} {\bibfnamefont {C.~M.}\ \bibnamefont {Will}},\
  }\bibfield  {title} {\bibinfo {title} {{On gravitational-wave spectroscopy of
  massive black holes with the space interferometer LISA}},\ }\href
  {https://doi.org/10.1103/PhysRevD.73.064030} {\bibfield  {journal} {\bibinfo
  {journal} {Phys.\ Rev.\ D}\ }\textbf {\bibinfo {volume} {73}},\ \bibinfo
  {pages} {064030} (\bibinfo {year} {2006}{\natexlab{a}})},\ \Eprint
  {https://arxiv.org/abs/gr-qc/0512160} {arXiv:gr-qc/0512160} \BibitemShut
  {NoStop}%
\bibitem [{\citenamefont {Cook}\ and\ \citenamefont
  {Zalutskiy}(2014)}]{Cook:2014cta}%
  \BibitemOpen
  \bibfield  {author} {\bibinfo {author} {\bibfnamefont {G.~B.}\ \bibnamefont
  {Cook}}\ and\ \bibinfo {author} {\bibfnamefont {M.}~\bibnamefont
  {Zalutskiy}},\ }\bibfield  {title} {\bibinfo {title} {{Gravitational
  perturbations of the Kerr geometry: High-accuracy study}},\ }\href
  {https://doi.org/10.1103/PhysRevD.90.124021} {\bibfield  {journal} {\bibinfo
  {journal} {Phys.\ Rev.\ D}\ }\textbf {\bibinfo {volume} {90}},\ \bibinfo
  {pages} {124021} (\bibinfo {year} {2014})},\ \Eprint
  {https://arxiv.org/abs/1410.7698} {arXiv:1410.7698 [gr-qc]} \BibitemShut
  {NoStop}%
\bibitem [{\citenamefont {Majumdar}\ and\ \citenamefont
  {Panchapakesan}(1989)}]{Majumdar:1989tzg}%
  \BibitemOpen
  \bibfield  {author} {\bibinfo {author} {\bibfnamefont {B.}~\bibnamefont
  {Majumdar}}\ and\ \bibinfo {author} {\bibfnamefont {N.}~\bibnamefont
  {Panchapakesan}},\ }\bibfield  {title} {\bibinfo {title} {{Schwarzschild
  black-hole normal modes using the Hill determinant}},\ }\href
  {https://doi.org/10.1103/PhysRevD.40.2568} {\bibfield  {journal} {\bibinfo
  {journal} {Phys.\ Rev.\ D}\ }\textbf {\bibinfo {volume} {40}},\ \bibinfo
  {pages} {2568} (\bibinfo {year} {1989})}\BibitemShut {NoStop}%
\bibitem [{\citenamefont {Casals}\ and\ \citenamefont
  {Ottewill}(2005)}]{Casals:2004zq}%
  \BibitemOpen
  \bibfield  {author} {\bibinfo {author} {\bibfnamefont {M.}~\bibnamefont
  {Casals}}\ and\ \bibinfo {author} {\bibfnamefont {A.~C.}\ \bibnamefont
  {Ottewill}},\ }\bibfield  {title} {\bibinfo {title} {{High frequency
  asymptotics for the spin-weighted spheroidal equation}},\ }\href
  {https://doi.org/10.1103/PhysRevD.71.064025} {\bibfield  {journal} {\bibinfo
  {journal} {Phys.\ Rev.\ D}\ }\textbf {\bibinfo {volume} {71}},\ \bibinfo
  {pages} {064025} (\bibinfo {year} {2005})},\ \Eprint
  {https://arxiv.org/abs/gr-qc/0409012} {arXiv:gr-qc/0409012} \BibitemShut
  {NoStop}%
\bibitem [{\citenamefont {Berti}\ \emph
  {et~al.}(2006{\natexlab{b}})\citenamefont {Berti}, \citenamefont {Cardoso},\
  and\ \citenamefont {Casals}}]{Berti:2005gp}%
  \BibitemOpen
  \bibfield  {author} {\bibinfo {author} {\bibfnamefont {E.}~\bibnamefont
  {Berti}}, \bibinfo {author} {\bibfnamefont {V.}~\bibnamefont {Cardoso}},\
  and\ \bibinfo {author} {\bibfnamefont {M.}~\bibnamefont {Casals}},\
  }\bibfield  {title} {\bibinfo {title} {{Eigenvalues and eigenfunctions of
  spin-weighted spheroidal harmonics in four and higher dimensions}},\ }\href
  {https://doi.org/10.1103/PhysRevD.73.109902} {\bibfield  {journal} {\bibinfo
  {journal} {Phys.\ Rev.\ D}\ }\textbf {\bibinfo {volume} {73}},\ \bibinfo
  {pages} {024013} (\bibinfo {year} {2006}{\natexlab{b}})},\ \bibinfo {note}
  {[Erratum: Phys.\ Rev.\ D 73, 109902 (2006)]},\ \Eprint
  {https://arxiv.org/abs/gr-qc/0511111} {arXiv:gr-qc/0511111} \BibitemShut
  {NoStop}%
\bibitem [{\citenamefont {Casals}\ \emph {et~al.}(2019)\citenamefont {Casals},
  \citenamefont {Ottewill},\ and\ \citenamefont {Warburton}}]{Casals:2018cgx}%
  \BibitemOpen
  \bibfield  {author} {\bibinfo {author} {\bibfnamefont {M.}~\bibnamefont
  {Casals}}, \bibinfo {author} {\bibfnamefont {A.~C.}\ \bibnamefont
  {Ottewill}},\ and\ \bibinfo {author} {\bibfnamefont {N.}~\bibnamefont
  {Warburton}},\ }\bibfield  {title} {\bibinfo {title} {{High-order asymptotics
  for the spin-weighted spheroidal equation at large real frequency}},\ }\href
  {https://doi.org/10.1098/rspa.2018.0701} {\bibfield  {journal} {\bibinfo
  {journal} {Proc.\ Roy.\ Soc.\ Lond.\ A}\ }\textbf {\bibinfo {volume} {475}},\
  \bibinfo {pages} {20180701} (\bibinfo {year} {2019})},\ \Eprint
  {https://arxiv.org/abs/1810.00432} {arXiv:1810.00432 [gr-qc]} \BibitemShut
  {NoStop}%
\bibitem [{\citenamefont {Detweiler}(1980)}]{Detweiler:1980gk}%
  \BibitemOpen
  \bibfield  {author} {\bibinfo {author} {\bibfnamefont {S.}~\bibnamefont
  {Detweiler}},\ }\bibfield  {title} {\bibinfo {title} {{Black holes and
  gravitational waves. III - The resonant frequencies of rotating holes}},\
  }\href {https://doi.org/10.1086/158109} {\bibfield  {journal} {\bibinfo
  {journal} {Astrophys.\ J.}\ }\textbf {\bibinfo {volume} {239}},\ \bibinfo
  {pages} {292} (\bibinfo {year} {1980})}\BibitemShut {NoStop}%
\bibitem [{\citenamefont {Brito}\ \emph {et~al.}(2015)\citenamefont {Brito},
  \citenamefont {Cardoso},\ and\ \citenamefont {Pani}}]{Brito:2015oca}%
  \BibitemOpen
  \bibfield  {author} {\bibinfo {author} {\bibfnamefont {R.}~\bibnamefont
  {Brito}}, \bibinfo {author} {\bibfnamefont {V.}~\bibnamefont {Cardoso}},\
  and\ \bibinfo {author} {\bibfnamefont {P.}~\bibnamefont {Pani}},\ }\href
  {https://doi.org/10.1007/978-3-319-19000-6} {\emph {\bibinfo {title}
  {{Superradiance}: {New Frontiers in Black Hole Physics}}}},\ Vol.\ \bibinfo
  {volume} {906}\ (\bibinfo  {publisher} {Springer},\ \bibinfo {year} {2015})\
  \Eprint {https://arxiv.org/abs/1501.06570} {arXiv:1501.06570 [gr-qc]}
  \BibitemShut {NoStop}%
\bibitem [{\citenamefont {Chandrashekhar}(1984)}]{Chandrashekhar1984}%
  \BibitemOpen
  \bibfield  {author} {\bibinfo {author} {\bibfnamefont {S.}~\bibnamefont
  {Chandrashekhar}},\ }\bibfield  {title} {\bibinfo {title} {On algebraically
  special perturbations of black holes},\ }\href
  {https://doi.org/10.1098/rspa.1984.0021} {\bibfield  {journal} {\bibinfo
  {journal} {Proc.\ Roy.\ Soc.\ Lond.\ A.}\ }\textbf {\bibinfo {volume}
  {392}},\ \bibinfo {pages} {1} (\bibinfo {year} {1984})}\BibitemShut {NoStop}%
\bibitem [{\citenamefont {Cook}\ and\ \citenamefont
  {Zalutskiy}(2016)}]{Cook:2016ngj}%
  \BibitemOpen
  \bibfield  {author} {\bibinfo {author} {\bibfnamefont {G.~B.}\ \bibnamefont
  {Cook}}\ and\ \bibinfo {author} {\bibfnamefont {M.}~\bibnamefont
  {Zalutskiy}},\ }\bibfield  {title} {\bibinfo {title} {{Modes of the Kerr
  geometry with purely imaginary frequencies}},\ }\href
  {https://doi.org/10.1103/PhysRevD.94.104074} {\bibfield  {journal} {\bibinfo
  {journal} {Phys.\ Rev.\ D}\ }\textbf {\bibinfo {volume} {94}},\ \bibinfo
  {pages} {104074} (\bibinfo {year} {2016})},\ \Eprint
  {https://arxiv.org/abs/1607.07406} {arXiv:1607.07406 [gr-qc]} \BibitemShut
  {NoStop}%
\bibitem [{\citenamefont {Yang}\ \emph
  {et~al.}(2013{\natexlab{a}})\citenamefont {Yang}, \citenamefont {Zhang},
  \citenamefont {Zimmerman}, \citenamefont {Nichols}, \citenamefont {Berti},\
  and\ \citenamefont {Chen}}]{Yang:2012pj}%
  \BibitemOpen
  \bibfield  {author} {\bibinfo {author} {\bibfnamefont {H.}~\bibnamefont
  {Yang}}, \bibinfo {author} {\bibfnamefont {F.}~\bibnamefont {Zhang}},
  \bibinfo {author} {\bibfnamefont {A.}~\bibnamefont {Zimmerman}}, \bibinfo
  {author} {\bibfnamefont {D.~A.}\ \bibnamefont {Nichols}}, \bibinfo {author}
  {\bibfnamefont {E.}~\bibnamefont {Berti}},\ and\ \bibinfo {author}
  {\bibfnamefont {Y.}~\bibnamefont {Chen}},\ }\bibfield  {title} {\bibinfo
  {title} {{Branching of quasinormal modes for nearly extremal Kerr black
  holes}},\ }\href {https://doi.org/10.1103/PhysRevD.87.041502} {\bibfield
  {journal} {\bibinfo  {journal} {Phys.\ Rev.\ D}\ }\textbf {\bibinfo {volume}
  {87}},\ \bibinfo {pages} {041502} (\bibinfo {year} {2013}{\natexlab{a}})},\
  \Eprint {https://arxiv.org/abs/1212.3271} {arXiv:1212.3271 [gr-qc]}
  \BibitemShut {NoStop}%
\bibitem [{\citenamefont {Yang}\ \emph
  {et~al.}(2013{\natexlab{b}})\citenamefont {Yang}, \citenamefont {Zimmerman},
  \citenamefont {Zengino\u{g}lu}, \citenamefont {Zhang}, \citenamefont
  {Berti},\ and\ \citenamefont {Chen}}]{Yang:2013uba}%
  \BibitemOpen
  \bibfield  {author} {\bibinfo {author} {\bibfnamefont {H.}~\bibnamefont
  {Yang}}, \bibinfo {author} {\bibfnamefont {A.}~\bibnamefont {Zimmerman}},
  \bibinfo {author} {\bibfnamefont {A.}~\bibnamefont {Zengino\u{g}lu}},
  \bibinfo {author} {\bibfnamefont {F.}~\bibnamefont {Zhang}}, \bibinfo
  {author} {\bibfnamefont {E.}~\bibnamefont {Berti}},\ and\ \bibinfo {author}
  {\bibfnamefont {Y.}~\bibnamefont {Chen}},\ }\bibfield  {title} {\bibinfo
  {title} {{Quasinormal modes of nearly extremal Kerr spacetimes: Spectrum
  bifurcation and power-law ringdown}},\ }\href
  {https://doi.org/10.1103/PhysRevD.88.044047} {\bibfield  {journal} {\bibinfo
  {journal} {Phys.\ Rev.\ D}\ }\textbf {\bibinfo {volume} {88}},\ \bibinfo
  {pages} {044047} (\bibinfo {year} {2013}{\natexlab{b}})},\ \Eprint
  {https://arxiv.org/abs/1307.8086} {arXiv:1307.8086 [gr-qc]} \BibitemShut
  {NoStop}%
\bibitem [{\citenamefont {Teo}(2003)}]{Teo:2003}%
  \BibitemOpen
  \bibfield  {author} {\bibinfo {author} {\bibfnamefont {E.}~\bibnamefont
  {Teo}},\ }\bibfield  {title} {\bibinfo {title} {{Spherical photon orbits
  around a Kerr black hole}},\ }\href {https://doi.org/10.1023/A:1026286607562}
  {\bibfield  {journal} {\bibinfo  {journal} {Gen.\ Rel.\ Grav.}\ }\textbf
  {\bibinfo {volume} {35}},\ \bibinfo {pages} {1909} (\bibinfo {year}
  {2003})}\BibitemShut {NoStop}%
\bibitem [{\citenamefont {Fiziev}(2010)}]{Fiziev:2009wn}%
  \BibitemOpen
  \bibfield  {author} {\bibinfo {author} {\bibfnamefont {P.~P.}\ \bibnamefont
  {Fiziev}},\ }\bibfield  {title} {\bibinfo {title} {{Classes of exact
  solutions to the Teukolsky master equation}},\ }\href
  {https://doi.org/10.1088/0264-9381/27/13/135001} {\bibfield  {journal}
  {\bibinfo  {journal} {Class.\ Quant.\ Grav.}\ }\textbf {\bibinfo {volume}
  {27}},\ \bibinfo {pages} {135001} (\bibinfo {year} {2010})},\ \Eprint
  {https://arxiv.org/abs/0908.4234} {arXiv:0908.4234 [gr-qc]} \BibitemShut
  {NoStop}%
\bibitem [{\citenamefont {Hatsuda}(2020)}]{Hatsuda:2020sbn}%
  \BibitemOpen
  \bibfield  {author} {\bibinfo {author} {\bibfnamefont {Y.}~\bibnamefont
  {Hatsuda}},\ }\bibfield  {title} {\bibinfo {title} {{Quasinormal modes of
  Kerr-de Sitter black holes via the Heun function}},\ }\href
  {https://doi.org/10.1088/1361-6382/abc82e} {\bibfield  {journal} {\bibinfo
  {journal} {Class.\ Quant.\ Grav.}\ }\textbf {\bibinfo {volume} {38}},\
  \bibinfo {pages} {025015} (\bibinfo {year} {2020})},\ \Eprint
  {https://arxiv.org/abs/2006.08957} {arXiv:2006.08957 [gr-qc]} \BibitemShut
  {NoStop}%
\bibitem [{\citenamefont {Giscard}\ and\ \citenamefont
  {Tamar}(2020)}]{Giscard:2020iqg}%
  \BibitemOpen
  \bibfield  {author} {\bibinfo {author} {\bibfnamefont {P.-L.}\ \bibnamefont
  {Giscard}}\ and\ \bibinfo {author} {\bibfnamefont {A.}~\bibnamefont
  {Tamar}},\ }\bibfield  {title} {\bibinfo {title} {{Elementary integral series
  for Heun functions. With an application to black-kole perturbation theory}},\
  }\href@noop {} {\  (\bibinfo {year} {2020})},\ \Eprint
  {https://arxiv.org/abs/2010.03919} {arXiv:2010.03919 [math-ph]} \BibitemShut
  {NoStop}%
\bibitem [{\citenamefont {Sasaki}\ and\ \citenamefont
  {Nakamura}(1982)}]{Sasaki:1981sx}%
  \BibitemOpen
  \bibfield  {author} {\bibinfo {author} {\bibfnamefont {M.}~\bibnamefont
  {Sasaki}}\ and\ \bibinfo {author} {\bibfnamefont {T.}~\bibnamefont
  {Nakamura}},\ }\bibfield  {title} {\bibinfo {title} {{Gravitational radiation
  from a Kerr black hole. I. Formulation and a method for numerical
  analysis}},\ }\href {https://doi.org/10.1143/PTP.67.1788} {\bibfield
  {journal} {\bibinfo  {journal} {Prog.\ Theor.\ Phys.}\ }\textbf {\bibinfo
  {volume} {67}},\ \bibinfo {pages} {1788} (\bibinfo {year}
  {1982})}\BibitemShut {NoStop}%
\bibitem [{\citenamefont {Nakamura}\ and\ \citenamefont
  {Nakano}(2016)}]{Nakamura:2016yjl}%
  \BibitemOpen
  \bibfield  {author} {\bibinfo {author} {\bibfnamefont {T.}~\bibnamefont
  {Nakamura}}\ and\ \bibinfo {author} {\bibfnamefont {H.}~\bibnamefont
  {Nakano}},\ }\bibfield  {title} {\bibinfo {title} {{How close can we approach
  the event horizon of the Kerr black hole from the detection of gravitational
  quasinormal modes?}},\ }\href {https://doi.org/10.1093/ptep/ptw026}
  {\bibfield  {journal} {\bibinfo  {journal} {Prog.\ Theor.\ Exp.\ Phys.}\
  }\textbf {\bibinfo {volume} {2016}},\ \bibinfo {pages} {041E01} (\bibinfo
  {year} {2016})},\ \Eprint {https://arxiv.org/abs/1602.02385}
  {arXiv:1602.02385 [gr-qc]} \BibitemShut {NoStop}%
\bibitem [{\citenamefont {Nakano}\ \emph {et~al.}(2016)\citenamefont {Nakano},
  \citenamefont {Nakamura},\ and\ \citenamefont {Tanaka}}]{Nakano:2016sgf}%
  \BibitemOpen
  \bibfield  {author} {\bibinfo {author} {\bibfnamefont {H.}~\bibnamefont
  {Nakano}}, \bibinfo {author} {\bibfnamefont {T.}~\bibnamefont {Nakamura}},\
  and\ \bibinfo {author} {\bibfnamefont {T.}~\bibnamefont {Tanaka}},\
  }\bibfield  {title} {\bibinfo {title} {{The detection of quasinormal mode
  with a/M = 0.95 would prove a sphere 99\% soaking in the ergoregion of the
  Kerr space-time}},\ }\href {https://doi.org/10.1093/ptep/ptw015} {\bibfield
  {journal} {\bibinfo  {journal} {Prog.\ Theor.\ Exp.\ Phys.}\ }\textbf
  {\bibinfo {volume} {2016}},\ \bibinfo {pages} {031E02} (\bibinfo {year}
  {2016})},\ \Eprint {https://arxiv.org/abs/1602.02875} {arXiv:1602.02875
  [gr-qc]} \BibitemShut {NoStop}%
\bibitem [{\citenamefont {Strand}\ and\ \citenamefont
  {Reinhardt}(1979)}]{StrandReinhardt1979}%
  \BibitemOpen
  \bibfield  {author} {\bibinfo {author} {\bibfnamefont {M.~P.}\ \bibnamefont
  {Strand}}\ and\ \bibinfo {author} {\bibfnamefont {W.~P.}\ \bibnamefont
  {Reinhardt}},\ }\bibfield  {title} {\bibinfo {title} {Semiclassical
  quantization of the low lying electronic states of ${{H}_2}^+$},\ }\href
  {https://doi.org/10.1063/1.437932} {\bibfield  {journal} {\bibinfo  {journal}
  {J.\ Chem.\ Phys.}\ }\textbf {\bibinfo {volume} {70}},\ \bibinfo {pages}
  {3812} (\bibinfo {year} {1979})}\BibitemShut {NoStop}%
\bibitem [{\citenamefont {Duan}\ \emph {et~al.}(1995)\citenamefont {Duan},
  \citenamefont {Yuan},\ and\ \citenamefont {Bao}}]{DuanYuanBao1995}%
  \BibitemOpen
  \bibfield  {author} {\bibinfo {author} {\bibfnamefont {Y.}~\bibnamefont
  {Duan}}, \bibinfo {author} {\bibfnamefont {J.-M.}\ \bibnamefont {Yuan}},\
  and\ \bibinfo {author} {\bibfnamefont {C.}~\bibnamefont {Bao}},\ }\bibfield
  {title} {\bibinfo {title} {Periodic orbits of the hydrogen molecular ion and
  their quantization},\ }\href {https://doi.org/10.1103/PhysRevA.52.3497}
  {\bibfield  {journal} {\bibinfo  {journal} {Phys.\ Rev.\ A}\ }\textbf
  {\bibinfo {volume} {52}},\ \bibinfo {pages} {3497} (\bibinfo {year}
  {1995})}\BibitemShut {NoStop}%
\bibitem [{\citenamefont {Duan}\ and\ \citenamefont
  {Yuan}(1999)}]{DuanYuan1999}%
  \BibitemOpen
  \bibfield  {author} {\bibinfo {author} {\bibfnamefont {Y.}~\bibnamefont
  {Duan}}\ and\ \bibinfo {author} {\bibfnamefont {J.-M.}\ \bibnamefont
  {Yuan}},\ }\bibfield  {title} {\bibinfo {title} {Periodic orbits of the
  hydrogen molecular ion},\ }\href {https://doi.org/10.1007/s100530050315}
  {\bibfield  {journal} {\bibinfo  {journal} {Eur.\ Phys.\ J.\ D}\ }\textbf
  {\bibinfo {volume} {6}},\ \bibinfo {pages} {319} (\bibinfo {year}
  {1999})}\BibitemShut {NoStop}%
\bibitem [{\citenamefont {Ezra}\ \emph {et~al.}(1991)\citenamefont {Ezra},
  \citenamefont {Richter}, \citenamefont {Tanner},\ and\ \citenamefont
  {Wintgen}}]{Ezra1991}%
  \BibitemOpen
  \bibfield  {author} {\bibinfo {author} {\bibfnamefont {G.~S.}\ \bibnamefont
  {Ezra}}, \bibinfo {author} {\bibfnamefont {K.}~\bibnamefont {Richter}},
  \bibinfo {author} {\bibfnamefont {G.}~\bibnamefont {Tanner}},\ and\ \bibinfo
  {author} {\bibfnamefont {D.}~\bibnamefont {Wintgen}},\ }\bibfield  {title}
  {\bibinfo {title} {Semiclassical cycle expansion for the helium atom},\
  }\href {https://doi.org/10.1088/0953-4075/24/17/001} {\bibfield  {journal}
  {\bibinfo  {journal} {J.\ Phys.\ B: At.\ Mol.\ Opt.\ Phys.}\ }\textbf
  {\bibinfo {volume} {24}},\ \bibinfo {pages} {L413} (\bibinfo {year}
  {1991})}\BibitemShut {NoStop}%
\bibitem [{\citenamefont {Wintgen}\ \emph {et~al.}(1992)\citenamefont
  {Wintgen}, \citenamefont {Richter},\ and\ \citenamefont
  {Tanner}}]{WintgenRichterTanner1992}%
  \BibitemOpen
  \bibfield  {author} {\bibinfo {author} {\bibfnamefont {D.}~\bibnamefont
  {Wintgen}}, \bibinfo {author} {\bibfnamefont {K.}~\bibnamefont {Richter}},\
  and\ \bibinfo {author} {\bibfnamefont {G.}~\bibnamefont {Tanner}},\
  }\bibfield  {title} {\bibinfo {title} {The semiclassical helium atom},\
  }\href {https://doi.org/10.1063/1.165920} {\bibfield  {journal} {\bibinfo
  {journal} {Chaos}\ }\textbf {\bibinfo {volume} {2}},\ \bibinfo {pages} {19}
  (\bibinfo {year} {1992})}\BibitemShut {NoStop}%
\bibitem [{\citenamefont {Carter}(1968)}]{Carter:1968rr}%
  \BibitemOpen
  \bibfield  {author} {\bibinfo {author} {\bibfnamefont {B.}~\bibnamefont
  {Carter}},\ }\bibfield  {title} {\bibinfo {title} {{Global structure of the
  Kerr family of gravitational fields}},\ }\href
  {https://doi.org/10.1103/PhysRev.174.1559} {\bibfield  {journal} {\bibinfo
  {journal} {Phys.\ Rev.}\ }\textbf {\bibinfo {volume} {174}},\ \bibinfo
  {pages} {1559} (\bibinfo {year} {1968})}\BibitemShut {NoStop}%
\bibitem [{\citenamefont {Berry}\ and\ \citenamefont
  {Tabor}(1976)}]{BerryTabor1976}%
  \BibitemOpen
  \bibfield  {author} {\bibinfo {author} {\bibfnamefont {M.~V.}\ \bibnamefont
  {Berry}}\ and\ \bibinfo {author} {\bibfnamefont {M.}~\bibnamefont {Tabor}},\
  }\bibfield  {title} {\bibinfo {title} {Closed orbits and the regular bound
  spectrum},\ }\href {https://doi.org/10.1098/rspa.1976.0062} {\bibfield
  {journal} {\bibinfo  {journal} {Proc.\ Roy.\ Soc.\ Lond.\ A.}\ }\textbf
  {\bibinfo {volume} {349}},\ \bibinfo {pages} {101} (\bibinfo {year}
  {1976})}\BibitemShut {NoStop}%
\bibitem [{\citenamefont {Berry}\ and\ \citenamefont
  {Tabor}(1977)}]{BerryTabor1977}%
  \BibitemOpen
  \bibfield  {author} {\bibinfo {author} {\bibfnamefont {M.~V.}\ \bibnamefont
  {Berry}}\ and\ \bibinfo {author} {\bibfnamefont {M.}~\bibnamefont {Tabor}},\
  }\bibfield  {title} {\bibinfo {title} {Calculating the bound spectrum by path
  summation in action-angle variables},\ }\href
  {https://doi.org/10.1088/0305-4470/10/3/009} {\bibfield  {journal} {\bibinfo
  {journal} {J.\ Phys.\ A: Math.\ Gen.}\ }\textbf {\bibinfo {volume} {10}},\
  \bibinfo {pages} {371} (\bibinfo {year} {1977})}\BibitemShut {NoStop}%
\bibitem [{\citenamefont {Poincar\'e}(1887)}]{Poincare1887}%
  \BibitemOpen
  \bibfield  {author} {\bibinfo {author} {\bibfnamefont {H.}~\bibnamefont
  {Poincar\'e}},\ }\bibfield  {title} {\bibinfo {title} {{Sur les r\'esidus des
  int\'egrales doubles}},\ }\href {https://doi.org/10.1007/BF02406742}
  {\bibfield  {journal} {\bibinfo  {journal} {Acta Math.}\ }\textbf {\bibinfo
  {volume} {9}},\ \bibinfo {pages} {321} (\bibinfo {year} {1887})}\BibitemShut
  {NoStop}%
\bibitem [{\citenamefont {Esposito}(2020)}]{Esposito:2020prr}%
  \BibitemOpen
  \bibfield  {author} {\bibinfo {author} {\bibfnamefont {G.}~\bibnamefont
  {Esposito}},\ }\bibfield  {title} {\bibinfo {title} {{Contour and surface
  integrals in potential scattering}},\ }\href
  {https://doi.org/10.1140/epjp/s13360-020-00514-5} {\bibfield  {journal}
  {\bibinfo  {journal} {Eur.\ Phys.\ J.\ Plus}\ }\textbf {\bibinfo {volume}
  {135}},\ \bibinfo {pages} {501} (\bibinfo {year} {2020})},\ \Eprint
  {https://arxiv.org/abs/2001.11217} {arXiv:2001.11217 [hep-th]} \BibitemShut
  {NoStop}%
\end{thebibliography}%

\end{document}